# Personalizable Models of Cardiac Elasticity with Physics-informed Symbolic Regression

Sophia Ohnemus[1,2,3], Kristin Fullerton[4], Leto L. Riebel[5], Mary M. Maleckar[6,7], Andrew D. McCulloch[9], Viviane Timmermann[†3,8], Gabriel Balaban[†6,10*]

[1]Institute for Experimental Cardiovascular Medicine, University Heart Center Freiburg – Bad Krozingen, Medical Faculty and Medical Center – University of Freiburg, Freiburg im Breisgau, Germany.
[2]Speemann Graduate School of Biology and Medicine, University of Freiburg, Freiburg im Breisgau, Germany.
[3]Faculty of Mathematics and Physics, University of Freiburg, Freiburg im Breisgau, Germany.
[4]Physiology, Biophysics, and Systems Biology Program, Weill Cornell Graduate School of Medical Sciences, New York, New York, United States.
[5]Department of Computer Science, University of Oxford, Oxford, United Kingdom.
[6]Department of Computational Physiology, Simula Research Laboratory, Oslo, Norway.
[7]Tulane University, School of Medicine, New Orleans, United States of America.
[8]Faculty of Medicine, University of Freiburg, Freiburg im Breisgau, Germany.
[9]Institute of Engineering in Medicine, University of California San Diego, La Jolla, California, United States of America.
[10]Faculty of Health Sciences and Technology, Kristiania University of Applied Sciences, Oslo, Norway.

*Corresponding author(s). E-mail(s): gabriel.balaban@kristiania.no;

† Viviane Timmermann and Gabriel Balaban contributed jointly as co-last authors.

Cardiac digital twins hold great promise for personalized medicine, but they currently depend on complex constitutive models of tissue mechanics that are often over-parameterized for the clinical context. To address this we introduce CHESRA (Cardiac Hyperelastic Evolutionary Symbolic Regression Algorithm), a penalized physics-informed machine learning framework that automatically derives simple strain energy functions from multiple experimental data sources. Using a normalizing loss function, CHESRA identified two new functions with only three and four parameters, respectively. These functions achieved high data fitting accuracy in experimental scenarios while enabling more consistent parameter estimation than state-of-the-art approaches, both in tissue benchmarks and 3D biventricular simulations. By combining biophysical constraints with data-driven discovery, CHESRA demonstrates how physics-informed learning can generate simple, personalizable models with potential application in cardiac digital twins and clinical decision-making.

## Introduction

The increasing amounts of healthcare data and computational power have accelerated the development of personalized computational models in cardiovascular medicine, with the cardiac digital twin emerging as one of the leading concepts [1, 2]. Similar to personalized simulation models, a cardiac digital twin is a digital version of a patient's heart that enables the testing and optimization of medical treatments in silico, before they are applied to the patient in the physical world [3]. However, the digital twin adds a dynamic data update component [4, 5] where the digital twin is "dynamically and continuously updated with data from its physical twin" [6]. The dynamic data updates are useful for patient monitoring [7], and the assessment of time-varying risk factors to guide the timing of medical interventions. For example, Kolk et al. trained a neural network that updates its prediction of malignant ventricular arrhythmias as new electrocardiograms from an implanted defibrillator become available [8]. A follow-up study extended the technique to wearable accelerometer data collected continuously over a year [9]. While dynamic data updates have been implemented in machine learning frameworks [8, 9], turning a physics-based 3D heart simulation model into a dynamically updated digital twin is an ongoing challenge. A key barrier is the need for suitable model updating procedures that can handle frequent data updates while maintaining a stable and faithful representation of the physical heart [4, 7].

Technically, the updating of 3D physics-based simulation models often takes the form of mathematical inverse problems, in which the models's internal parameters are updated from the available data [4]. Thus, a good 3D physics-based cardiac digital twin should meet a dual set of requirements: being complex enough to *accurately* model a patient's cardiac physiology (low bias), while also being *uniquely parameterizable* (low variance) to reflect a particular patient's physiology. However, the bias-variance trade-off from machine learning tells us that the two requirements are in opposition [10]. Increasing digital twin complexity can improve physical accuracy (reducing bias), at the cost of higher parameter variance, which means that the digital twin becomes more sensitive to small fluctuations in the data and hence more difficult to personalize. Conversely, simpler digital twins with lower parameter variance may have higher bias, and thereby miss key biophysical features needed for accurate medical decision-making. This issue is particularly salient for heart tissue elasticity models, which are key components of the personalized cardiac simulation frameworks [11, 12] that enable cardiac digital twins, and are crucial to understanding heart disease states, such as fibrosis [11], myocardial infarction [12], and heart failure with preserved ejection fraction [13].

Traditionally, resting myocardial elasticity has been modeled using hyperelastic strain energy functions (SEFs) [14–17] that quantify the amount of energy stored during a deformation. The data that inform these SEFs [18–21] have shown that healthy heart tissue possesses a sophisticated microstructure, characterized by interconnected cells with a locally prevailing orientation, typically modelled as "fibres", as well as cell bundles organized into sheets. These microstructural features cause orthotropic and nonlinear elastic behavior at the tissue scale [14]. Consequently, human-expert designed SEFs are quite complex [14–17], consisting of seven to twelve tunable material parameters that are connected to strain measures via nonlinear formulations involving exponential functions. Unfortunately, the same complexity that enables the expert-designed SEFs to closely fit experimental data has proven to be problematic in cardiac simulation studies, where the available patient data is typically noisy and sparse. Indeed, previous studies have reported difficulties in uniquely identifying cardiac SEF parameters due to parameter correlations [22] or high parameter variance [12]. This has led to simplified representations of cardiac elasticity being used in medical applications; including transversely isotropic SEFs [12, 23], or orthotropic SEFs, for which only a reduced set of parameters is fitted [24–28]. While such ad-hoc model simplifications increase the feasibility of parameter identification, the utility of the resulting models for reproducing measured data is an open question. Furthermore, fitting only a reduced set of a SEF's parameters in a digital twin setting involves ethical dilemmas. Any choice of default values for the remaining parameters risks creating systematic bias toward particular patient groups.

Inspired by the shortcomings of human expert-designed SEFs, researchers have begun to develop machine learning methods that automatically design cardiac SEFs directly from experimental data [29–31]; including spline methods [29], and constitutive neural networks [30, 31]. SEFs designed using splines and constitutive neural networks have achieved excellent model-data fits [29–32], highlighting the accuracy of data-driven cardiac SEFs for fitting experimental data. Also, in contrast to the human expert-designed cardiac SEFs [14–17], the complexity of data-driven SEFs can be optimized via penalized regression to learn simple models with relatively few parameters [30, 31]. While promising, these initial results lack validation in external experimental datasets, leaving open the question of model generalization. A further issue relates to the form of the data-derived SEFs, as the SEFs designed by constitutive neural networks follow a prescribed additive form [30, 31]. Accordingly, the intriguing possibility of designing even simpler SEFs by exploring a broader search space remains unresolved. Finally, previous data-driven SEF studies [29–31] are missing systematic evaluations of how well the data-driven SEF parameters can be estimated in 3D organ-level models, which is crucial for the personalization of cardiac digital twins.

In this study we introduce the Cardiac Hyperelastic Evolutionary Symbolic Regression Algorithm (CHESRA), a novel physics informed machine learning (PI-ML) framework for automatically designing simple hyperelastic SEFs. Important for CHESRA's approach is an innovative normalizing loss function that enables CHESRA to simultaneously learn from multiple experimental datasets, while maintaining model parsimony via the use of a function length penalty. Unlike previous automated discovery approaches [30, 31] that prioritize fitting single, high-fidelity datasets, the CHESRA framework is designed to discover parsimonious SEFs from heterogeneous experiments. This multi-source training is intended to serve as a form of regularization, reducing the effect of lab-specific artifacts and helping identify mathematical forms with greater generalizability, which we consider a prerequisite for developing reliable cardiac digital twins in diverse patient populations. Using CHESRA, we identified two novel simple polynomial cardiac SEFs ($\psi_{\mathrm{CH}_1}, \psi_{\mathrm{CH}_2}$) that fit data from four experimental studies [18–21] with high accuracy, while using fewer free parameters than state-of-the-art orthotropic SEFs [14–17, 30]. We also confirmed the advantage of $\psi_{\mathrm{CH}_1}, \psi_{\mathrm{CH}_2}$ for parameter estimation in an experimental tissue-data benchmark, and for $\psi_{\mathrm{CH}_1}$ in a biventricular simulation benchmark based on cardiac magnetic resonance imaging (MRI). Overall, our results show that CHESRA can be used to automatically design simple SEFs from experimental data and that CHESRA's SEF $\psi_{\mathrm{CH}_1}$ improves parameter estimation in organ-scale simulations. This has implications for the field of healthcare digital twins as a whole, paving the way for digital twin models that are designed a-priori to satisfy an optimal balance between accuracy and personalizability.

# Results

## Evolutionary symbolic regression algorithm CHESRA

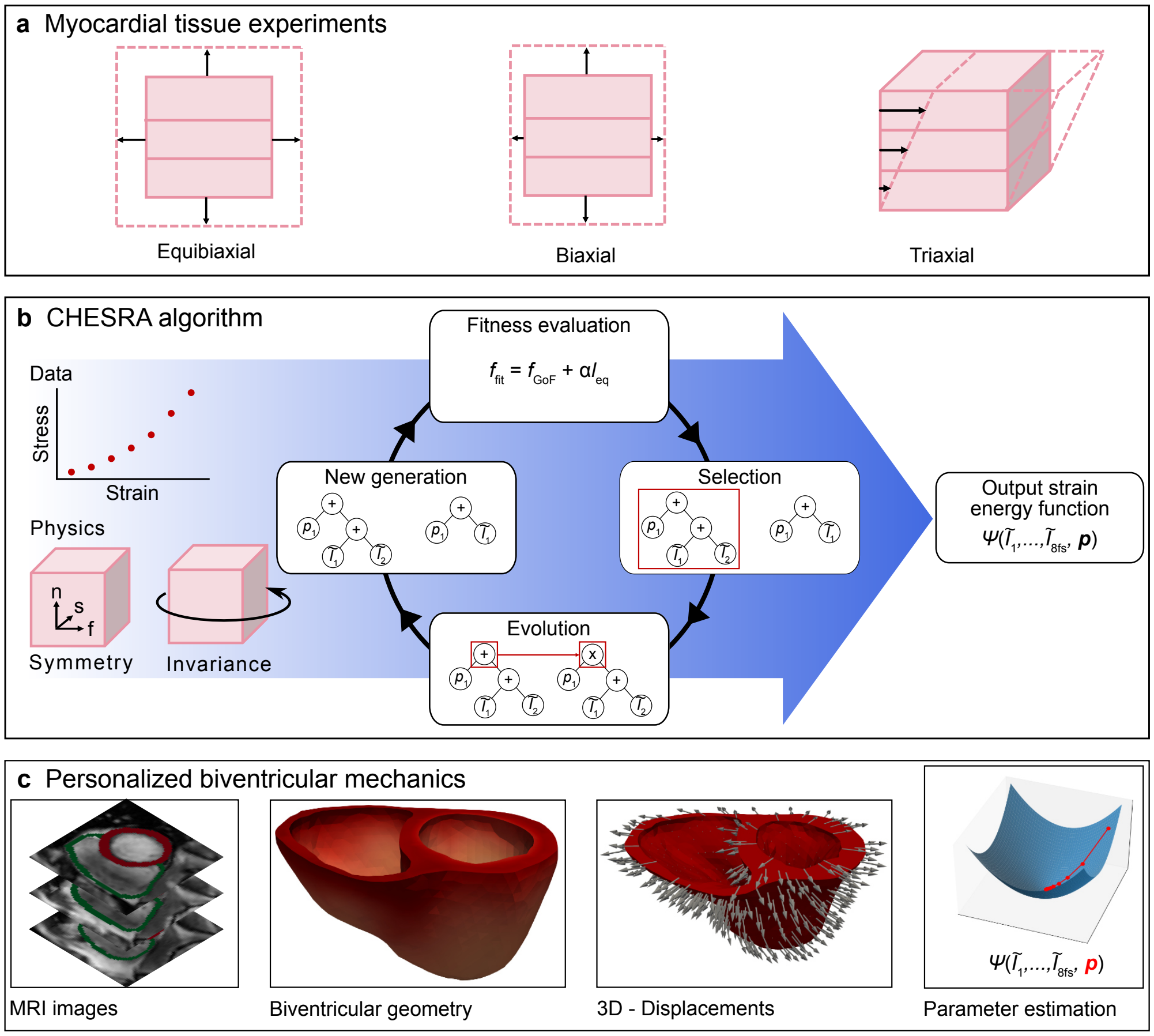

**Fig. 1: Overview of the Cardiac Hyperelastic Evolutionary Symbolic Regression Algorithm (CHESRA) and its validation in a personalized biventricular mechanics framework. a** CHESRA takes as input experimental datasets of myocardial mechanics with *equibiaxial*, *biaxial*, and *triaxial* deformation protocols. **b** CHESRA combines the mechanical data with the physical principle of material symmetry in the orthotropic fibre (f), sheet (s) and sheet-normal (n) system with frame invariance to design a parsimonious strain energy function ($\psi$). The strain energy function is designed using function trees with symbols representing material parameters $\boldsymbol{p} = (p_1, p_2, \dots)$ and strain invariants $(\tilde{I}_1, \tilde{I}_2, \dots, \tilde{I}_{8\text{fs}})$. During each CHESRA iteration, functions are selected based on a *fitness* criterion ($f_{\text{fit}}$) consisting of function length ($l_{\text{eq}}$) and goodness of fit ($f_{\text{GoF}}$) balanced by the length penalty ($\alpha$). The selected functions are then *evolved*, i.e. mated, mutated, reduced, and extended to form a *new generation* of functions. After a pre-set number of generations, CHESRA stops and outputs the optimal strain energy function. **c** Personalized cardiac biventricular mechanics pipeline to validate the computational efficacy of the optimal strain energy functions, involving an MRI derived biventricular geometry and the estimation of material parameters $\boldsymbol{p}$ using 3D synthetic displacement data to assess parameter identifiability. Note that the MRI shown is a representative example, author's own scan, used for illustration only.

CHESRA is a physics-informed evolutionary framework that manipulates symbolic representations of cardiac SEFs to fit experimental observations (Figure 1) while minimizing SEF complexity [33]. CHESRA takes one or more experimental datasets of myocardial stress-strain relations as input (Supplementary Table S1) and evolves a population of SEFs according to the fitness function

$$f_{\text{fit}} = f_{\text{GoF}}(\boldsymbol{p}) + \alpha l_{\text{eq}}. \tag{1}$$

Here, $f_{\text{GoF}}$ quantifies the goodness of fit of each SEF's stress-strain relation to the experimental data, with each SEF's parameters $\boldsymbol{p}$ optimized via weighted least squares (Equation 8 in Methods). Throughout the optimization, the physical principles of frame-invariance and material symmetry are automatically incorporated into CHESRA's SEFs via the use of strain invariant functions (see Methods *Invariant-based framework for designing elastic energy functions*). The term $\alpha$ penalizes the function length $l_{\text{eq}}$, favoring simpler SEFs with fewer parameters, fewer mechanical invariants, and simpler forms. A low fitness value therefore indicates a simple function with a good fit, with the relative weight of the two influenced by $\alpha$. Evolution in CHESRA occurs via mating, mutation, extension, or reduction, with probabilities $p_{\text{mate}}$, $p_{\text{mutate}}$, $p_{\text{reduce}}$, and $p_{\text{extend}}$ (detailed in Methods *Evolutionary changes* and Supplementary Table S2). To handle multiple datasets, CHESRA utilizes a standardized sum of squared residuals term for $f_{\text{GoF}}$ (Equation 10 in Methods), where each dataset's error is normalized by its squared deviations from the mean. This allows CHESRA to compare errors across datasets and evolve SEFs adaptable to diverse data.

## CHESRA's penalty controls strain energy function complexity

We conducted a systematic hyperparameter test using the Dokos [21] shear dataset (Supplementary Table S1) to evaluate the efficacy of the length penalty $\alpha$ and identify optimal genetic hyperparameters ($p_{\text{mate}}$, $p_{\text{mutate}}$, $p_{\text{reduce}}$, $p_{\text{extend}}$) for further experiments. This was done in two steps. First, we tested eight $\alpha$ values (0.0, $10^{-4}$, $2\times10^{-4}$, $3\times10^{-4}$, $5\times10^{-4}$, $10^{-3}$, $5\times10^{-3}$, $10^{-2}$) and set the genetic hyperparameters to values from the set $(0, 0.25, 0.5, 0.75)$. This resulted in $4^4 \times 8 = 2048$ hyperparameter combinations. Pre-set values were used for other CHESRA hyperparameters (Supplementary Table S2). The results, plotted in Figure 2a, show a clear trend: increasing $\alpha$ reduced SEF complexity ($l_{\text{eq}}$) but worsened the goodness of fit $f_{\text{GoF}}$. For $\alpha$ between $1\times10^{-4}$ and $1\times10^{-3}$, SEFs clustered in the lower left corner of the plot, with near-optimal $f_{\text{GoF}} \leq 0.01$ and modest complexity ($l_{\text{eq}} \leq 30$). In contrast, $\alpha \geq 5\times10^{-3}$ yielded simple SEFs ($l_{\text{eq}} \leq 10$) but poorer fits ($f_{\text{GoF}} \geq 0.01$), while $\alpha = 0$ (no penalty) produced more complex SEFs (median $l_{\text{eq}} = 35$).

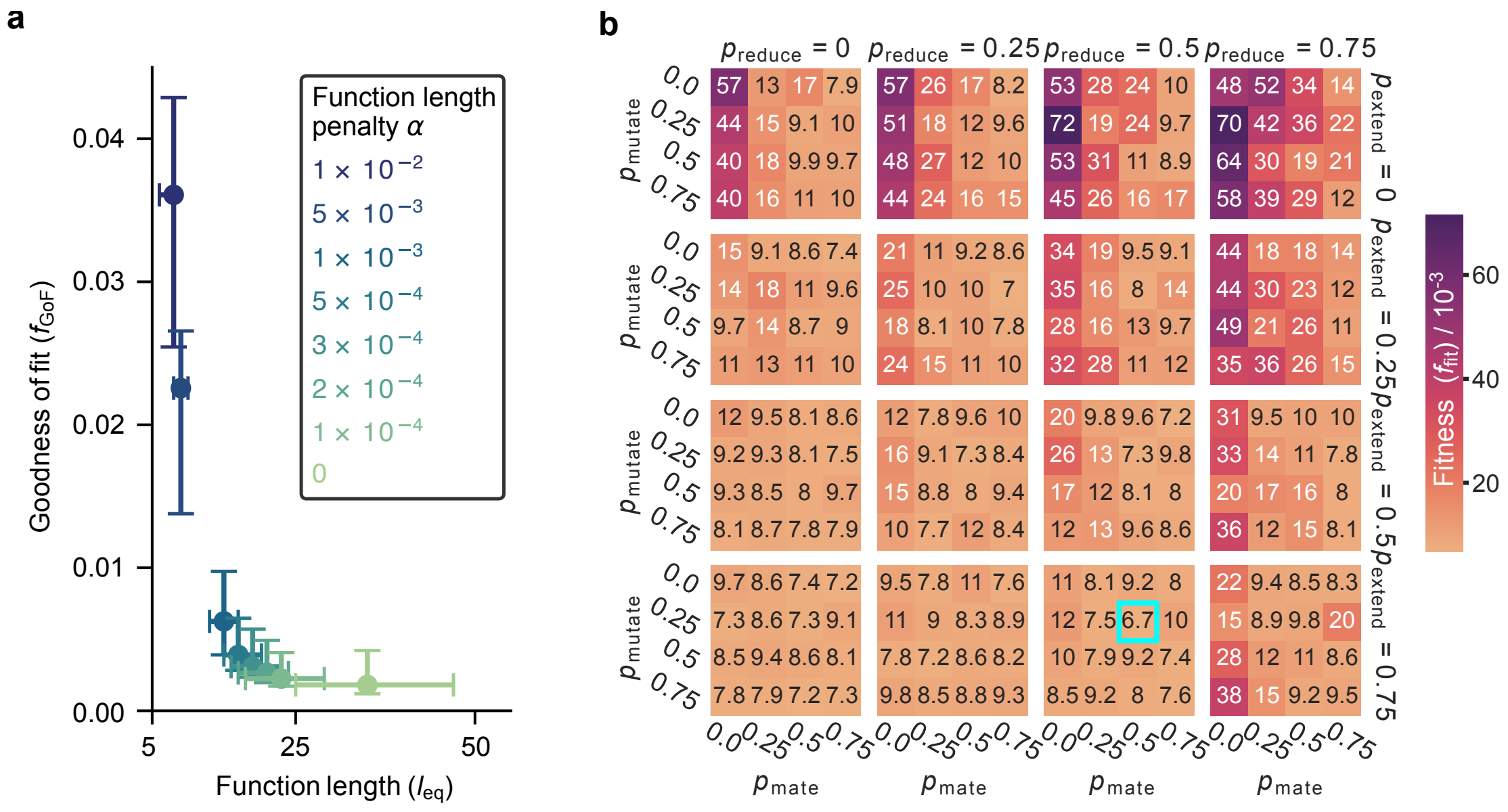

**Fig. 2: Effect of systematic CHESRA hyperparameter variation on strain energy functions generated using the Dokos [21] shear dataset.** **a** Distribution of goodness of fit ($f_{\text{GoF}}$) and function length ($l_{\text{eq}}$) resulting from 2048 genetic hyperparameter and function length penalty ($\alpha$) combinations. Circles indicate the median, and error bars show the interquartile range. **b** Heatmap of average fitness scores ($f_{\text{fit}} = f_{\text{GoF}} + \alpha l_{\text{eq}}$) for 256 combinations of the genetic parameters ($p_{\text{mate}}$, $p_{\text{mutate}}$, $p_{\text{reduce}}$, $p_{\text{extend}}$) with fixed function length penalty $\alpha = 2\times10^{-4}$. Larger fitness values indicate worse agreement with the data or larger functions. The cyan box highlights the best average fitness.

Next, we identified optimal genetic hyperparameter combinations with $\alpha$ fixed at $2\times10^{-4}$, chosen from the lower-left corner of Figure 2a to balance SEF complexity and goodness of fit. We ran CHESRA three times for each of the $4^4 = 256$ genetic parameter combinations and recorded the average $f_{\text{GoF}}$ (Figure 2b). Combinations with $p_{\text{reduce}} < p_{\text{extend}}$ (lower left triangle of the heatmap) and $p_{\text{mate}} > 0$ (three rightmost columns in each subplot) tended to improve fitness. The best average fitness was achieved with ($p_{\text{mate}} = 0.5, p_{\text{mutate}} = 0.25, p_{\text{reduce}} = 0.5, p_{\text{extend}} = 0.75$), which we used in subsequent experiments. These results demonstrate that optimizing CHESRA's genetic hyperparameters improves SEF fitness and that a well-chosen $\alpha$ can yield excellent model-data fits while effectively limiting function complexity.

## Cross-validation tests with CHESRA's strain energy functions

To assess the generalizability of CHESRA's SEFs, we conducted cross-validation tests under two distinct scenarios: (1) a *leave-one-out* setup, where the data were partitioned into five training sets and one held-out test set, and (2) a *single-fit* setup, employing one training set and five test sets. Across these scenarios, we evaluated 24 experimental configurations, spanning all six possible training and test data variations of datasets from Supplementary Table S1 and four length penalty values $\alpha \in \{0, 10^{-4}, 10^{-3}, 10^{-2}\}$. For each cross-validation fold, we ran CHESRA five times, with the best-performing SEF obtained from the training data used for goodness of fit calculations with the test data.

First, we compared the out-of-distribution performance of leave-one-out and single-fit SEFs by computing the median $f_{\text{GoF}}$ across withheld datasets (Figure 3a). For $\alpha = 0$, median errors were negligible ($f_{\text{GoF}} < 0.01$) in both cases. However, at $\alpha = 10^{-2}$, single-fit SEFs exhibited substantially higher errors ($f_{\text{GoF}} = 0.53$), whereas leave-one-out SEFs maintained strong performance ($f_{\text{GoF}} = 0.021$). We further evaluated SEF generalizability by comparing leave-one-out SEFs to single-fit SEFs, where the latter were trained and tested on the same dataset (Figure 3b–d). SEFs derived under weak length penalties ($\alpha \leq 10^{-4}$) achieved superior fits ($f_{\text{GoF}} \leq 0.026$) compared to those with stronger penalties ($\alpha \geq 10^{-3}$), a trend supported by visual inspection of the fitted curves (Figure 3c, d; Supplementary Figure S1). Nevertheless, the gap between single-fit and leave-one-out $f_{\text{GoF}}$ was reduced for the largest length penalty $\alpha = 10^{-2}$, indicating that shorter SEFs were less specialized to their particular training datasets. Also, the Yin Biaxial dataset [18] exhibited the highest fitting errors, likely due to missing low-strain data [14]. Overall, the results suggest that the relatively longer SEFs with many free parameters could generalize effectively and that the shorter SEFs with fewer free parameters ($\alpha = 10^{-2}$) could also generalize if multiple datasets were used to derive them.

The functional forms of CHESRA's SEFs are detailed in Supplementary Table S3. For $\alpha \geq 10^{-3}$, leave-one-out SEFs exhibited greater consistency in structure compared to single-fit SEFs, reflecting CHESRA's robustness when trained on diverse datasets. This is further supported by the convergence of leave-one-out SEF lengths around the mean length for $\alpha > 0$ (Figure 3b). Additionally, while single-fit SEFs occasionally incorporated exponential terms, leave-one-out SEFs with $\alpha \geq 10^{-3}$ did not, underscoring their structural consistency. Taken together, these findings demonstrate that CHESRA can produce simple yet generalizable SEFs when optimized with an appropriate length penalty $\alpha$ and sufficient training data.

## CHESRA generates simple strain energy functions

In order to fully exploit the information contained within the experimental data (Supplementary Table S1) we fed all of the datasets collectively into CHESRA and set the function length penalty to a range of values (Section 10). For every value of the penalty, we ran the CHESRA three times, each time starting with a different randomly generated set of functions. For each set of runs, we recorded the SEFs that achieved the best fitness score (Supplementary Table S4). As expected, we observed that increases in the function length penalty $\alpha$ led to shorter and simpler functions, with a corresponding rise in fitting error. Surprisingly, the decreases in function length were quite substantial, as the function length of 24 at $\alpha = 0$ was reduced to 7 at $\alpha = 10^{-3}$. Furthermore, for $\alpha \leq 5\times10^{-3}$ the models fit closely to the data, with only minor discrepancies shown in Figure 4 and Supplementary Figure S2. For $\alpha \leq 5\times10^{-3}$ the output SEFs contained three or more mechanical invariants, which meant that they could support two independent axes of anisotropy, suitable for modeling orthotropic cardiac tissue. Of these SEFs, we selected the two most parsimonious for further testing, corresponding to $\alpha = 5\times10^{-3}$ and $\alpha = 10^{-3}$. We analyzed the expanded form of these two functions (Supplementary Section *Interpretation of the material parameters in the CHESRA models*) and conducted a parameter sensitivity study of each parameter in uniaxial, biaxial and simple shear loading (Supplementary Figure S3). Based on

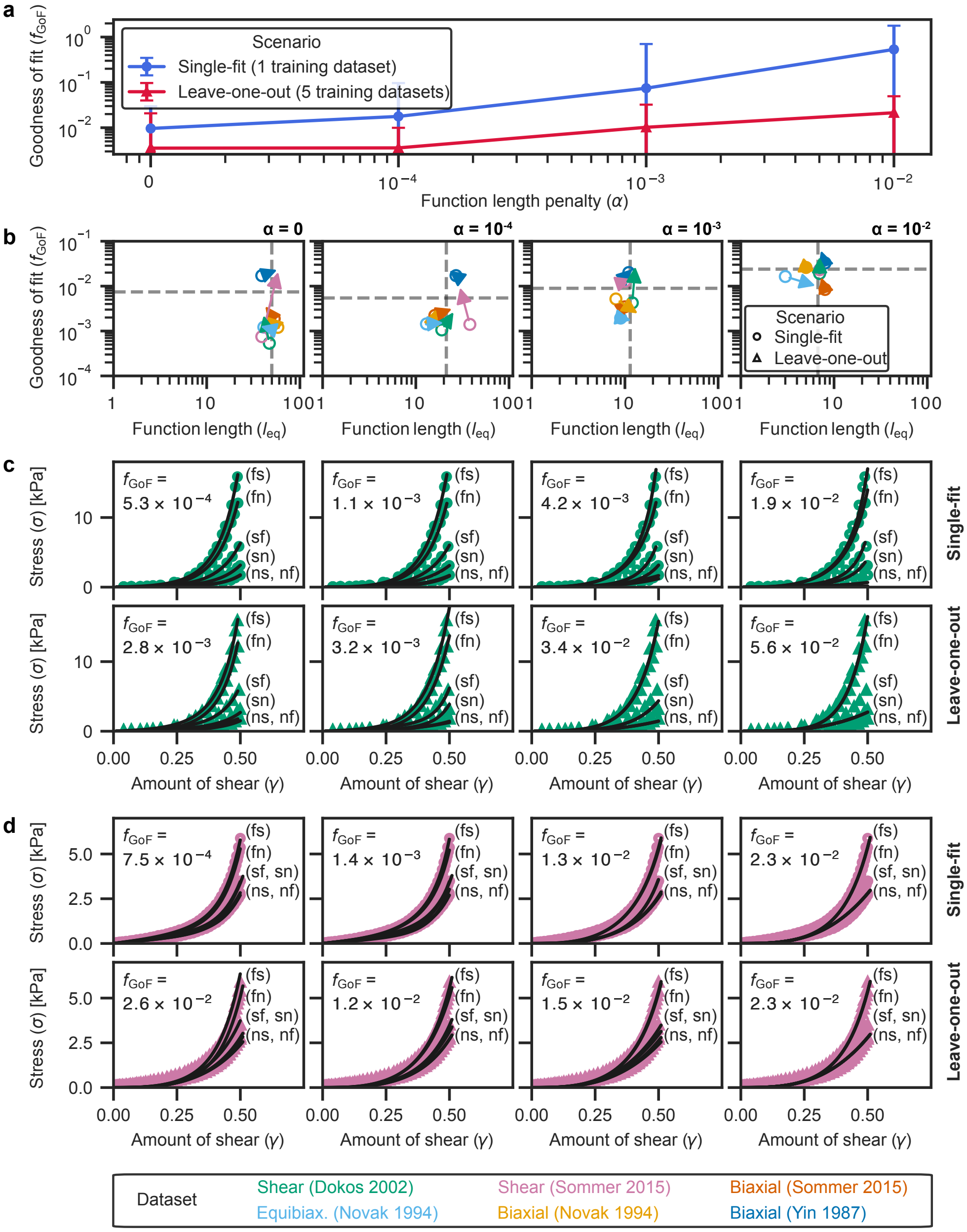

**Fig. 3: Comparison of strain energy function fits to experimental datasets in the cross validated leave-one-out and single-fit scenarios. a** Function length penalty ($\alpha$) versus goodness of fit ($f_{\mathrm{GoF}}$) evaluated on withheld data. Markers represent the median and the error bars the interquartile ranges resulting from five CHESRA runs per length penalty. **b** Function length ($l_{\mathrm{eq}}$) versus goodness of fit evaluated on the single-fit training data and leave-one-out test data. Dashed lines indicate median values. Here, the goodness of fit for single fit functions is evaluated on the training data to help assess the generalizability gap. **c** Model (black lines) vs data (markers) fits for the Dokos [21] shear data. **d** Model (black lines) vs data (markers) fits for the Sommer [20] shear data. $\sigma$ = Cauchy stress, $\gamma$ = simple shear angle. The labels (fs, fn, sf, sn, ns, nf) indicate the direction of shear, with the first coordinate indicating the normal of the tissue cube face that is translated by the shear, and the second the shearing direction in which that face is shifted. Directions (f) = fibre, (s) = sheet, (n) = sheet-normal.

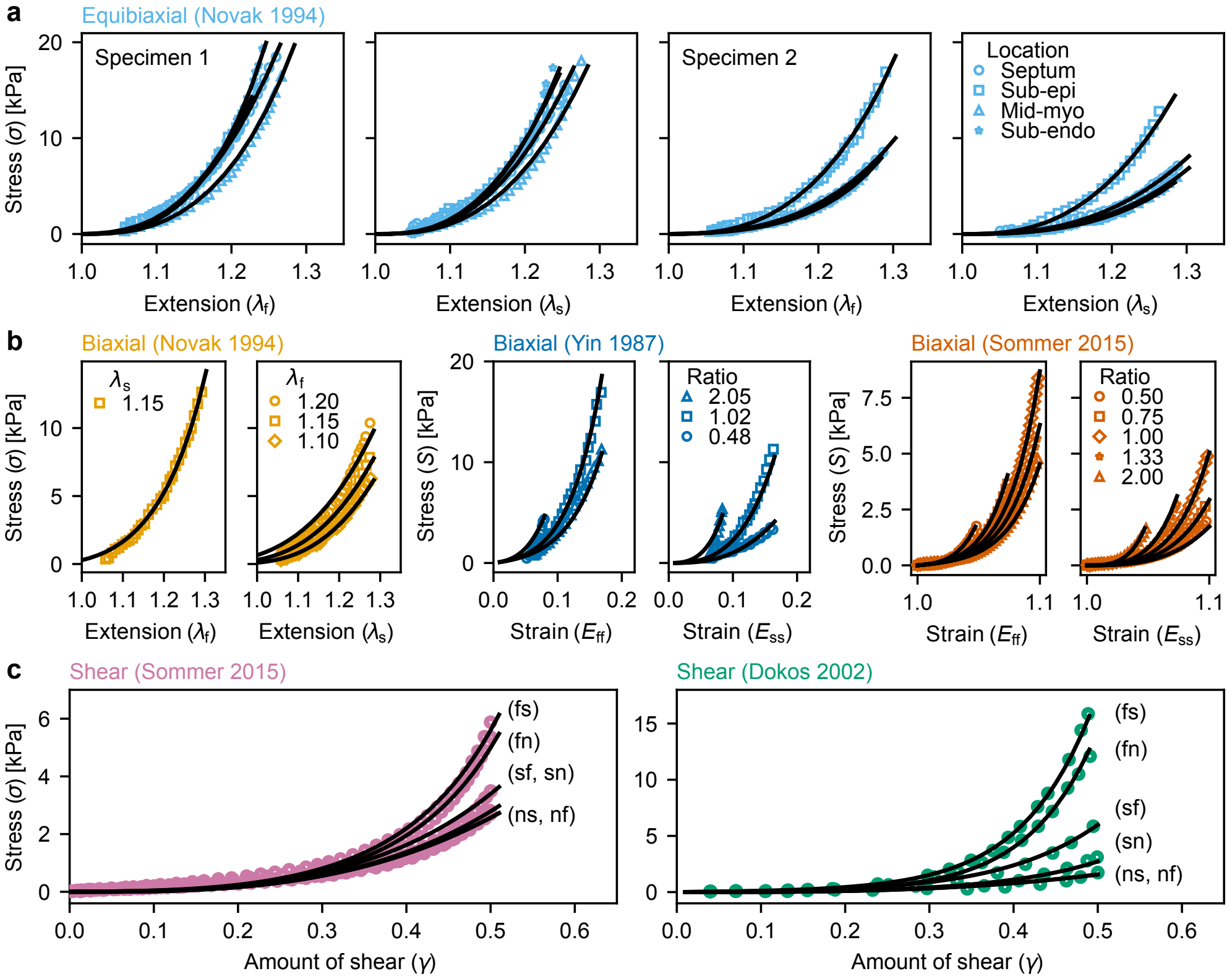

**Fig. 4: Model-data fits with the first CHESRA strain energy function $\psi_{\mathrm{CH}_1}$.** **a** Equibiaxial stretch [19], **b** biaxial stretch [18–20], and **c** triaxial shear [20, 21] data. Solid lines represent model results whereas markers indicate experimental data. For the Dokos shear [21] dataset, the (nf) model curve is located close to the (sn) data, a discrepancy which is not apparent in the visualization, but which helps explain why the fitting error is larger for $\psi_{\mathrm{CH}_1}$ than $\psi_{\mathrm{CH}_2}$. $\sigma$ = Cauchy stress, $S$ = Second Piola-Kirchhoff stress, $E$ = Green-Lagrange strain, $\lambda$ = extension ratio, $\gamma$ = simple shear angle. The labels (fs, fn, sf, sn, ns, nf) indicate the direction of shear, with the first coordinate indicating the normal of the tissue cube face that is translated by the shear, and the second the shearing direction in which that face is shifted. Directions (f) = fibre, (s) = sheet, (n) = sheet-normal.

these results we chose suitable names for the SEFs' parameters. The two SEFs were

$$\begin{aligned}
\psi_{\mathrm{CH}_1} &= \left(p_{\text{f-fs}} + \tilde{I}_1\right)\left(p_{\text{iso}} + p_{\text{coup}}\left[\tilde{I}_{8\text{fs}} + \tilde{I}_{5\text{f}}\right]\right) - p_{\text{f-fs}}\,p_{\text{iso}}, \\
\psi_{\mathrm{CH}_2} &= p_{\text{glob}}\left(p_{\text{s-iso}} + \tilde{I}_{5\text{f}}\right)\left(p_{\text{f-fs}} + \tilde{I}_1\right)\left(p_{\text{f-io}} + \tilde{I}_{5\text{s}}\right) - p_{\text{glob}}\,p_{\text{s-iso}}\,p_{\text{f-s}}\,p_{\text{f-iso}},
\end{aligned} \tag{2}$$

and are shown here with normalization terms for the sake of completeness. The SEFs $\psi_{\mathrm{CH}_1}$ and $\psi_{\mathrm{CH}_2}$ are notably simpler than the state-of-the-art SEFs listed in Methods *State-of-the-art strain energy functions*. Unlike those more complex SEFs, $\psi_{\mathrm{CH}_1}$ and $\psi_{\mathrm{CH}_2}$ include no exponential terms and use fewer invariants (Supplementary Table S5). Specifically, $\psi_{\mathrm{CH}_1}$ relies on just three invariants and three material parameters, the minimum required for a fully adjustable orthotropic SEF, while $\psi_{\mathrm{CH}_2}$ introduces only one additional parameter. In contrast, widely used SEFs such as the Holzapfel-Ogden model $\psi_{\mathrm{HO}}$ typically involve five or more parameters and additional complexity (Methods *State-of-the-art strain energy functions*). Most importantly, $\psi_{\mathrm{CH}_1}$ and $\psi_{\mathrm{CH}_2}$ were locally optimal, as removing any single term from their symbolic forms resulted in a worsening of their goodness of fit (Supplementary Table S6).

Though the material parameters of $\psi_{\mathrm{CH}_1}$ and $\psi_{\mathrm{CH}_2}$ were algorithmically designed without explicit consideration for interpretability, two of the parameters could be be interpreted relatively unambiguously. In $\psi_{\mathrm{CH}_1}$ the parameter $p_{\text{iso}}$ multiplies the isotropic invariant $\tilde{I}_1$ and can therefore be related to the myocardial non-collagenous and non-muscular matrix [34]. In $\psi_{\mathrm{CH}_2}$, the parameter $p_{\text{glob}}$ multiples the rest of the function, hence it has a global effect that equally effects all components of the stress. The interpretation of the other parameters involves ambiguity, which we have resolved as best as we could in our parameter naming scheme (Supplementary Results *Interpretation of the material parameters in the CHESRA models*).

## Parameter variability benchmark with tissue experiment data

We compared the material parameter variability of our CHESRA SEFs ($\psi_{\mathrm{CH}_1}, \psi_{\mathrm{CH}_2}$) to state-of-the-art SEFs ($\psi_{\mathrm{MA}}, \psi_{\mathrm{CL}}, \psi_{\mathrm{HO}}, \psi_{\mathrm{SFL}}, \psi_{\mathrm{PZL}}$), defined in the Methods (see *State-of-the-art strain energy functions*) using a benchmark experiment with the Sommer [20] shear and Dokos [21] shear datasets (Supplementary Table S1). We chose these datasets because they involve 3D tissue samples that enable the identification of orthotropic material parameters, unlike the 2D biaxial datasets (Supplementary Table S1). For each SEF, we varied the initial value of $\boldsymbol{p}$ in the least squares fitting procedure (Equation 13) using 100 samples from a Latin hypercube design with parameter bounds (0, 100). We quantified the parameter variability using the coefficient of variation $c_p = \frac{s_p}{\overline{p}}$, where $s_p$ is the standard deviation of parameter $p$ and $\overline{p}$ its mean. This allowed for a direct comparison of parameter variability across different SEFs. For each SEF-dataset combination, we also computed the average $c_p$, denoted $\overline{c}_p$.

The benchmark results (Figure 5, Supplementary Figure S4) highlight the advantages of $\psi_{\mathrm{CH}_1}$ and $\psi_{\mathrm{CH}_2}$ for parameter estimation. Both CHESRA SEFs yielded unique parameter estimates from the Dokos [21] shear and Sommer [20] shear datasets, with all 100 random initializations converging to the same optimal values (consistent to five decimal places; Supplementary Table S7). In contrast, the state-of-the-art SEFs ($\psi_{\mathrm{MA}}, \psi_{\mathrm{CL}}, \psi_{\mathrm{HO}}, \psi_{\mathrm{SFL}}, \psi_{\mathrm{PZL}}$) exhibited significant parameter variability, with $\overline{c}_p$ values ranging from 0.9 to 8.9. Moreover, all of the state-of-the-art SEFs showed variable goodness of fit (Figure 5b, Supplementary Figure S4b), with $f_{\mathrm{GoF}}$ distributions including values $\geq 10^{-1}$.

The best fitting parameterizations of each SEF provided close model-data fits (Supplementary Figure S5), with the exception of $\psi_{\mathrm{MA}}$, which fit the data of Sommer et al. [20] well, but provided only a coarse approximation to the shear data of Dokos et al. [21]. This is somewhat expected as only the Sommer et al. data was used to derive the Martonová SEF $\psi_{\mathrm{MA}}$ [30]. In contrast, the quality of the worst fitting parameterizations differed depending on the specific SEF. Even the worst fitting parameterizations of $\psi_{\mathrm{CH}_1}$ and $\psi_{\mathrm{CH}_2}$ fit the data closely, whereas the worst fitting parameterizations of the state-of-the-art SEFs ($\psi_{\mathrm{MA}}, \psi_{\mathrm{CL}}, \psi_{\mathrm{HO}}, \psi_{\mathrm{SFL}}, \psi_{\mathrm{PZL}}$) had substantial model-data mismatch (second and fourth column of Figure 5a and Supplementary Figure S4a). This suggests the presence of sub-optimal local minima for state-of-the-art SEFs, a problem avoided by the simpler $\psi_{\mathrm{CH}_1}$ and $\psi_{\mathrm{CH}_2}$.

The non-uniqueness of state-of-the-art SEF parameters is further illustrated in Figure 6, which compares the goodness of fit landscapes (Methods *Quality assessment of strain energy functions*) of $\psi_{\mathrm{CH}_1}$, $\psi_{\mathrm{CH}_2}$,$\psi_{\mathrm{MA}}$, $\psi_{\mathrm{CL}}$, and $\psi_{\mathrm{HO}}$. The fitting landscapes for $\psi_{\mathrm{CH}_1}$ and $\psi_{\mathrm{CH}_2}$ exhibit sharp minima, while those for $\psi_{\mathrm{MA}}, \psi_{\mathrm{CL}}$ and $\psi_{\mathrm{HO}}$ contain flat regions where entire parameter ranges yield equivalent fits. Similar flat regions can be observed for $\psi_{\mathrm{PZL}}$ and $\psi_{\mathrm{SFL}}$ in Supplementary Figure S6. Together, these results demonstrate that the CHESRA generated SEFs ($\psi_{\mathrm{CH}_1}, \psi_{\mathrm{CH}_2}$) provide more consistent parameter estimations in the tissue data setting as compared to the state-of-the-art SEFs ($\psi_{\mathrm{CL}}$, $\psi_{\mathrm{MA}}$, $\psi_{\mathrm{HO}}$, $\psi_{\mathrm{PZL}}$, $\psi_{\mathrm{SFL}}$).

## Parameter variability benchmark with 3D simulation models

We conducted a second parameter variability benchmark using eight 3D ventricular simulation models, each testing one of the SEFs $\psi_{\mathrm{CH}_1}$, $\psi_{\mathrm{CH}_2}$, $\psi_{\mathrm{HO}}$ or $\psi_{\mathrm{MA}}$ in one of two scenarios, ex vivo or in vivo (Methods *In vivo and ex vivo simulation scenarios*). The ex vivo scenario involved laboratory pressure-volume measurements spanning a wide pressure range under controlled conditions [35], whereas the in vivo scenario involved clinically derived pressure and volume data over a narrower physiological range with more complex in vivo conditions [36]. In the benchmark, we selected $\psi_{\mathrm{HO}}$ and $\psi_{\mathrm{MA}}$ as representatives for the state-of-the-art SEFs (Methods *State-of-the-art strain energy functions*) due the recent development of $\psi_{\mathrm{MA}}$ and widespread use of $\psi_{\mathrm{HO}}$ at the time of our study. The geometry of the simulation models was based on a mid-diastolic patient MRI, including the left and right ventricles (Figure 1c) and rule-based fiber and fiber-sheet directions (Supplementary Figure S7).

In the first stage of our benchmark we estimated a separate pressure-free geometry for each SEF-scenario combination (Methods *Estimation of pressure-free geometries*), to enable the consistent calculation of displacements from a stress-free reference. We also simultaneously obtained benchmark values for all SEF parameter sets. Both pressure-free geometries and benchmark SEF parameters were obtained by optimizing each simulation model's left ventricular volumes either to the patient data [36] (in vivo scenario), or the Klotz [35] empirical relationship (ex vivo scenario). The resulting pressure-free geometries are shown in Figure 7b for the in vivo scenario, and the benchmark material parameters are provided in Supplementary Table S8. In the ex vivo scenario, we note that all four SEFs ($\psi_{\mathrm{CH}_1}$, $\psi_{\mathrm{CH}_2}$, $\psi_{\mathrm{HO}}$ or $\psi_{\mathrm{MA}}$) were able to closely replicate the target volume curve (Figure 7a). In the in vivo scenario all four SEFs were within 3 ml of the target volumes. However, only the Holzapfel-Ogden SEF $\psi_{\mathrm{HO}}$ was able to replicate the concave shape of the in vivo diastolic volume curve. This could be due to the $\psi_{\mathrm{HO}}$ SEF's relatively complex form with eight material parameters, which allowed for more flexible behaviour than the more simple SEFs $\psi_{\mathrm{CH}_1}$, $\psi_{\mathrm{CH}_2}$, and $\psi_{\mathrm{MA}}$.

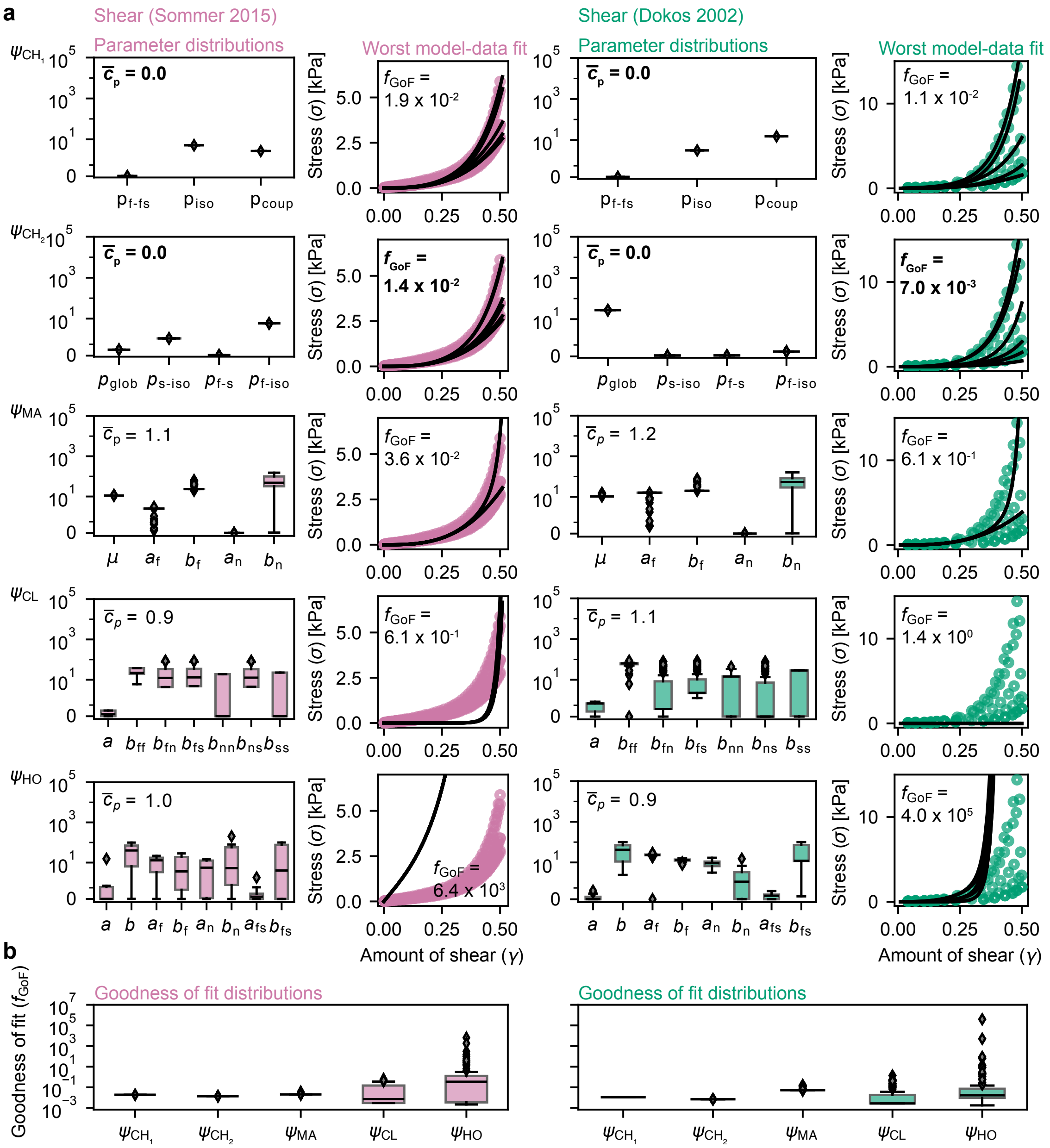

**Fig. 5: Parameter variability benchmark with experimental tissue data comparing the CHESRA strain energy functions (SEFs) $\psi_{\mathrm{CH}_1}$ and $\psi_{\mathrm{CH}_2}$ to the state-of-the-art SEFs ($\psi_{\mathrm{MA}}, \psi_{\mathrm{CL}}, \psi_{\mathrm{HO}}$). a** Distributions of estimated material parameters resulting from least-squares fitting to shear data with 100 random initializations (first and third column), along with goodness of fit ($f_{\mathrm{GoF}}$) for the worst fitting sets of model parameters (second and forth column). In the parameter distribution plots, the $x$-axis shows the material parameters of the corresponding SEF. The distributions of $\psi_{\mathrm{CH}_1}$ and $\psi_{\mathrm{CH}_2}$ are unique (up to five decimal places), whereas the parameters of the other SEFs show substantial variability. Optimal results are highlighted in bold-face. Solid lines represent model results whereas the markers indicate experimental data. **b** Corresponding goodness of fit distributions. $\overline{c}_p$ = average model parameter coefficient of variation, $\sigma$ = Cauchy stress, $\gamma$ = simple shear angle.

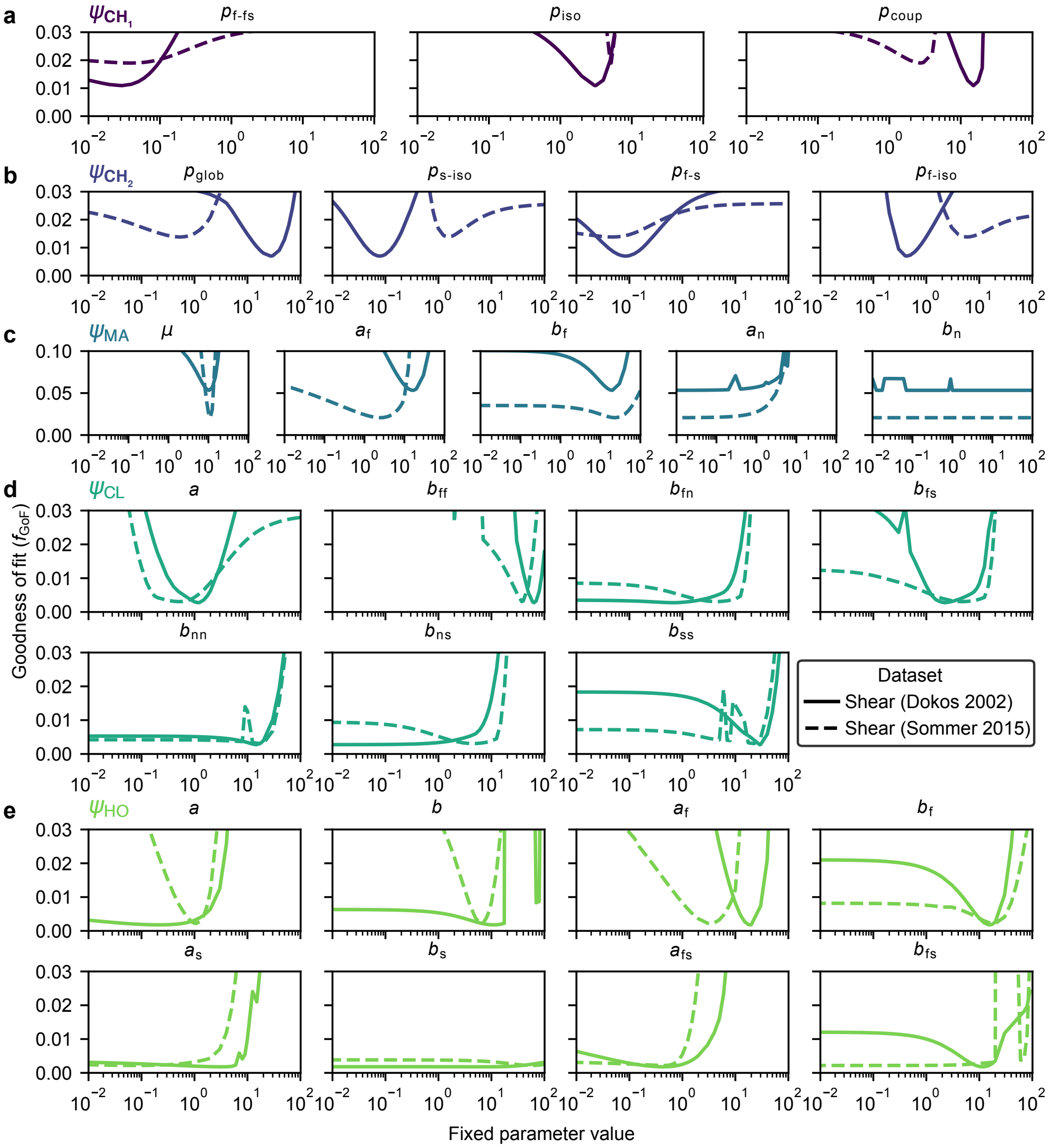

**Fig. 6: Goodness of fit landscapes.** Goodness of fit ($f_{\mathrm{GoF}}$) for the CHESRA strain energy functions (SEFs) **a** $\psi_{\mathrm{CH}_1}$ and **b** $\psi_{\mathrm{CH}_2}$ as compared to the state-of-the-art SEFs **c** $\psi_{\mathrm{MA}}$, **d** $\psi_{\mathrm{CL}}$, and **e** $\psi_{\mathrm{HO}}$. Within each plot $f_{\mathrm{GoF}}$ is calculated with one material parameter fixed and all other parameters optimized. The fixed material parameter is indicated in the subfigure title. Note that all of the CHESRA SEF parameters have unique minima, while some of the $\psi_{\mathrm{MA}}$, $\psi_{\mathrm{CL}}$, $\psi_{\mathrm{HO}}$ parameters have flat fit landscapes and/or multiple minima.

Next we generated synthetic diastolic displacement data as optimization targets (Methods *3D ventricular mechanics finite element framework*), using the pressure free geometries and benchmark material parameter values. To simulate real-world noise, we corrupted the displacements by adding random noise vectors with magnitudes equal to 50% of the local displacement. We then optimized each SEF's material parameters to the corresponding ex vivo and in vivo noisy displacements using twenty parameter initializations (Methods *In silico parameter variability benchmark with 3D ventricular simulation models*). We sampled each set of initial points using a Latin hypercube design, with bounds spanning one order of magnitude around the benchmark parameter values. Figure 7c shows the results. Of the four

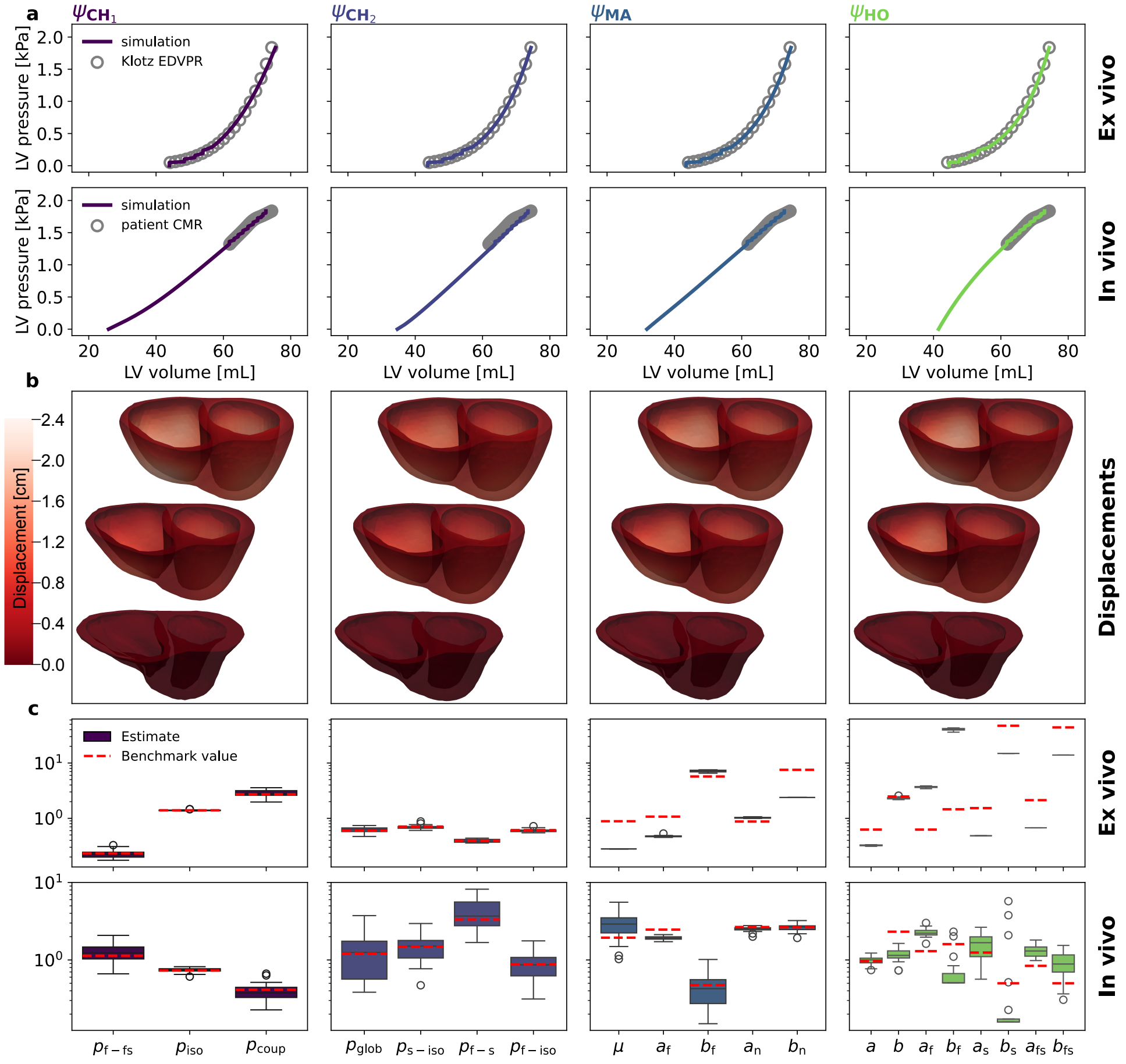

**Fig. 7: Parameter variability benchmark with 3D biventricular simulation models comparing the CHESRA strain energy functions (SEFs) $\psi_{CH_1}$ and $\psi_{CH_2}$ to the state-of-the-art SEFs ($\psi_{MA}$, $\psi_{HO}$).** **a** Simulated versus measured left ventricular (LV) pressure-volume data used to calculate pressure-free reference geometries and benchmark material parameters using the augmented Sellier's method with variable material properties (*Algorithm 1* in Methods). **b** Example displacement magnitudes (in vivo case) calculated using the pressure free geometries and benchmark material parameters from the previous step. The three geometries within each box correspond to the pressure free configuration (bottom position), mid-diastolic pressure (mid position, 1.32 kPa LV pressure) and end-diastolic pressure (top position, 2.19 kPa LV pressure). **c** Estimated vs benchmark material parameters obtained from optimizations starting from twenty random material parameter initializations. Each optimization attempted to match the estimated displacements to the respective target displacements by changing the material parameters, with the pressure-free reference geometry fixed throughout the optimization.

SEFs, $\psi_{CH_1}$ provided the most consistent and accurate parameter estimates, which we expected due to $\psi_{CH_1}$ having the least number of parameters to fit. For $\psi_{CH_2}$ the parameter estimates were accurate in the ex vivo scenario, but exhibited greater variability in the in vivo scenario. Nevertheless, the median parameter estimates for $\psi_{CH_2}$ were relatively close to their benchmark values, which meant that they could be approximated by aggregating data from many optimizations. In contrast, the estimated parameters of $\psi_{MA}$ and $\psi_{HO}$ could consistently differ from their benchmark values, indicating the presence of local minima in the corresponding optimization landscapes that hinder the acquisition of accurate parameter estimates, even after aggregating results from multiple optimizations.

## Computational Efficiency Analysis

To compare the computational efficiency of the SEFs $\psi_{\mathrm{CH}_1}, \psi_{\mathrm{CH}_2}, \psi_{\mathrm{MA}}$, and $\psi_{\mathrm{HO}}$ we collected run-time data for the tissue and 3D biventricular simulation benchmarks (Fig 8). The tissue benchmark was performed in serial on a MacBook Air (Apple M1 chip, 8-core CPU, 16 GB memory), wheras the 3D simulation benchmark was performed in serial on a high-performance computing node equipped with an AMD EPYC 7542 32-core processor (2.9 GHz, 64 logical CPU) and 252 GB RAM, running Ubuntu 24.04.2 LTS.

In the tissue benchmark $\psi_{\mathrm{CH}_1}$ was the most efficient, using on average only 33 Levenberg-Marquardt iterations and 2 s of run-time or less, with low variability. In terms of median efficiency, $\psi_{\mathrm{CH}_2}$ and $\psi_{\mathrm{HO}}$ ranked after $\psi_{\mathrm{CH}_1}$, with $\psi_{\mathrm{CH}_2}$ being more robust due to fewer high computational cost outliers than $\psi_{\mathrm{HO}}$. The other SEFs ranked $\psi_{\mathrm{MA}}, \psi_{\mathrm{CL}}, \psi_{\mathrm{SFL}}, \psi_{\mathrm{PZL}}$ in terms of run-time. Overall, the run-time scaled with the number of iterations for all SEFs except for $\psi_{\mathrm{CL}}$, reflecting relatively stable run-time costs per Levenberg-Marquardt iteration.

In the ex vivo 3D simulation benchmark the run-times did not scale with the number of optimization iterations, possibly due to the differing cost of solving the virtual-work equation (21), which involved a variable amount of Newton iterations. Indeed, in the ex vivo benchmark the CHESRA functions $\psi_{\mathrm{CH}_1}$ and $\psi_{\mathrm{CH}_2}$ ran an order of magnitude faster than $\psi_{\mathrm{MA}}$ and $\psi_{\mathrm{HO}}$, despite using a greater number of optimization iterations. This indicates that the virtual-work equation (21) was most likely solved with relatively fewer Newton iterations for $\psi_{\mathrm{CH}_1}$ and $\psi_{\mathrm{CH}_2}$, which involved only polynomial non-linearity as opposed to the exponential formulations $\psi_{\mathrm{MA}}$ and $\psi_{\mathrm{HO}}$. Note that the use of the adjoint calculations in calculating the loss gradient for the 3D simulations meant that the cost of gradient calculation was independent of the number of material parameters. Thus, in the 3D simulations the SEFs did not differ in the number of forward passes required to evaluate the loss function (29) or its gradient, thereby facilitating comparability between SEFs. In the case of the in vivo 3D simulation benchmark, the total run time once again scaled with the number of optimization iterations. The SEF $\psi_{\mathrm{MA}}$ had the lowest median run-time, followed by $\psi_{\mathrm{HO}}$ and $\psi_{\mathrm{CH}_2}$. The run-time performance of $\psi_{\mathrm{CH}_1}$ was more variable, with the number of iterations and run-time ranging between the values of $\psi_{\mathrm{HO}}$ and $\psi_{\mathrm{MA}}$.

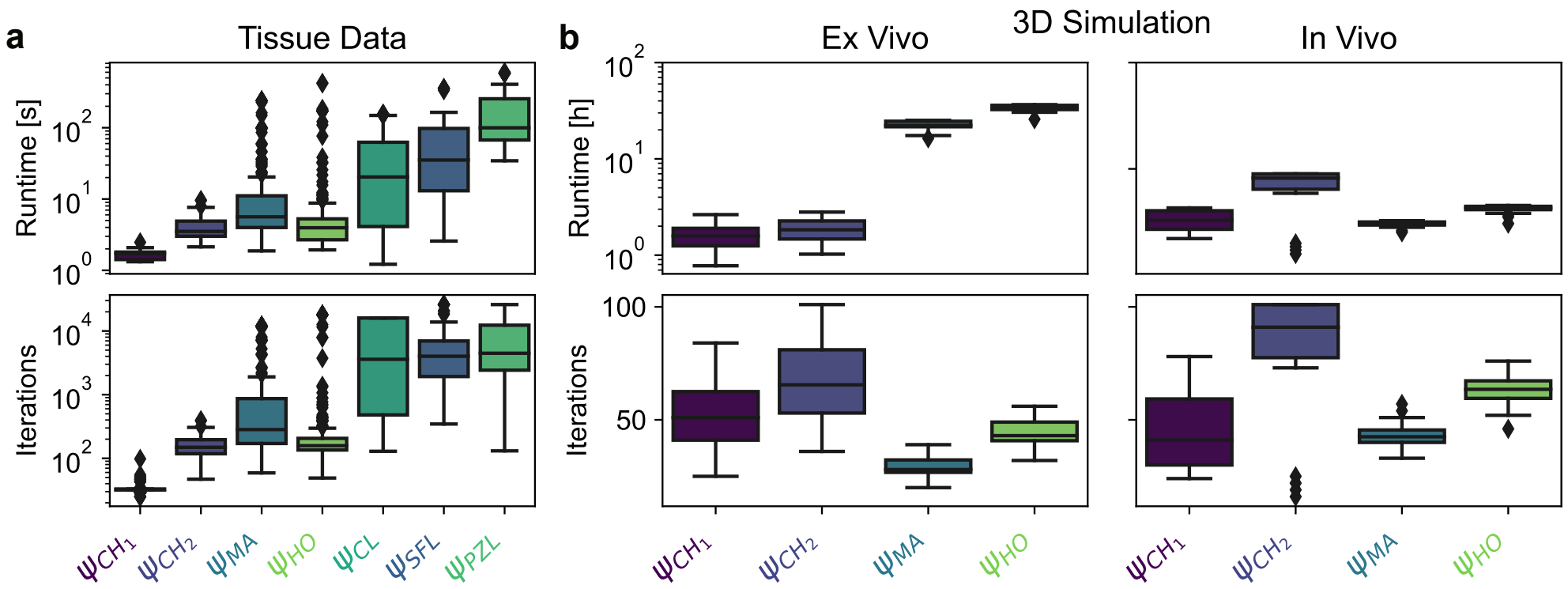

**Fig. 8: Computational efficiency in the parameter estimation benchmarks.** Number of optimization iterations and run-time for fitting the parameters of **a** the strain energy functions ($\psi_{\mathrm{CH}_1}, \psi_{\mathrm{CH}_2}, \psi_{\mathrm{MA}}, \psi_{\mathrm{HO}}, \psi_{\mathrm{CL}}, \psi_{\mathrm{SFL}}, \psi_{\mathrm{PZL}}$) in the tissue benchmark and **b** ($\psi_{\mathrm{CH}_1}, \psi_{\mathrm{CH}_2}, \psi_{\mathrm{MA}}, \psi_{\mathrm{HO}}$) in the 3D simulation benchmarks.

## Discussion

This study introduces the Cardiac Hyperelastic Evolutionary Symbolic Regression Algorithm (CHESRA) as a novel physics informed machine learning (PI-ML) framework for discovering parsimonious constitutive laws in cardiac tissue mechanics. CHESRA unites biophysical modeling with data-driven discovery by embedding the physical principles of frame invariance, material symmetry, and hyperelastic energy conservation directly into a symbolic regression framework. Through this integration, CHESRA restricts the space of possible SEFs to physically admissible models while learning SEF forms directly from experimental data. Applying this framework to cardiac tissue mechanics, CHESRA uncovered two previously unreported SEFs, $\psi_{\mathrm{CH}_1}$ and $\psi_{\mathrm{CH}_2}$, that balance parsimony with predictive accuracy. Both SEFs reproduced experimental stress-strain data from multiple studies [18–21] with fidelity comparable to, or exceeding, that of established orthotropic SEFs [14–17, 30], despite relying on only three or four material parameters. Systematic cross-validation confirmed the capacity of CHESRA's SEFs to generalize across independent datasets (Figure 3), while benchmark analyses (Figure 5 and Figure 6) revealed sharply defined parameter landscapes indicative of unique and stable parameter estimates. When embedded in 3D biventricular mechanics simulations, $\psi_{\mathrm{CH}_1}$ demonstrated markedly reduced parameter variability relative to state-of-the-art SEFs (Figure 7), underscoring its potential to enhance the robustness and personalizability of cardiac digital twins, where frequent parameter updates are required [6].

A key strength of our study is the simplicity of CHESRA's SEFs ($\psi_{\mathrm{CH}_1}, \psi_{\mathrm{CH}_2}$), which involve only three to four tunable material parameters and have simple polynomial forms. Such simple low-parameter models already have a well-established history of use within cardiac electrophysiology [37, 38]. Here, we have algorithmically derived similar models specifically for cardiac mechanics. In contrast, the state-of-the-art SEFs [14–17, 30] have five or more parameters and more complex mathematical forms involving exponential functions (see *State-of-the-art strain energy functions* in Methods). This complexity arises partly from differences in how these SEFs were conceived, whether through human expertise or data-driven approaches. The human expert-designed SEFs [14–17] were designed to fit experimental data and the designer's choice of biophysical considerations, but lacked an automated simplification mechanism. As a consequence, the SEFs [14–17] have issues with high dimensionality and parameter interdependence, limiting their identifiability from patient data [12]. Conversely, the purely data-driven SEFs [29] can fit experimental observations but can lack interpretability [29] and risk overfitting to their training datasets [29–31]. Recognizing these limitations, CHESRA utilizes a physics-informed evolutionary process to learn mathematically interpretable closed-form SEFs from multiple datasets, while also balancing goodness of fit and model simplicity.

While high-parameter SEFs like the Holzapfel-Ogden SEF [14] can be made identifiable by fixing certain parameters to literature values [25–30], this approach introduces a dual burden. First, it requires a-priori sensitivity analysis to decide which parameters to fix, making the modeling pipeline more complex and less automated. Second, fixing parameters to pre-determined values introduces a structural bias that may lead to unequal healthcare outcomes among differing patient groups. In contrast, for the CHESRA-derived SEFs ($\psi_{\mathrm{CH}_1}, \psi_{\mathrm{CH}_2}$) the parsimonious functional forms promote unique estimation of all parameters, working towards digital twins with fully personalized material properties.

CHESRA builds upon previous work on evolutionary symbolic regression (ESR) for SEF design [39, 40], and represents the first application of ESR to the discovery of cardiac SEFs. Alternative data-driven approaches include sparse regression [41], constitutive neural networks [30, 31], and data-driven reduction from pre-specified forms [42], each of which is constrained to either linear combinations of pre-defined basis functions or a-priori specified functional forms. By contrast, ESR traverses a substantially broader space of mathematical structures, combining basic elements through products, nesting, and transcendental functions in ways that competing methods [30, 31, 41, 42] do not consider. It is worth noting that while the discovered SEFs ($\psi_{\mathrm{CH}_1}, \psi_{\mathrm{CH}_2}$) maintained low parameter counts, they preserved the essential nonlinearities of the myocardium. Simplification to a linear elastic form was rejected by the CHESRA framework during the evolutionary process, as linear SEFs are unable to reconcile the low-stiffness "toe region" with the high-stiffness "fiber recruitment region" observed across all five experimental datasets.

In the cross-validation tests (Figure 3), we observed that the generalizability of CHESRA's functions improved when we used experimental data from multiple sources [18–21]. This is in-line with pure machine learning approaches for classification and regression, which typically see improvements in validation accuracy when the amount and variety of their training data is increased. Our results confirm this trend for the first time in PI-ML algorithms for cardiac SEF design, with training data intentionally spanning porcine [21], canine [18, 19], and human [20] sources, thereby relaxing the standard i.i.d. assumption of cross-validation in favour of cross-species diversity. The degree to which the observed generalizability reflects the learning of species-invariant physical principles warrants further investigation, but it suggests promise for translation into diverse clinical populations. In the future, the collection of further high-quality experimental datasets of tissue mechanics could allow for further refinements of CHESRA's SEFs, with potential increases in SEF generalizability and data fitting accuracy. In particular, including data from pathological tissues into CHESRA would be an interesting future direction, as it would allow for more broadly validated SEFs and pave the way for pathology-specific clinical applications.

Beyond generalizability, another essential aspect of model reliability is parameter uniqueness, which ensures that fitted parameters reflect true tissue properties rather than artifacts of the optimization process. As shown in Figure 5, the CHESRA-derived functions ($\psi_{\mathrm{CH}_1}$ and $\psi_{\mathrm{CH}_2}$) produced unique and stable parameter estimates across repeated optimizations to shear experiment data [20, 21]. This is in contrast to traditional SEF formulations [14–17] that exhibited broad parameter variability, consistent with previously reported experiments [42]. By visualizing the goodness of fit landscapes (Figure 6), we demonstrate the source of the parameter variability: while the state-of-the-art SEFs possess flat, non-unique optimization landscapes, the CHESRA-discovered SEFs exhibit well-defined minima (Figure 6). This suggests that symbolic regression can successfully right-size the constitutive law to the available data, a critical step for robust clinical personalization. Indeed, the CHESRA functions' improved parameter identifiability addresses long-standing concerns in cardiac modeling, where parameter non-uniqueness has been linked to poor reproducibility and uncertainty in model personalization [43]. By constraining SEF complexity through a function-length penalty and a physics-informed invariant framework, CHESRA effectively narrows the solution landscape, reducing the prevalence of local minima and ill-conditioned parameter spaces.

In the 3D biventricular simulation setting, we confirmed that the CHESRA function $\psi_{\mathrm{CH}_1}$ had less parameter variability than state-of-the-art SEFs when fit to synthetic 3D displacement data (Figure 7). From a clinical perspective, reduced parameter variability translates directly into improved personalization reliability and predictive stability. Indeed, in personalized cardiac mechanics models, small variations in estimated tissue parameters can produce large differences in simulated ventricular deformation or pressure–volume relationships [44], potentially confounding diagnostic or prognostic conclusions. The parsimony and identifiability of CHESRA-derived SEFs mitigate this risk, paving the way for digital twins that can be confidently individualized to each patient's imaging and hemodynamic data. Such individualized digital twins have potential applications in disease prediction, prevention, and management; for example, assessing the progression of myocardial fibrosis [12, 45], detecting cardiomyopathy [46], optimizing ventricular assist device settings [47], forecasting response to cardiac resynchronization therapy [48] or in silico drug safety assessment [49]. Furthermore, CHESRA's invariant-based symbolic framework can be directly applied to other soft tissues and disease contexts where mechanics play a diagnostic or prognostic role, such as vascular remodeling in hypertension [50], lung compliance in chronic obstructive pulmonary disease [51], tumor stiffness in oncology [52], or cerebral tissue deformation in traumatic brain injury [53]. By enabling the derivation of parsimonious and interpretable constitutive laws from limited data, CHESRA offers a generalizable PI-ML approach for building dedicated SEFs for digital twins across multiple organ systems. These SEFs can support inverse problem solving, and the design of personalized therapeutic strategies. Real-time simulations and real-time data digital twin updates are another exciting possibility, which could be facilitated by switching the current slow and expensive FEM solver to faster neural network based solvers [54].

We note that model fidelity is an important issue for cardiac digital twins, as missing crucial aspects of a patient's physiology could potentially lead to suboptimal medical decisions. In our study, we balanced this consideration with the need for benchmark simplicity for comparing SEFs. Indeed, the 3D biventricular simulations presented here follow the established practice of using simplified organ-level geometries to validate material law stability and identifiability [26, 55]. In the ex vivo benchmark, our modeling assumptions closely matched the experimental conditions [35]. However, in the in vivo benchmark, our setup was not able to incorporate the complex interactions between the ventricles, pericardium, and atria that contribute to physiological in vivo motion. Consequently, we included unphysiological low tissue compressibility in the in vivo simulations for numerical reasons. We note that the ability of the Holzapfel-Ogden SEF to match the in vivo pressure-volume curve (Figure 7) demonstrates a key advantage of high-parameter SEFs, namely that they can more flexibly accommodate for mis-specified modeling assumptions. Nevertheless, this flexibility may come at a cost in the clinical setting: a high-parameter SEF that absorbs modeling errors into its material parameters risks conflating tissue mechanics with unmodeled physiology.

Before concluding, we acknowledge several limitations of our study and the CHESRA algorithm. Firstly, we were not able to guarantee that any boundary value problems involving CHESRA's SEFs will have unique solutions. This issue can be addressed in future studies by explicitly designing polyconvex SEFs that also satisfy a coercivity condition [56]. While CHESRA does not enforce polyconvexity, it does, however, promote convex-like behaviour by using squared strain invariant functions, and via restrictions on the functional forms of the SEFs; namely that all the material parameters are positive, and the binary operators are restricted to "+" and "×". Nevertheless, CHESRA's SEFs satisfy several other desirable theoretical properties, including material frame indifference, material symmetry, and stress-free undeformed configurations. A further limitation of CHESRA is that the material parameters $\boldsymbol{p}$ lack a-priori physiological interpretations. This could be addressed in the future by pre-specifying terms with physiological significance in the symbolic regression. The sparsity of available experimental datasets limited our ability to thoroughly assess the generalizability of CHESRA's output functions. In particular, we only had access to two 3D datasets [20, 21], and the tissue experiment benchmark results (Figure 5) lacked validation in datasets not used to derive the CHESRA SEFs ($\psi_{\mathrm{CH}_1}, \psi_{\mathrm{CH}_2}$). If additional 3D experimental data were available, then CHESRA's output SEFs could be further refined. Currently, the lack of clinical tissue experiment data means that only the parameters of CHESRA's SEFs can be personalized rather than the entire forms of the SEFs. However, this limitation may be alleviated in the future by extension of the CHESRA algorithm to clinical data, or advances in human stem cell derived cardiomyocyte technologies [57] that will allow for individualized patient tissue mechanics data collection. Finally, the data updates in personalized 3D biventricular models were limited to a single MRI study, estimated pressure values, and synthetic displacement data. However, the use of synthetic data allowed for ground truth parameter values and parameter error calculations. Future research could test SEFs in more complex digital-twin scenarios with multiple data updates and medical image-derived motion data.

Additionally, we acknowledge limitations related to clinical data sparsity in the 3D biventricular simulations. First, the in vivo pressure-volume targets were not recorded simultaneously for the specific patient but were fitted from established hemodynamic curves. Second, the absence of CINE-MRI data precluded a direct comparison between predicted and actual geometric displacements, limiting the validation of the 3D biventricular model to global volume and pressure targets. Third, as this proof-of-concept was conducted on a single biventricular geometry, further studies across diverse heart shapes and pathologies are required to confirm generalizability. Finally, the myocardial fiber architecture was generated using rule-based methods rather than patient-specific diffusion tensor imaging, which may have affected the accuracy of our material parameter estimates. While these constraints reflect common limitations in clinical data availability, they represent key hurdles that must be addressed to transition to a fully validated clinical digital twin. Finally, we note that the uncertainties inherent in the model generation process were not formally assessed in terms of their impact on the SEFs' performance. While the parsimony of the discovered SEFs was intended to provide a buffer against model uncertainties, the sensitivity of the resulting SEF parameters to such uncertainties warrants future investigation. Integrating symbolic regression with probabilistic or Bayesian uncertainty quantification frameworks [44], along with Markov Chain Monte Carlo techniques, can help to fully characterize the reliability of discovered SEFs in the presence of real-world clinical data errors.

In summary, CHESRA embodies the principles of physics-informed machine learning for personalized digital medicine. By merging symbolic learning with physical laws of elasticity, it achieves a balance between accuracy, generalizability, and interpretability that is crucial for the clinical uptake of biomedical digital twins. More broadly, CHESRA demonstrates how embedding mechanistic insight into machine learning can accelerate the development of transparent, data-efficient, and clinically actionable digital health technologies.

# Methods

## Invariant framework for designing energy functions

We modeled the myocardium as an orthotropic, hyperelastic material with three orthonormal basis vectors specifying the directions of anisotropy [14]. The orthonormal vectors were: the "fiber" axis (f) corresponding to the locally prevailing cell orientation; the "sheet" axis (s) associated with the direction in the plane of the cell sheets orthogonal to (f); and the "sheet-normal" axis (n) orthogonal to (f) and (s). To ensure objectivity, we built our SEFs using principal invariants of the right Cauchy-Green tensor $\boldsymbol{C} = \boldsymbol{F}^T\boldsymbol{F}$, with $\boldsymbol{F}$ being the deformation gradient [34]. These invariants were defined as

$$\begin{aligned} &I_1 = \operatorname{tr}\boldsymbol{C}, && I_2 = \frac{1}{2}[I_1^2 - \operatorname{tr}\boldsymbol{C}^2], && I_3 = \det\boldsymbol{C}, \\ &I_{4i} = \boldsymbol{e}_i^T(\boldsymbol{C}\boldsymbol{e}_i), && I_{5i} = \boldsymbol{e}_i^T(\boldsymbol{C}^2\boldsymbol{e}_i), && I_{8ij} = \boldsymbol{e}_i^T(\boldsymbol{C}\boldsymbol{e}_j), \end{aligned} \tag{3}$$

with $i, j \in \{\text{f,s,n}\}$ and $\boldsymbol{e}_i$ the corresponding basis vectors. Similar to the approach used in [14], we normalized the invariants to guarantee that they vanished at zero strain. Moreover, we squared them to ensure positivity,

$$\begin{aligned} &\tilde{I}_1 = (I_1 - 3)^2, && \tilde{I}_2 = (I_2 - 3)^2, && \\ &\tilde{I}_{4i} = (I_{4i} - 1)^2, && \tilde{I}_{5i} = (I_{5i} - 1)^2, && \tilde{I}_{8ij} = I_{8ij}^2. \end{aligned} \tag{4}$$

Here we have dropped the third invariant as volumetric changes were not reported in the experimental data [18–21] that we used for parameter estimation. The five types of invariants in Equation 4, together with material parameters

$$\boldsymbol{p} = p_1, .., p_k, \tag{5}$$

formed the basic building blocks of our SEFs

$$\psi(\tilde{I}_1, .., \tilde{I}_{8\text{fs}}, \boldsymbol{p}), \tag{6}$$

where we learned the functional form of each SEF $\psi$ from experimental data using the CHESRA algorithm.

## Tissue experiment data sources

To algorithmically design SEFs, we used data from four tissue studies of cardiac mechanics [18–21]. All four studies used small tissue samples with minimal variation in the local fiber and or sheet directions. By modeling these variations as uniform [14], we were able to directly evaluate the invariants (Equation 4) for each dataset, allowing for a computationally efficient comparison of stress and strains between SEF and data. Three of the tissue studies included (equi-)biaxial stretch protocols [18–20], and two studies included triaxial shear protocols [20, 21]. The equibiaxial protocols involved uniform stretching of the myocardium in both the fiber and cross-fiber directions [19], while the biaxial protocols employed distinct stretch ratios for each direction [18–20]. For the triaxial shear protocols, simple shear was applied in either the fiber/sheet, fiber/sheet-normal, or sheet/sheet-normal planes [20, 21]. The biaxial, equibiaxial, and triaxial protocols are demonstrated in Figure 1a. Two of our data-source studies [19, 20] included two distinct experimental protocols, which gave a total of six distinct experimental studies and mechanical protocol combinations, which we henceforth refer to as *datasets*. The human myocardial tissue data were obtained from Sommer et al. [20], which received approval from the Ethical Committee of the Medical University of Graz, Austria (approval numbers 24–003 and 24–107).

To obtain numerical data for each dataset, we digitized the data plots found in the experimental reports [18–21], using WebPlotDigitizer software. We did not normalize or post-process any of the digitized data, as CHESRA performed normalization internally (section *Fitness evaluation*). Supplementary Table S1 summarizes the technical details regarding each dataset, including the stress and strain measures used. Note that the Novak equibiaxial dataset [19] contains independent measurements from four regions of the heart (sub-endocardium, mid-myocardium, sub-epicardium, and mid-septum) of two specimens. This information is reflected in the *subsets* column of Supplementary Table S1, with the Novak equibiaxial dataset containing eight distinct subsets, and every other dataset containing only a single subset. To account for the differences in material properties among subsets, we calculated separate sets of SEF material parameters $\boldsymbol{p}$ for each data subset within the CHESRA algorithm.

## CHESRA Algorithm

**Initialization** We initialized CHESRA by generating a set of random SEFs. Each SEF had a function tree representation, with each tree node being either an operator, exponential function, material parameter symbol $p_1, .., p_k$, or invariant symbol $\tilde{I}_1, \ldots, \tilde{I}_{8fs}$ (Equation 4). All operators were chosen from the set $\{+, \times\}$ to promote function convexity. Each initial SEF was grown from a single symbol node, representing either a material parameter or an invariant, decided with equal probability. The SEF was then extended for a maximum number of $l_{\text{init}}$ iterations. Each extension added either an exponential function (probability 0.2) or a random operator plus material parameter (probability 0.4), or a random operator plus invariant (probability 0.4).

**Fitness evaluation** We evaluated the fitness $f_{\text{fit}}$ of each candidate SEF $\psi$ using a formula based on the goodness of fit $f_{\text{GoF}}$ and a function length penalty $l_{\text{eq}}$

$$f_{\text{fit}}(\psi, \boldsymbol{x}, \boldsymbol{y}, \alpha) = f_{\text{GoF}}(\psi, \boldsymbol{x}, \boldsymbol{y}) + \alpha l_{\text{eq}}(\psi), \tag{7}$$

where $\boldsymbol{x}, \boldsymbol{y}$ represent the respective strain and stress data from the input datasets. Furthermore, the function length $l_{\text{eq}}(\psi)$ counts the total number of nodes in the tree representation of $\psi$, and the hyperparameter $\alpha$ is a balance term which controls the relative emphasis on data fitting versus function simplicity. We define the data-fit term $f_{\text{GoF}}$ as

$$f_{\text{GoF}}(\psi, \boldsymbol{x}, \boldsymbol{y}) = \sum_{i=1}^{N} w_i f_{\text{sRSS}}(\psi, \boldsymbol{x_i}, \boldsymbol{y_i}), \tag{8}$$

where $i$ is the dataset index, and $N$ the total number of datasets. Within CHESRA, we set the data weights $w_i$ to give equal weight to biaxial and shear data

$$w_i = \frac{1}{2n_i} \quad n_i \in \{n_{\text{biax}}, n_{\text{shear}}\}, \tag{9}$$

with $n_{\text{biax}}$ and $n_{\text{shear}}$ representing the number of inputted biaxial and shear datasets, respectively. Furthermore, we defined the standardized sum of squared residuals $f_{\text{sRSS}}$ as

$$f_{\text{sRSS}}(\psi, \boldsymbol{x}_i, \boldsymbol{y}_i) = \frac{\sum_{j=1}^{N_i} \left(\boldsymbol{y}_{ij} - Y_i(\psi(\boldsymbol{x}_{ij}, \boldsymbol{p}_i))\right)^2}{\sum_{j=1}^{N_i} \left(\boldsymbol{y}_{ij} - \bar{\boldsymbol{y}}_i\right)^2}, \tag{10}$$

where $j$ denotes the data point number in dataset $i$. Here $Y_i$ denotes the stress operator that calculates the relevant component of the stress tensor (the Cauchy stress $\boldsymbol{\sigma}$ or the second Piola-Kirchhoff stress $\boldsymbol{S}$) at strain value $\boldsymbol{x}_{ij}$, and material parameter vector $\boldsymbol{p}_i$. We calculated the Cauchy stress tensor $\boldsymbol{\sigma}$ of a SEF $\psi$ using the formula

$$\boldsymbol{\sigma} = \boldsymbol{F} \sum_k \frac{\partial \psi}{\partial I_k} \frac{\partial I_k}{\partial \boldsymbol{F}} - p\boldsymbol{I}, \tag{11}$$

with the assumption $\det \boldsymbol{F} = 1$ since none of the tissue experiment datasets [18–21] measured volumetric changes. Note that the pressure term $p\boldsymbol{I}$ does not appear in our calculations since the 3D datasets [20, 21] measured shear stresses that appear off the main diagonal of the stress tensor $\boldsymbol{\sigma}$, and the other datasets were 2D. Similarly, we calculated the second Piola-Kirchhoff stress $\boldsymbol{S}$ using

$$\boldsymbol{S} = \boldsymbol{F}^{-1} \boldsymbol{\sigma} \boldsymbol{F}^{-T}. \tag{12}$$

Furthermore, we standardized the squared error term of the numerator of Equation 10 by the denominator, which included the term $\overline{\boldsymbol{y}}_i$, the average stress of dataset $i$. In this way, $f_{\mathrm{sRSS}}$ provided a standardized, dimensionless measure of the relative fitting error of the SEF $\psi$ for each input dataset.

To properly evaluate the stress $Y_i$, we needed to determine the parameter values $\boldsymbol{p}_i = p_{i1}, \ldots, p_{ik}$ which best reflected the material properties of the tissue used to generate a particular dataset. We accomplished this by using the Levenberg-Marquart algorithm implemented in the Python package lmfit, using the default settings (function tolerance $1.5 \times 10^{-8}$, solution tolerance $1.5 \times 10^{-8}$, and gradient tolerance 0). Levenberg-Marquart is a gradient-based optimization method that minimizes the sum of squared residuals between observed data points and model output, which in our study was

$$\min_{\boldsymbol{p}_i} \sum_{j=1}^{n_i} \left( \boldsymbol{y}_{ij} - Y_i(\psi(\boldsymbol{x}_{ij}, \boldsymbol{p}_i)) \right)^2 . \tag{13}$$

Within CHESRA, we initialized the Levenberg-Marquart algorithm with the vector $\boldsymbol{p}_i = (1, .., 1)$, as we noticed that the gradient at $\boldsymbol{p}_i = (0, .., 0)$ could vanish for some SEFs. To promote SEF convexity, a lower bound of 0 was enforced for $\boldsymbol{p}_i$, with no upper bound. For some particularly ill-suited SEFs, the Levenberg-Marquart algorithm could experience a lack of convergence. In order to prevent such SEF from excessively increasing the run-time of CHESRA we set a hard limit of 200 seconds to evaluate Equation 13. If this limit was reached, the Levenberg-Marquart algorithm was terminated, and a fitness of $f_{\mathrm{fit}} = \infty$ was assigned to effectively remove the SEF from the next CHESRA iteration.

Note that parsimony is enforced within the CHESRA framework using the complexity-weighted fitness function (7). This function balances the goodness-of-fit against the total number of mathematical operations (nodes) in the expression tree, functioning as an evolutionary $L_0$-regularizer. Crucially, this regularization is decoupled from the parameter optimization process; while the symbolic search is derivative-free, the subsequent least-squares calibration of material constants (13) utilizes standard gradient-based techniques, ensuring that the complexity penalty does not interfere with the convergence of the squared residuals (10).

**Selection** During each iteration, CHESRA selected a new SEF population of size $n_{\mathrm{ind}}$ for breeding by using a combination of elitism and random tournament. Selection by elitism meant that the best $n_{\mathrm{hof}}$ SEFs (with the lowest fitness scores) were automatically selected for the next generation. This ensured the preservation of the best solutions. To promote genetic variability, CHESRA selected a further $n_{\mathrm{evo}} = n_{\mathrm{ind}} - n_{\mathrm{hof}}$ individuals by random tournament, with each tournament consisting of randomly choosing two SEFs and selecting the SEF with the best fitness.

**Evolutionary changes** After selection, the $n_{\mathrm{evo}}$ SEFs selected by tournament were randomly modified by evolutionary operators to generate new candidate SEFs and thereby explore the space of SEF designs. The evolutionary operators were mating, mutation, reduction, and extension (Supplementary Figure S8), applied in this order and each occurring with probability $p_{\mathrm{mate}}$, $p_{\mathrm{mutate}}$, $p_{\mathrm{reduce}}$, and $p_{\mathrm{extend}}$, respectively.

The mating operator randomly swapped subtrees from two SEFs. For mutations, an operator $(\times, +)$ or symbolic parameter (invariants and material parameters) was replaced with a random node of the same type. A function reduction resulted in an operator or exponential function node being randomly chosen and deleted along with its dependent subtree. Finally, we extended functions by adding either an exponential, an operator with a symbolic parameter, or an operator with an invariant with probabilities 0.2, 0.4, 0.4, respectively.

**New Generation** During each CHESRA iteration, a new generation of SEFs was formed by combining the $n_{\mathrm{evo}}$ evolved SEFs together with the $n_{\mathrm{hof}}$ elite SEFs. Afterwards, we replaced duplicate SEFs with new random SEFs to ensure genetic diversity. The next iteration of CHESRA would then start again at fitness evaluation until a total of $n_{\mathrm{gen}}$ SEF generations had been reached.

**Output strain energy function and post-processing** After reaching $n_{\mathrm{gen}}$ iterations, CHESRA outputted the SEF that yielded the lowest fitness score. We then algebraically simplified the output SEF using the simplify function of the Python package sympy, and manually unified any redundant material parameters (e.g. $\psi = I_1 + p_1 + p_2 \rightarrow \psi = I_1 + p_1$). For the SEFs $\psi_{\mathrm{CH}_1}$ and $\psi_{\mathrm{CH}_2}$ we added a purely material parameter-dependent normalization term to ensure zero energy at zero strain. In principle, such a normalization can be done for all SEFs generated by CHESRA, but are often omitted in our results for the sake of brevity.

## CHESRA hyperparameter selection

CHESRA contained a number of hyperparameters that needed to be set before running the algorithm. These hyperparameters included the genetic algorithm parameters $n_{\text{gen}}, n_{\text{ind}}, n_{\text{hof}}$, function initialization size $l_{\text{init}}$, genetic operation probabilities $p_{\text{mate}}, p_{\text{mutate}}, p_{\text{reduce}}, p_{\text{extend}}$, and function length penalty parameter $\alpha$. We pre-set the values of $n_{\text{gen}}, n_{\text{ind}}, n_{\text{hof}}, l_{\text{init}}$ and determined the other hyperparameters via systematic experimentation (Results *CHESRA's penalty parameter effectively controls the complexity of elastic energy functions*). Our chosen hyperparameter values and value ranges are summarized in Supplementary Table S2.

We gave the function length penalty $\alpha$ special consideration, as we observed that it had a large effect on the simplicity of the output SEF. We therefore considered several orders of magnitude of $\alpha$ values, ranging from $10^{-4}$ to $10^{-2}$, and including 0 (no penalty). We chose $\alpha$ values more densely around $10^{-4}$ as we observed a good trade-off between function length and data fitting for these values during our hyperparameter experiments (Results *CHESRA's penalty parameter effectively controls the complexity of elastic energy functions*).

## State-of-the-art strain energy functions

In our tissue parameter variability benchmark, we included orthotropic SEFs from the literature, which are structurally based, relating to the fiber, sheet, and normal directions of myocardium [14]. The SEFs we tested were the Costa Law $\psi_{\text{CL}}$ [16], Separable Fung-type Law $\psi_{\text{SFL}}$ [15], the Pole-Zero Model $\psi_{\text{PZL}}$ [17], the widely used Holzapfel-Ogden Model $\psi_{\text{HO}}$ [14], and the recently established Martonová Three-Term Model $\psi_{\text{MA}}$, which was found using a constitutive neural network [30].

**Pole-Zero Model (1997)** The Pole-Zero model is one of the earliest orthotropic cardiac SEFs and is based on the Green-Lagrange strain tensor $\boldsymbol{E}$. It was proposed by Hunter et al in 1997 [17], motivated by a set of (equi)-biaxial experiments [58]. The definition of the Pole-Zero model [15] is

$$\begin{aligned}\psi_{\text{PZL}} = &\frac{k_{\text{ff}}E_{\text{ff}}}{|a_{\text{ff}} - |E_{\text{ff}}||^2} + \frac{k_{\text{ss}}E_{\text{ss}}}{|a_{\text{ss}} - |E_{\text{ss}}||^2} + \frac{k_{\text{nn}}E_{\text{nn}}}{|a_{\text{nn}} - |E_{\text{nn}}||^2} \\ &+ \frac{k_{\text{fs}}E_{\text{fs}}}{|a_{\text{fs}} - |E_{\text{fs}}||^2} + \frac{k_{\text{fn}}E_{\text{fn}}}{|a_{\text{fn}} - |E_{\text{fn}}||^2} + \frac{k_{\text{sn}}E_{\text{sn}}}{|a_{\text{sn}} - |E_{\text{sn}}||^2},\end{aligned} \tag{14}$$

with twelve material parameters $k_{ij}$ and $a_{ij}$, where $i, j$ are the six combinations of {f,s,n}, accounting for the symmetry of $\boldsymbol{E}$.

**Costa Law (2001)** The Costa Law, first proposed in 2001, is a Fung-type exponential strain energy function with seven material parameters $a$ and $b_{ij}$ with $i, j \in$ {f,s,n}. Mathematically, the Costa Law is

$$\psi_{\text{CL}} = \frac{1}{2}a(\exp Q - 1) \tag{15}$$

where

$$Q = b_{\text{ff}}E_{\text{ff}}^2 + b_{\text{ss}}E_{\text{ss}}^2 + b_{\text{nn}}E_{\text{nn}}^2 + b_{\text{fs}}E_{\text{fs}}^2 + b_{\text{fn}}E_{\text{fn}}^2 + b_{\text{sn}}E_{\text{sn}}^2. \tag{16}$$

**Separable Fung-type Law (2006)** The Separable Fung-type Law $\psi_{\text{SFL}}$ is a decoupled version of the Costa Law created by Schmid et al. in 2006 [15]. It includes multiple exponential functions, and twelve material parameters $a_{ij}, b_{ij}$ with $i, j \in$ {f,s,n},

$$\begin{aligned}\psi_{\text{SFL}} = &\frac{1}{2}(a_{\text{ff}}[\exp\{b_{\text{ff}}E_{\text{ff}}^2\} - 1] + a_{\text{ss}}[\exp\{b_{\text{ss}}E_{\text{ss}}^2\} - 1] + a_{\text{nn}}[\exp\{b_{\text{nn}}E_{\text{nn}}^2\} - 1] + a_{\text{fs}}[\exp\{b_{\text{fs}}E_{\text{fs}}^2\} - 1] \\ &+ a_{\text{sf}}[\exp\{b_{\text{sf}}E_{\text{sf}}^2\} - 1] + a_{\text{sn}}[\exp\{b_{\text{sn}}E_{\text{sn}}^2\} - 1]).\end{aligned} \tag{17}$$

**Holzapfel-Odgen Model (2009)** After its introduction in 2009, the Holzapfel-Ogden model quickly became popular due to its simple invariant-based formulation with only eight material parameters, and explicit consideration of theoretical material stability requirements useful for 3D finite element analysis. The Holzapfel-Ogden model is

$$\psi_{\mathrm{HO}} = \frac{a}{2b}\exp[b(I_1-3)] + \sum_{i=\mathrm{f,s}} \frac{a_i}{2b_i}\{\exp[b_i(I_{4i}-1)^2]-1\} + \frac{a_{\mathrm{fs}}}{2b_{\mathrm{fs}}}\left[\exp\left(b_{\mathrm{fs}}I_{8\mathrm{fs}}^2\right)-1)\right] \tag{18}$$

where the $I_{4f}$ term vanishes if $I_{4f} \leq 1$ due to physiological considerations [14]. The eight material parameters of the Holzapfel-Ogden model are $a, b, a_{\mathrm{f}}, b_{\mathrm{f}}, a_{\mathrm{s}}, b_{\mathrm{s}}, a_{\mathrm{fs}}, b_{\mathrm{fs}}$.

**Martonová Three-Term Model (2024)** Recently, Martonová et al. applied a constitutive neural network to derive SEFs for human cardiac tissue [30] using the triaxial shear and biaxial stretch data obtained by Sommer et al. [20] to train the network. The simplest orthotropic SEF found in [30] was

$$\psi_{\mathrm{MA}} = \frac{1}{2}\mu(I_2-3)^2 + \frac{a_{\mathrm{f}}}{2b_{\mathrm{f}}}\{\exp[b_f(I_{4\mathrm{f}}^*-1)^2]-1\} + \frac{a_{\mathrm{n}}}{2b_{\mathrm{n}}}\{\exp[b_{\mathrm{n}}(I_{4\mathrm{n}}^*-1)^2]-1\} \tag{19}$$

with $I_{4i}^* = \max\{I_{4i}, 1\}$. Thus, the five material parameters of the three-term Martonová model are $\mu, a_{\mathrm{f}}, a_{\mathrm{n}}, b_{\mathrm{f}}, b_{\mathrm{n}}$.

## Quality assessment of strain energy functions

We evaluated the quality of SEFs based on their utility for fitting the experimental data (Supplementary Table S1) and their complexity. Our aim was to find the SEF with the best possible fit (Equation 8) and the least possible complexity. For quantifying SEF complexity, we considered a variety of metrics related to the SEF's function tree representation; namely, the total number of nodes $l_{\mathrm{eq}}$, the number of material parameters $n_p$, the number of exponential functions $n_{\mathrm{exp}}$, and for the invariant-based SEFs ($\psi_{\mathrm{HO}}, \psi_{\mathrm{MA}}$, and CHESRA functions) the number of unique invariants $n_I$. Finally, we also considered goodness of fit landscapes in analogy to a profile likelihood analysis [59–61]. Here, we fixed the value of one parameter $p_{i,k}$ of dataset $i$ in turn at a value in the range [0, 100], including the original parameter estimate, and re-optimized the other parameters

$$\min_{\boldsymbol{p}_{i,l\neq k}} \sum_{j=1}^{n_i} \left(\boldsymbol{y}_{ij} - Y_i(\psi(\boldsymbol{x}_{ij}, \boldsymbol{p}_i))\right)^2 . \tag{20}$$

This optimization problem is derived from Equation 13. We used the goodness of fit landscapes to assess parameter identifiability, with sharp valleys indicating identifiable parameters, and flat landscapes indicating that parameters may not be identifiable from the data, providing a computationally efficient alternative to probabilistic sampling methods like Markov Chain Monte Carlo.

## Patient CMR biventricular geometry

To build our 3D simulation models, we used a biventricular mesh geometry [36] that was derived from a patient's end-diastolic cardiac MRI images collected at the National Heart Centre of Singapore. The patient gave written and informed consent to the research use of their data in accordance with the Declaration of Helsinki. The patient's ventricular geometry was cut below the LV valve plane, and discretized using 2396 vertices and 7362 linear tetrahedral elements, with edge lengths between 1.2-0.25 cm. Within the geometry, we generated synthetic fiber $\boldsymbol{e}_f$ and sheet $\boldsymbol{e}_s$ directions using the Laplace-Dirichlet algorithm [62] with fiber angles 60° and −60° on the endo- and epicardium respectively [36], and sheet angles −65° and 25° on the endo- and epicardium, respectively [62]. The synthetic fiber and sheet microstructures of the patient ventricle model are visualized in Supplementary Figure S7. The left and right ventricular pressures waveforms of the patient were estimated based on preset healthy waveforms, adjusted to the patient via a cuff-pressure measurement [36].

## In vivo and ex vivo simulation scenarios

To thoroughly test our SEF's, we considered two separate simulation scenarios, representing in vivo and ex vivo conditions. The in vivo simulations were calibrated to match LV pressures and CMR-derived LV volumes from late diastole [36]. In contrast, the ex vivo simulations were calibrated to match the empirically observed end-diastolic-pressure volume relation (EDPVR) of Klotz et al [35]. Calculating the Klotz EDPVR required a single end-diastolic pressure and volume measurement, which we obtained from the in vivo data. We then calculated the rest of the EDVPR by using the curve fitting method outlined by Klotz et al [35]. Note that the in vivo LV pressure-volume data is limited to a narrow range of physiological blood pressures observed in diastole, whereas the ex vivo EDPVR contains LV volumes spanning a wider pressure range, from the end-diastolic value of 2.19 kPa all the way to 0 kPa. The in vivo data, therefore, represented a more challenging simulation scenario, as the true in vivo zero-pressure LV volume was unknown, and the data were influenced by more complex physical conditions, including the effects of atrial contraction and acceleration forces. In contrast, the ex vivo Klotz EDPVR represented controlled laboratory conditions where the left ventricle was slowly inflated via an external fluid pump with the valve plane completely fixed [35], thereby excluding atrial and acceleration forces. We simulated the motion of the ventricles in both in vivo and ex vivo simulation scenarios using the same elastic model, albeit with different boundary conditions and compressibility properties tailored to each scenario. The elastic model is explained in more detail in the next section.

## 3D ventricular mechanics finite element framework

In our 3D simulations we calculated the ventricular wall displacement $\boldsymbol{u}$ at each point of the cardiac cycle using a previously published cardiac mechanics framework [36]. This framework was based on the quasi-static principle of virtual work using Lagrangian coordinates (Chapter 8.2 of [34])

$$\int_{\Omega} \boldsymbol{P} : \mathrm{Grad}\delta\boldsymbol{u}\; dV + \sum_{i\in\{\mathrm{lv},\,\mathrm{rv}\}} p_i \int_{\partial\Omega_i} J\boldsymbol{F}^{-T}\boldsymbol{N}\cdot\delta\boldsymbol{u}\; dS = 0. \tag{21}$$

Here the first term represents the internal work done by the elastic forces of the heart, whereas the last two terms represent the external work done by the left and right ventricular pressures $p_{\mathrm{lv}}$ and $p_{\mathrm{rv}}$. Furthermore, $\delta\boldsymbol{u}$ is a virtual variation in $\boldsymbol{u}$, $J = \det\boldsymbol{F}$, and $\boldsymbol{N}$ represents the boundary normal. The internal virtual work is evaluated over the heart walls $\Omega$, whereas the other integrals involve the left and right ventricular cavity walls $\partial\Omega_{\mathrm{lv}}$ and $\partial\Omega_{\mathrm{rv}}$.

In Equation 21 the first Piola-Kirchoff stress $\boldsymbol{P}$ is given by

$$\boldsymbol{P} = \frac{\partial\psi_{\mathrm{tot}}}{\partial\boldsymbol{F}}, \tag{22}$$

where we split the total elastic energy $\psi_{\mathrm{tot}}$ into deviatoric and volumetric components for numerical stability

$$\psi_{\mathrm{tot}} = \psi_{\mathrm{dev}}(J^{-\frac{1}{3}}\boldsymbol{F}) + \psi_{\mathrm{vol}}(J). \tag{23}$$

For the deviatoric component $\psi_{\mathrm{dev}}$ we used energy functions obtained from the literature and from CHESRA, whereas the volumetric component was tailored to the ex vivo and in vivo simulation scenarios. For the ex vivo scenario, we employed a three parameter Yeoh model [63]

$$\psi_{\mathrm{vol}} = \sum_{i=1}^{3} \frac{\kappa_i}{2}(J-1)^{2i}, \tag{24}$$

with the parameters set to $\kappa_1 = 7.31\mathrm{kPa}$, $\kappa_2 = 5.12\mathrm{kPa}$, and $\kappa_3 = 2.49 \times 10^6$ based on the human myocardium compression experiments of McEvoy et al [64]. With the Yeoh model configured this way, the compression stress increases moderately up to circa 95% compression, and increases rapidly thereafter (Figure 2C [64]). For the in vivo scenario, numerical difficulties with the Yeoh model necessitated the use of a simpler compressibility model

$$\psi_{\mathrm{vol}}(J) = \frac{\kappa}{2}(J-1)^2, \tag{25}$$

with $\kappa = 10\,\mathrm{kPa}$. This increased compressibility facilitated the calculation of stress-free reference configurations, which proved more demanding in the in vivo scenario owing to the larger volume changes compared to the ex vivo case.

To simulate the fixed valve plane in the ex vivo filling experiments [35], we fixed the ventricular base in all directions using a Dirichlet boundary condition. For the in vivo scenario, we instead employed a more flexible spring boundary condition to facilitate larger volume changes

$$\int_{\partial\Omega_{base}} k\boldsymbol{u} : \delta\boldsymbol{u}\, dS, \tag{26}$$

with $k = 10$kPa, in line with the compressible model (25). The term (26) was applied to the left ventricular base $\partial\Omega_{\text{base}}$, and contributed to the virtual work (21) in the vivo scenario to allow some in-plane motion at the valves. Similarly to the ex vivo case, the out-of-plane motion at the valves was eliminated in the in vivo scenario via a Dirichlet boundary condition.

We solved the virtual work equation (21) with a linear finite element discretization, and a Newton trust region solver implemented in the package PETSc-SNES. We set the absolute and relative tolerances of the Newton solver to $10^{-4}$, and used LU factorization for linear system solution. The linearized system was automatically derived by symbolic derivation of the virtual work (21) and automatically implemented into code via the FEniCS form compiler [65]. In cases of Newton solver divergence, the current pressure increments were halved, and the wall displacements were reset to their previous values before restarting the Newton algorithm.

## Estimation of pressure-free geometries

Calculating displacements within the Lagrangian framework (21) required that we specify a pressure-free reference geometry $\Omega$. However, from the CMR images, we only had knowledge of an in vivo pressurized geometry, which we denote $\omega$. We therefore estimated $\Omega$ from $\omega$ by applying an unloading algorithm based on the augmented Sellier's inverse method [66], which has previously been explored for the uniform scaling of SEF parameters [27]. In our version of the algorithm, we enabled a general, multi-parameter search for optimal SEF parameter values. This nested optimization adds an additional material parameter estimation loop so that the updated algorithm simultaneously estimates the pressure-free geometry $\Omega$ and the SEF parameters $\boldsymbol{p}$, which is essential for simulation model personalization, ensuring that the fitted SEF reflects an individual patient's physiology.

Our extended augmented Sellier method consisted of a two-step procedure, with an initial line search to obtain a good guess for $\boldsymbol{p}$, followed by a more exhaustive Nelder-Mead optimization [67]. Both methods are gradient-free, which was important because the volume-square loss (27) was insensitive to small material parameter changes, thereby invalidating finite-difference approximations of the volume-loss gradient. For the initial line search, we tested the uniform parameter values $\boldsymbol{p} \in \{0.5, 0.6, 0.7, 0.8, 0.9, 1, 1.5, 2, 2.5, 3\}$, and then used the resulting optimal $\boldsymbol{p}$ to initialize the more costly Nelder-Mead algorithm, where each component of $\boldsymbol{p}$ could be varied independently. For each SEF, we set an empirically determined parameter lower bound (Supplementary Table S8) to prevent non-convergence during Nelder-Mead optimization. The goal of both optimization steps was to minimize the left ventricular volume square loss

$$\mathcal{L}_{\text{vol}}(\boldsymbol{p}) = \sum_{i=1}^{N} \left[ V_{\text{lv}}^i - \tilde{V}_{\text{lv}}^i(\psi(\boldsymbol{p}), p_{\text{lv}}^i, p_{\text{rv}}^i) \right]^2, \tag{27}$$

where $V_{\text{lv}}^i, \tilde{V}_{\text{lv}}^i$ represented the target and simulated left ventricular volume at level $i$, respectively. For the in vivo scenarios, the index $i$ ran over $N = 24$ levels, with the exact volume and pressure values $V_{\text{lv}}^i, p_{\text{lv}}^i, p_{\text{rv}}^i$ obtained via quadratic interpolation from a base set of nine mid-diastolic values [36, Case 1]. For the ex vivo case, we used $N = 20$ points sampled from the fitted Klotz EDPVR [35], sampled with equal volume increments. At each volume-pressure level $i$, we calculated the simulated volume $\tilde{V}_{\text{lv}}^i$ using the divergence theorem

$$\tilde{V}_{\text{lv}}^i(\psi(\boldsymbol{p}), p_{\text{lv}}^i, p_{\text{rv}}^i) = -\int_{\partial\Omega_{\text{lv}}} \frac{1}{3}(\boldsymbol{X} + \boldsymbol{u}^i)J\boldsymbol{F}^{-T} \cdot \boldsymbol{N} dS, \tag{28}$$

where $\boldsymbol{X}$ represents the reference coordinates of $\Omega$, and $\boldsymbol{u}^i = \boldsymbol{u}^i(\psi(\boldsymbol{p}), p_{\text{lv}}^i, p_{\text{rv}}^i)$ the deformation field resulting from the in vivo loading conditions at pressure level $i$.

Each time $\boldsymbol{p}$ was updated, we re-estimated the pressure-free geometry $\Omega$ using the original augmented Sellier method, which we present here again in an adapted form (Algorithm 1). Our inputs to the Sellier method were: the parameterized SEF $\psi(\boldsymbol{p})$, the in vivo MRI geometry at end-diastole $\omega$, and the end-diastolic pressures $p_{\mathrm{lv}}^{\mathrm{ED}}, p_{\mathrm{rv}}^{\mathrm{ED}}$, so that $\Omega = \Omega(\omega, \psi(\boldsymbol{p}), p_{\mathrm{lv}}^{\mathrm{ED}}, p_{\mathrm{rv}}^{\mathrm{ED}})$. For $\psi_{\mathrm{CH}_1}, \psi_{\mathrm{HO}}, \psi_{\mathrm{MA}}$ we set the Sellier tolerance to 0.1 and the Nelder-Mead tolerance to 0.05. For $\psi_{\mathrm{CH}_2}$ we adjusted these values to 0.05 and 0.1, respectively, to prevent a premature convergence that we experienced during pressure-free geometry estimation.

**Algorithm 1** Augmented Sellier's Inverse Method with Variable Material Properties

**input data**
- In vivo mesh coordinate vector $\boldsymbol{x}^0 \in \omega$
- Strain energy function $\psi$ and material parameters $\boldsymbol{p}$
- In vivo left and right end-diastolic ventricular pressures $p_{\mathrm{lv}}^{\mathrm{ED}}, p_{\mathrm{rv}}^{\mathrm{ED}}$
- Tolerance $\epsilon$

**initialization**
- Reference mesh coordinate vector $\boldsymbol{X}^1 \leftarrow \boldsymbol{x}^0$
- $k \leftarrow 0$
- $\alpha = 0.5$

**while** max $||\boldsymbol{R}^k|| > \epsilon$ **do**
- Update counter, $k \leftarrow k + 1$
- Solve forward problem (21) for displacement $\boldsymbol{u}^k \left(\boldsymbol{X}^k, \psi(\boldsymbol{p}), p_{\mathrm{lv}}^{\mathrm{ED}}, p_{\mathrm{rv}}^{\mathrm{ED}}\right)$
- Set $\boldsymbol{x}^k \leftarrow \boldsymbol{X}^k + \boldsymbol{u}^k$
- Calculate nodal error vector, $\boldsymbol{R}^k = \boldsymbol{x}^k - \boldsymbol{x}^0$
- **if** $k > 1$ **then**
  - Update alpha, $\alpha \leftarrow -\alpha \frac{\boldsymbol{R}^{k-1} : \left[\boldsymbol{R}^k - \boldsymbol{R}^{k-1}\right]}{\left[\boldsymbol{R}^k - \boldsymbol{R}^{k-1}\right] : \left[\boldsymbol{R}^k - \boldsymbol{R}^{k-1}\right]}$
- **end if**
- Update reference vector, $\boldsymbol{X}^{k+1} \leftarrow \boldsymbol{X}^k - \alpha \boldsymbol{R}^k$

**end while**

**output**
- Pressure-free reference coordinates $\boldsymbol{X}^k$ of the geometry $\Omega$.

## Parameter variability benchmark with 3D simulation models

In the 3D benchmark, we optimized the material parameters of each SEF to match sparse and noisy diastolic displacement data for both the ex vivo and in vivo scenarios. We performed each optimization using the truncated Newton method [68] available from the Python package Scipy as the "TNC" solver, with convergence tolerance set to $10^{-6}$. During each TNC iteration, we calculated the gradient of the target function (Equation 29) using an efficient adjoint-gradient method [69], which we implemented using the finite element framework FEniCS [65]. For each SEF, we used a set of twenty random initialization points obtained from a Latin hypercube sample, with lower and upper bounds set to span an order of magnitude around the values obtained from the pressure-free geometry estimation. In particular, if $\boldsymbol{p}(\psi)$ were the material parameters estimated during the pressure-free geometry calculation with SEF $\psi$, then to initialize the Nelder-Mead search, we generated initial guesses using the Latin hypercube sample within a continuous range defined by $10^{\log_{10} \boldsymbol{p}(\psi) \pm 0.5}$. This logarithmic window ensured a broad search across an order of magnitude for each material parameter.

The target function for the 3D parameter estimations was the displacement square loss

$$\mathcal{L}_{\mathrm{disp}}(\boldsymbol{p}) = \sum_{i=1}^{N_{\mathrm{opt}}} \int_{\Omega(\psi, S)} \left(\boldsymbol{u}_{\mathrm{target}}^i - \boldsymbol{u}_{\mathrm{sim}}^i(\boldsymbol{p})\right)^2 dV, \tag{29}$$

where represents $\Omega(\psi, S)$ the pressure-free geometry calculated for the SEF $\psi$ in scenario $S \in \{\text{in vivo, ex vivo}\}$, $\boldsymbol{u}_{\mathrm{target}}^i$ the target displacements, and $\boldsymbol{u}_{\mathrm{sim}}^i(\boldsymbol{p})$ the simulated displacements that depended on the material parameters $\boldsymbol{p}$. In the in vivo case the displacements were evaluated at $N_{\mathrm{opt}} = 3$ points in the cardiac cycle; consisting of mid and end-diastole, with a third point in-between. The left ventricular pressures at these three points were 1.3, 1.8, and 2.2 kPa, respectively, with corresponding right ventricular pressures of 0.19, 0.46, 0.72 kPa. These points were chosen to represent the ventricular passive filling phase of the cardiac cycle, during which elastic forces dominate. For the ex vivo case, we selected twenty points spanning the entire fitted Klotz EDPVR [35], with equally spaced volume increments.

We pre-calculated the target displacements $\boldsymbol{u}^i_{\text{target}}$ using benchmark material parameter values obtained from the pressure-free geometry estimation. To reflect real-world conditions, we added random Gaussian noise to the target displacements, with magnitudes equal to 50% of the local displacement norm. Taken together, the sparsely sampled noisy displacements created challenging inverse parameter estimation scenarios, mimicking real-world conditions present in clinical cardiac imaging data for our in vivo case. In contrast, the ex vivo case represented a more tightly controlled scenario with laboratory conditions and more abundant data, thereby enabling comparisons of in vivo and ex vivo conditions.

## Software

All custom code, including CHESRA, was implemented in Python 3.10.6, using the packages DEAP, lmfit, sympy, multiprocessing, and matplotlib. These packages were used for genetic algorithm coordination, parameter fitting, parallel processing, and data plotting, respectively. We created 3D visualisations using ParaView, and edited our figure layouts using Inkscape. Automated function simplifications were performed with sympy. We used WebPlotDigitizer to digitize the experimental data listed in Supplementary Table S1. 3D ventricular mechanics simulations were performed using FEniCS 2019.1.0 [65].

## Data Availability

Raw data for figures are available at https://doi.org/10.6084/m9.figshare.29544404 [70]. Digitized data summarized in Supplementary Table S1, and personalized biventricular cardiac mechanics data are available at https://github.com/GabrielBalabanResearch/CHESRA.

## Code Availability

The source code for CHESRA is available at https://github.com/GabrielBalabanResearch/CHESRA.

## Author Contributions

SO contributed to the study design, implemented CHESRA software, ran experiments, analysed results, wrote substantial parts of the manuscript and provided feedback and commentary. KF contributed to the study design, implemented CHESRA software, ran experiments, analysed results, contributed to early versions of the manuscript, and provided feedback and commentary. LLR contributed to the study design, implemented CHESRA software, provided technical support for running experiments, contributed to early versions of the manuscript and provided feedback and commentary. MMM provided feedback on the study design during key time-points, supervised GB (2022- 2023) and provided manuscript feedback and commentary. ADM provided feedback on the study design during key time-points and provided manuscript feedback and commentary. VT contributed to the study design and data analysis, and provided feedback and commentary on the manuscript. GB contributed to study design, implemented FEM software and ran and analysed the 3D biventricular simulation model benchmark, assisted with data analysis for other experiments, wrote substantial parts of the manuscript, and provided feedback and commentary. VT and GB contributed jointly to supervision.

## Acknowledgements

Special thanks to Kimberly McCabe and Nickolas Forsch for organizing the Simula Summer School for Computational Physiology 2022 that kick-started the project which resulted in this paper, and to Henrik Finsberg and Lik Chuan for assistance with the preparation of the 3D biventricular geometry. Special thanks to Peter Kohl, Jens Timmer and Bogdan Marculescu for reviewing the manuscript. SO was supported by the Joachim Herz Foundation, the Baden-Württemberg Stiftung under project ArrhythMEC, and is a member of SFB1425, funded by the Deutsche Forschungsgemeinschaft (DFG, German Research Foundation) - Project #422681845. Moreover, funding was provided by a BBSRC PhD iCASE (BB/V509395/1) and Russell Studentship Agreement with AstraZeneca (R67719/CN001) to LLR. VT is funded by the Deutsche Forschungsgemeinschaft (DFG, German Research Foundation) – Project-ID 403222702 – SFB 1381, the Cluster of Excellence 'Centre for Integrative Biological Signalling Studies' (CIBSS) by the DFG under Germany's Excellence Strategy – EXC-2189 – 390939984, by the Baden-Württemberg Stiftung under project ArrhythMEC, and by the Hans A. Krebs Medical Scientist Program, Faculty of Medicine, University of Freiburg. ADM is a co-founder and equity-holder in Vektor Medical, Inc., which was not involved in this research. MMM and GB received funding from the Research Council of Norway via ProCardio, Senter for Innovativ Behandling av Hjertesykdommer, SFI IV 2020–2028, project number 309762. During the preparation of this work, the authors used ChatGPT (OpenAI), Perplexity AI (Perplexity AI, Inc.), Claude (Anthropic), Gemini (Google AI), and DeepL (DeepL SE) in order to improve readability and language. After using this service, the authors reviewed and edited the content as needed and take full responsibility for the content of the publication.

## Competing Interests

ADM is a co-founder and equity holder in Vektor Medical Inc. and is required by UC San Diego to disclose this relationship. Vektor Medical had no involvement whatsoever in any aspect of the conception, funding or conduct of this research. All other authors declare no financial or non-financial competing interests.

# References

1. Corral-Acero, J. *et al.* The 'Digital Twin' to enable the vision of precision cardiology. *Eur. Heart J.* **41,** 4556–4564 (2020).

2. Li, L., Camps, J., Rodriguez, B. & Grau, V. Solving the inverse problem of electrocardiography for cardiac digital twins: A survey. *IEEE Rev. Biomed. Eng.* **18,** 316–336 (2024).
3. Coorey, G. *et al.* The health digital twin to tackle cardiovascular disease—a review of an emerging interdisciplinary field. *npj Digit. Med.* **5,** 126 (2022).
4. National Academies of Sciences, Engineering, and Medicine. *Foundational Research Gaps and Future Directions for Digital Twins* (The National Academies Press, Washington, DC, 2024).
5. Grieves, M. & Vickers, J. *Digital twin: mitigating unpredictable, undesirable emergent behavior in complex systems* in *Transdisciplinary perspectives on complex systems: new findings and approaches* 85–113 (Springer, 2016).
6. Sel, K. *et al.* Building digital twins for cardiovascular health: from principles to clinical impact. *J. Am. Heart Assoc.* **13,** e031981 (2024).
7. Sarani Rad, F., Bitaraf, E., Jafarpour, M. & Li, J. Technologies, clinical applications, and implementation barriers of digital twins in precision cardiology: systematic review. *JMIR Cardio* **10,** e78499 (2026).
8. Kolk, M. Z. H. *et al.* Dynamic prediction of malignant ventricular arrhythmias using neural networks in patients with an implantable cardioverter-defibrillator. *EBioMedicine* **99** (2024).
9. Kolk, M. Z. H. *et al.* Artificial intelligence–enhanced wearable technology enables ventricular arrhythmia prediction. *Eur. Heart J. Digit. Health* **7,** ztae069 (2026).
10. Von Luxburg, U. & Schölkopf, B. *Statistical learning theory: Models, concepts, and results* in *Handbook of the History of Logic* (eds Gabbay, D. M., Hartmann, S. & Woods, J.) 651–706 (Elsevier, 2011).
11. Lunde, I. G., Rypdal, K. B., Van Linthout, S., Diez, J. & González, A. Myocardial fibrosis from the perspective of the extracellular matrix: mechanisms to clinical impact. *Matrix Biol.* (2024).
12. Balaban, G. *et al.* In vivo estimation of elastic heterogeneity in an infarcted human heart. *Biomech. Model. Mechanobiol.* **17,** 1317–1329 (2018).
13. Mandinov, L., Eberli, F. R., Seiler, C. & Hess, O. M. Diastolic heart failure. *Cardiovasc. Res.* **45,** 813–825 (2000).
14. Holzapfel, G. A. & Ogden, R. W. Constitutive modelling of passive myocardium: a structurally based framework for material characterization. *Philos. Trans. R. Soc. A* **367,** 3445–3475 (2009).
15. Schmid, H., Nash, M., Young, A. & Hunter, P. Myocardial material parameter estimation—a comparative study for simple shear. *J. Biomech. Eng.* **128,** 742–750 (2006).
16. Costa, K. D., Holmes, J. W. & McCulloch, A. D. Modelling cardiac mechanical properties in three dimensions. *Philos. Trans. R. Soc. A* **359,** 1233–1250 (2001).
17. Hunter, P. J. Computational electromechanics of the heart. *Computational Biology of the Heart,* 345–407 (1997).
18. Yin, F. C., Strumpf, R. K., Chew, P. H. & Zeger, S. L. Quantification of the mechanical properties of noncontracting canine myocardium under simultaneous biaxial loading. *J. Biomech.* **20,** 577–589 (1987).
19. Novak, V. P., Yin, F. & Humphrey, J. Regional mechanical properties of passive myocardium. *J. Biomech.* **27,** 403–412 (1994).
20. Sommer, G. *et al.* Biomechanical properties and microstructure of human ventricular myocardium. *Acta Biomater.* **24,** 172–192 (2015).
21. Dokos, S., Smaill, B. H., Young, A. A. & LeGrice, I. J. Shear properties of passive ventricular myocardium. *Am. J. Physiol. Heart Circ. Physiol.* **283,** H2650–H2659 (2002).
22. Remme, E. W. *et al.* Development of an in vivo method for determining material properties of passive myocardium. *J. Biomech.* **37,** 669–678 (2004).
23. Hadjicharalambous, M. *et al.* Analysis of passive cardiac constitutive laws for parameter estimation using 3D tagged MRI. *Biomech. Model. Mechanobiol.* **14,** 807–828 (2015).
24. Nasopoulou, A. *et al.* Improved identifiability of myocardial material parameters by an energy-based cost function. *Biomech. Model. Mechanobiol.* **16,** 971–988 (2017).
25. Krishnamurthy, A. *et al.* Patient-specific models of cardiac biomechanics. *J. Comput. Phys.* **244,** 4–21 (2013).
26. Sack, K. L. *et al.* Construction and validation of subject-specific biventricular finite-element models of healthy and failing swine hearts from high-resolution DT-MRI. *Front. Physiol.* **9,** 539 (2018).
27. Marx, L. *et al.* Robust and efficient fixed-point algorithm for the inverse elastostatic problem to identify myocardial passive material parameters and the unloaded reference configuration. *J. Comput. Phys.* **463.** ISSN: 10902716 (Aug. 2022).
28. Shi, L., Chen, I. Y., Takayama, H. & Vedula, V. An optimization framework to personalize passive cardiac mechanics. *Comput. Methods Appl. Mech. Eng.* **432,** 117401 (2024).

29. Latorre, M. & Montáns, F. J. WYPiWYG hyperelasticity without inversion formula: Application to passive ventricular myocardium. *Comput. Struct.* **185,** 47–58 (2017).
30. Martonová, D. *et al.* Automated model discovery for human cardiac tissue: discovering the best model and parameters. *Comput. Methods Appl. Mech. Eng.* **428,** 117078 (2024).
31. Martonová, D., Leyendecker, S., Holzapfel, G. A. & Kuhl, E. Discovering dispersion: How robust is automated model discovery for human myocardial tissue? D. Martonová et al. *Biomech. Model. Mechanobiol.* **24,** 1–15 (2025).
32. Gültekin, O. *et al.* A Physics-Informed Neural Network Model for the Anisotropic Hyperelasticity of the Human Passive Myocardium. *Int. J. Numer. Methods Eng.* **126,** e70067 (2025).
33. Ludwicki, K. *et al. An Automated Cardiac Constitutive Modelling Framework with Evolutionary Strain Energy Functions* in *Computational Physiology: Simula Summer School 2022 – Student Reports* (ed Sundnes, J.) 1–17 (Springer Nature Switzerland, 2023).
34. Holzapfel, G. A. *Nonlinear solid mechanics: a continuum approach for engineering science* (Kluwer Academic Publishers Dordrecht, 2002).
35. Klotz, S. *et al.* Single-beat estimation of end-diastolic pressure-volume relationship: a novel method with potential for noninvasive application. *Am. J. Physiol. Heart Circ. Physiol.* **291,** H403–H412 (2006).
36. Finsberg, H. *et al.* Efficient estimation of personalized biventricular mechanical function employing gradient-based optimization. *Int. J. Numer. Methods Biomed. Eng.* **34,** e2982 (2018).
37. Aliev, R. R. & Panfilov, A. V. A simple two-variable model of cardiac excitation. *Chaos Solitons Fractals* **7,** 293–301 (1996).
38. Mitchell, C. C. & Schaeffer, D. G. A two-current model for the dynamics of cardiac membrane. *Bull. Math. Biol.* **65,** 767–793 (2003).
39. Abdusalamov, R., Hillgärtner, M. & Itskov, M. Automatic generation of interpretable hyperelastic material models by symbolic regression. *Int. J. Numer. Methods Eng.* **124,** 2093–2104 (2023).
40. Hou, J., Chen, X., Wu, T., Kuhl, E. & Wang, X. Automated data-driven discovery of material models based on symbolic regression: A case study on the human brain cortex. *Acta Biomater.* **188,** 276–296 (2024).
41. Moon, H. *et al.* Physics-informed neural network-based discovery of hyperelastic constitutive models from extremely scarce data. *Comput. Methods Appl. Mech. Eng.* **446,** 118258 (2025).
42. Guan, D. *et al.* On the AIC-based model reduction for the general Holzapfel–Ogden myocardial constitutive law. *Biomech. Model. Mechanobiol.* **18,** 1213–1232 (2019).
43. Hadjicharalambous, M. *et al.* Investigating the reference domain influence in personalised models of cardiac mechanics: Effect of unloaded geometry on cardiac biomechanics. *Biomech. Model. Mechanobiol.* **20,** 1579–1597 (2021).
44. Lazarus, A., Dalton, D., Husmeier, D. & Gao, H. Sensitivity analysis and inverse uncertainty quantification for the left ventricular passive mechanics. *Biomech. Model. Mechanobiol.* **21,** 953–982 (2022).
45. Mojsejenko, D. *et al.* Estimating passive mechanical properties in a myocardial infarction using MRI and finite element simulations. *Biomech. Model. Mechanobiol.* **14,** 633–647 (2015).
46. Hadjicharalambous, M. *et al.* Non-invasive model-based assessment of passive left-ventricular myocardial stiffness in healthy subjects and in patients with non-ischemic dilated cardiomyopathy. *Ann. Biomed. Eng.* **45,** 605–618 (2017).
47. Ahmad Bakir, A., Al Abed, A., Stevens, M. C., Lovell, N. H. & Dokos, S. A multiphysics biventricular cardiac model: Simulations with a left-ventricular assist device. *Front. Physiol.* **9,** 1259 (2018).
48. Lee, A. W., Costa, C. M., Strocchi, M., Rinaldi, C. A. & Niederer, S. A. Computational modeling for cardiac resynchronization therapy. *J. Cardiovasc. Transl. Res.* **11,** 92–108 (2018).
49. Priego, L. *et al.* Integration of Electrophysiological and Mechanical Biomarkers in Cardiac Risk Assessment Models. *Comput. Methods Programs Biomed.,* 108896 (2025).
50. Holzapfel, G. A., Niestrawska, J. A., Ogden, R. W., Reinisch, A. J. & Schriefl, A. J. Modelling non-symmetric collagen fibre dispersion in arterial walls. *J. R. Soc. Interface* **12,** 20150188 (2015).
51. Papandrinopoulou, D., Tzouda, V. & Tsoukalas, G. Lung compliance and chronic obstructive pulmonary disease. *Pulm. Med.* **2012,** 542769 (2012).
52. Fahmy, Y., Trabia, M. B., Ward, B., Gallup, L. & Froehlich, M. Development of an Anisotropic Hyperelastic Material Model for Porcine Colorectal Tissues. *Bioengineering* **11,** 64 (2024).

53. Mihai, L. A., Budday, S., Holzapfel, G. A., Kuhl, E. & Goriely, A. A family of hyperelastic models for human brain tissue. *J. Mech. Phys. Solids* **106,** 60–79 (2017).
54. Gültekin, O. *et al.* A Physics-Informed Neural Network Model for the Anisotropic Hyperelasticity of the Human Passive Myocardium. *Int. J. Numer. Methods Eng.* **126,** e70067 (2025).
55. Pfaller, M. R. *et al.* The importance of the pericardium for cardiac biomechanics: from physiology to computational modeling: MR Pfaller et al. *Biomech. Model. Mechanobiol.* **18,** 503–529 (2019).
56. Ball, J. M. Convexity conditions and existence theorems in nonlinear elasticity. *Arch. Ration. Mech. Anal.* **63,** 337–403 (1976).
57. Campostrini, G. *et al.* Generation, functional analysis and applications of isogenic three-dimensional self-aggregating cardiac microtissues from human pluripotent stem cells. *Nat. Protoc.* **16,** 2213–2256 (2021).
58. Smaill, B. & Hunter, P. *Structure and function of the diastolic heart: material properties of passive myocardium* in *Theory of Heart: Biomechanics, Biophysics, and Nonlinear Dynamics of Cardiac Function* (eds Glass, L., Hunter, P. & McCulloch, A.) 1–29 (Springer, 1991).
59. Wieland, F.-G., Hauber, A. L., Rosenblatt, M., Tönsing, C. & Timmer, J. On structural and practical identifiability. *Curr. Opin. Syst. Biol.* **25,** 60–69 (2021).
60. Raue, A. *et al.* Structural and practical identifiability analysis of partially observed dynamical models by exploiting the profile likelihood. *Bioinformatics* **25,** 1923–1929 (2009).
61. Kreutz, C., Raue, A. & Timmer, J. Likelihood based observability analysis and confidence intervals for predictions of dynamic models. *BMC Syst. Biol.* **6,** 1–9 (2012).
62. Bayer, J. D., Blake, R. C., Plank, G. & Trayanova, N. A. A novel rule-based algorithm for assigning myocardial fiber orientation to computational heart models. *Ann. Biomed. Eng.* **40,** 2243–2254 (2012).
63. Yeoh, O. H. Some forms of the strain energy function for rubber. *Rubber Chem. Technol.* **66,** 754–771 (1993).
64. McEvoy, E., Holzapfel, G. A. & McGarry, P. Compressibility and anisotropy of the ventricular myocardium: experimental analysis and microstructural modeling. *J. Biomech. Eng.* **140,** 081004 (2018).
65. Alnæs, M. *et al.* The FEniCS project version 1.5. *Arch. Numer. Softw.* **3** (2015).
66. Rausch, M. K., Genet, M. & Humphrey, J. D. An augmented iterative method for identifying a stress-free reference configuration in image-based biomechanical modeling. *J. Biomech.* **58,** 227–231 (2017).
67. Gao, F. & Han, L. Implementing the Nelder-Mead simplex algorithm with adaptive parameters. *Comput. Optim. Appl.* **51,** 259–277 (2012).
68. Nash, S. G. Newton-type minimization via the Lanczos method. *SIAM J. Numer. Anal.* **21,** 770–788 (1984).
69. Balaban, G., Alnæs, M. S., Sundnes, J. & Rognes, M. E. Adjoint multi-start-based estimation of cardiac hyperelastic material parameters using shear data. *Biomech. Model. Mechanobiol.* **15,** 1509–1521 (2016).
70. Ohnemus, S. *et al.* *Raw data* https://doi.org/10.6084/m9.figshare.29544404.v1. July 2025.

# Supplementary Information

## Interpretation of the CHESRA energy function parameters

To interpret the material parameters in the two CHESRA strain energy functions (SEFs), we considered how the parameters related to the invariants in expanded forms of the SEFs. We supplemented this analysis with loading experiments (Supplementary Figure S3) to give more meaningful names to the parameters. Starting with $\psi_{\mathrm{CH}_1}$, we can see that its expanded form is given by

$$\psi_{\mathrm{CH}_1} = p_1 p_3 (\tilde{I}_{8\mathrm{fs}} + \tilde{I}_{5\mathrm{f}}) + p_2 \tilde{I}_1 + p_3 \tilde{I}_1 (\tilde{I}_{8\mathrm{fs}} + \tilde{I}_{5\mathrm{f}}), \tag{S1}$$

isolating each of the material parameters individually we get

$$\begin{aligned}\psi_{\mathrm{CH}_1} &= p_1 \left(p_3 (\tilde{I}_{8\mathrm{fs}} + \tilde{I}_{5\mathrm{f}})\right) && + p_2 \tilde{I}_1 + p_3 \tilde{I}_1 (\tilde{I}_{8\mathrm{fs}} + \tilde{I}_{5\mathrm{f}}) \\ &= p_2 \tilde{I}_1 && + p_1 p_3 (\tilde{I}_{8\mathrm{fs}} + \tilde{I}_{5\mathrm{f}}) + p_3 \tilde{I}_1 (\tilde{I}_{8\mathrm{fs}} + \tilde{I}_{5\mathrm{f}}) \\ &= p_3 \left(p_1 (\tilde{I}_{8\mathrm{fs}} + \tilde{I}_{5\mathrm{f}}) + \tilde{I}_1 (\tilde{I}_{8\mathrm{fs}} + \tilde{I}_{5\mathrm{f}})\right) && + p_2 \tilde{I}_1 .\end{aligned} \tag{S2}$$

Here, $p_1$ affects the transversely isotropic invariant $\tilde{I}_{5\mathrm{f}}$ and the coupling invariant $\tilde{I}_{8\mathrm{fs}}$ and can therefore be related to fiber-stretch and fiber-sheet shear. We can also see that the parameter $p_2$ acts exclusively on the isotropic invariant $\tilde{I}_1$ and can be interpreted as an isotropic parameter. Finally, $p_3$ acts on all thee invariants $\{\tilde{I}_1, \tilde{I}_{5\mathrm{f}}, \tilde{I}_{8\mathrm{fs}}\}$ in $\psi_{\mathrm{CH}_1}$, and couples the fiber-stretch and fiber-sheet shear response with the overall isotropic strain level. Examining the impact of each material parameter on the stress-strain curves (Supplementary Figure S3), we can see that $p_1$ most strongly influences the fiber direction under uniaxial and equibiaxial loading, while $p_2$ impacts all directions under uniaxial and equibiaxial loading (Supplementary Figure S3a). In contrast, $p_3$ has the strongest effect under shear loading. Putting everything together, we suggest the following names for the parameters of $\psi_{\mathrm{CH}_1}$

$$\begin{aligned} p_1 &\rightarrow p_{\text{f-fs}}, \\ p_2 &\rightarrow p_{\text{iso}}, \\ p_3 &\rightarrow p_{\text{coup}}. \end{aligned} \tag{S3}$$

For $\psi_{\mathrm{CH}_2}$, the expanded form is given by

$$\psi_{\mathrm{CH}_2} = p_1 (p_2 p_4 \tilde{I}_1 + p_3 p_4 \tilde{I}_{5\mathrm{f}} + p_4 \tilde{I}_1 \tilde{I}_{5\mathrm{f}} + p_2 p_3 \tilde{I}_{5\mathrm{s}} + p_2 \tilde{I}_1 \tilde{I}_{5\mathrm{s}} + p_3 \tilde{I}_{5\mathrm{f}} \tilde{I}_{5\mathrm{s}} + \tilde{I}_1 \tilde{I}_{5\mathrm{f}} \tilde{I}_{5\mathrm{s}}). \tag{S4}$$

isolating each of the material parameters individually we get

$$\begin{aligned}\psi_{\mathrm{CH}_2} &= p_1 \left(p_2 (p_4 \tilde{I}_1 + p_3 \tilde{I}_{5\mathrm{s}} + \tilde{I}_1 \tilde{I}_{5\mathrm{s}}) + p_3 p_4 \tilde{I}_{5\mathrm{f}} + p_4 \tilde{I}_1 \tilde{I}_{5\mathrm{f}} + p_3 \tilde{I}_{5\mathrm{f}} \tilde{I}_{5\mathrm{s}} + \tilde{I}_1 \tilde{I}_{5\mathrm{f}} \tilde{I}_{5\mathrm{s}}\right) \\ &= p_2 \left(p_1 (p_4 \tilde{I}_1 + p_3 \tilde{I}_{5\mathrm{s}} + \tilde{I}_1 \tilde{I}_{5\mathrm{s}})\right) && + p_1 p_3 p_4 \tilde{I}_{5\mathrm{f}} + p_1 p_4 \tilde{I}_1 \tilde{I}_{5\mathrm{f}} + p_1 p_3 \tilde{I}_{5\mathrm{f}} \tilde{I}_{5\mathrm{s}} + p_1 \tilde{I}_1 \tilde{I}_{5\mathrm{f}} \tilde{I}_{5\mathrm{s}} \\ &= p_3 \left(p_1 (p_2 \tilde{I}_{5\mathrm{s}} + p_4 \tilde{I}_{5\mathrm{f}} + \tilde{I}_{5\mathrm{f}} \tilde{I}_{5\mathrm{s}})\right) && + p_1 p_2 p_4 \tilde{I}_1 + p_1 p_4 \tilde{I}_1 \tilde{I}_{5\mathrm{f}} + p_1 p_2 \tilde{I}_1 \tilde{I}_{5\mathrm{s}} + p_1 \tilde{I}_1 \tilde{I}_{5\mathrm{f}} \tilde{I}_{5\mathrm{s}} \\ &= p_4 \left(p_1 (p_2 \tilde{I}_1 + p_3 \tilde{I}_{5\mathrm{f}} + \tilde{I}_1 \tilde{I}_{5\mathrm{f}})\right) && + p_1 p_2 p_3 \tilde{I}_{5\mathrm{s}} + p_1 p_2 \tilde{I}_1 \tilde{I}_{5\mathrm{s}} + p_1 p_3 \tilde{I}_{5\mathrm{f}} \tilde{I}_{5\mathrm{s}} + p_1 \tilde{I}_1 \tilde{I}_{5\mathrm{f}} \tilde{I}_{5\mathrm{s}} .\end{aligned} \tag{S5}$$

Here, $p_1$ multiplies every term in the SEF, so it can be related to the overall global stiffness. In contrast, $p_2$ and $p_4$ only appear in terms containing the the isotropic invariant $\tilde{I}_1$ and/or the transversely isotropic invariant $\tilde{I}_{5\mathrm{s}}$ or $\tilde{I}_{5\mathrm{f}}$, respectively. Lastly, $p_3$ appears only in terms that contain $\tilde{I}_{5\mathrm{s}}$ and/or $\tilde{I}_{5\mathrm{f}}$. In the loading experiments (Supplementary Figure S3b), we noted that effect of the material parameters on the deformation modes was intertwined: all four parameters affected the response in uniaxial and biaxial loading, whereas $p_1$ and $p_4$ had the most pronounced effect in shear experiments. Based on these considerations, we suggest the following names for the parameters of $\psi_{\mathrm{CH}_2}$

$$\begin{aligned} p_1 &\rightarrow p_{\text{glob}}, \\ p_2 &\rightarrow p_{\text{s-iso}}, \\ p_3 &\rightarrow p_{\text{f-s}}, \end{aligned}$$

$$p_4 \rightarrow p_{\text{f-iso}}. \tag{S6}$$

| Protocol | First Author | Year | Stress | Strain | Species | Region | Subsets | Reference |
|---|---|---|---|---|---|---|---|---|
| Biaxial | Yin | 1987 | $\boldsymbol{S}$ | $\boldsymbol{E}$ | Dog | LVFW (sub-epicardium) | 1 | Fig. 4 [1] |
| Biaxial | Novak | 1994 | $\boldsymbol{\sigma}$ | $\lambda$ | Dog | LVFW (sub-epicardium) | 1 | Fig. 4 [2] |
| Equibiaxial | Novak | 1994 | $\boldsymbol{\sigma}$ | $\lambda$ | Dog | LVFW (sub-endo., mid-myo., sub-epi.), mid-septum | 8 | Fig. 1-2 [2] |
| Biaxial | Sommer | 2015 | $\boldsymbol{\sigma}$ | $\lambda$ | Human | LVFW (sub-endo., mid-myo., sub-epi.), mid-septum | 1 | Fig. 9 [3] |
| Shear | Dokos | 2002 | $\boldsymbol{\sigma}$ | $\gamma$ | Pig | LVFW (mid-myocardium) | 1 | Fig. 6 [4] |
| Shear | Sommer | 2015 | $\boldsymbol{\sigma}$ | $\lambda$ | Human | LVFW (sub-endo., mid-myo., sub-epi.), mid-septum | 1 | Fig. 13 [3] |

**Supplementary Table S1: Summary of the experimental datasets used in CHESRA.** The datasets included various mechanical protocols, animal species, and heart regions, as well as measurements recorded using various stress and strain measures. The strain measures were simple shear amount $\gamma$, tissue extension $\lambda$, and Green-Lagrange strain $\boldsymbol{E} = \frac{1}{2}\boldsymbol{C} - \boldsymbol{I}$. The stress measures were Cauchy stress $\boldsymbol{\sigma}$ and 2nd Piola-Kirchoff Stress $\boldsymbol{S}$. Heart regions comprise the left ventricular free wall (LVFW) and septum. Note that the equibiaxial dataset of Novak et al. contained eight distinct subsets reflecting tissue samples with varying material properties.

| Hyperparameter | Value(s) | Description |
|---|---|---|
| $n_{\text{gen}}$ | 50, (*30 for hyperparameter search*) | Number of SEF generations |
| $n_{\text{ind}}$ | 200, (*50 for hyperparameter search*) | Number of SEFs per generation |
| $n_{\text{hof}}$ | 20 | Number of SEFs selected by elitism |
| $l_{\text{init}}$ | 5 | Number of initial function extensions |
| $\alpha$ | $0, 1 \times 10^{-4}, 2 \times 10^{-4}, 3 \times 10^{-4}, 5 \times 10^{-4}, 1 \times 10^{-3}, 5 \times 10^{-3}, 1 \times 10^{-2}$ | SEF length penalty |
| $p_{\text{mate}}$ | $0, 0.25, 0.5, 0.75$ | SEF mating probability |
| $p_{\text{mutate}}$ | $0, 0.25, 0.5, 0.75$ | SEF mutation probability |
| $p_{\text{reduce}}$ | $0, 0.25, 0.5, 0.75$ | SEF reduction probability |
| $p_{\text{extend}}$ | $0, 0.25, 0.5, 0.75$ | SEF extension probability |

**Supplementary Table S2: CHESRA Hyperparameters and their value ranges.** The parameters $n_{\text{gen}}$, $n_{\text{ind}}$, $n_{\text{hof}}$, and $l_{\text{init}}$ were pre-determined, whereas the other parameters were chosen experimentally from the considered value ranges. For the hyperparameter experiments in the Results *CHESRA's penalty parameter effectively controls the complexity of elastic energy functions*, the values of $n_{\text{gen}}$ and $n_{\text{ind}}$ were reduced for computational efficiency.

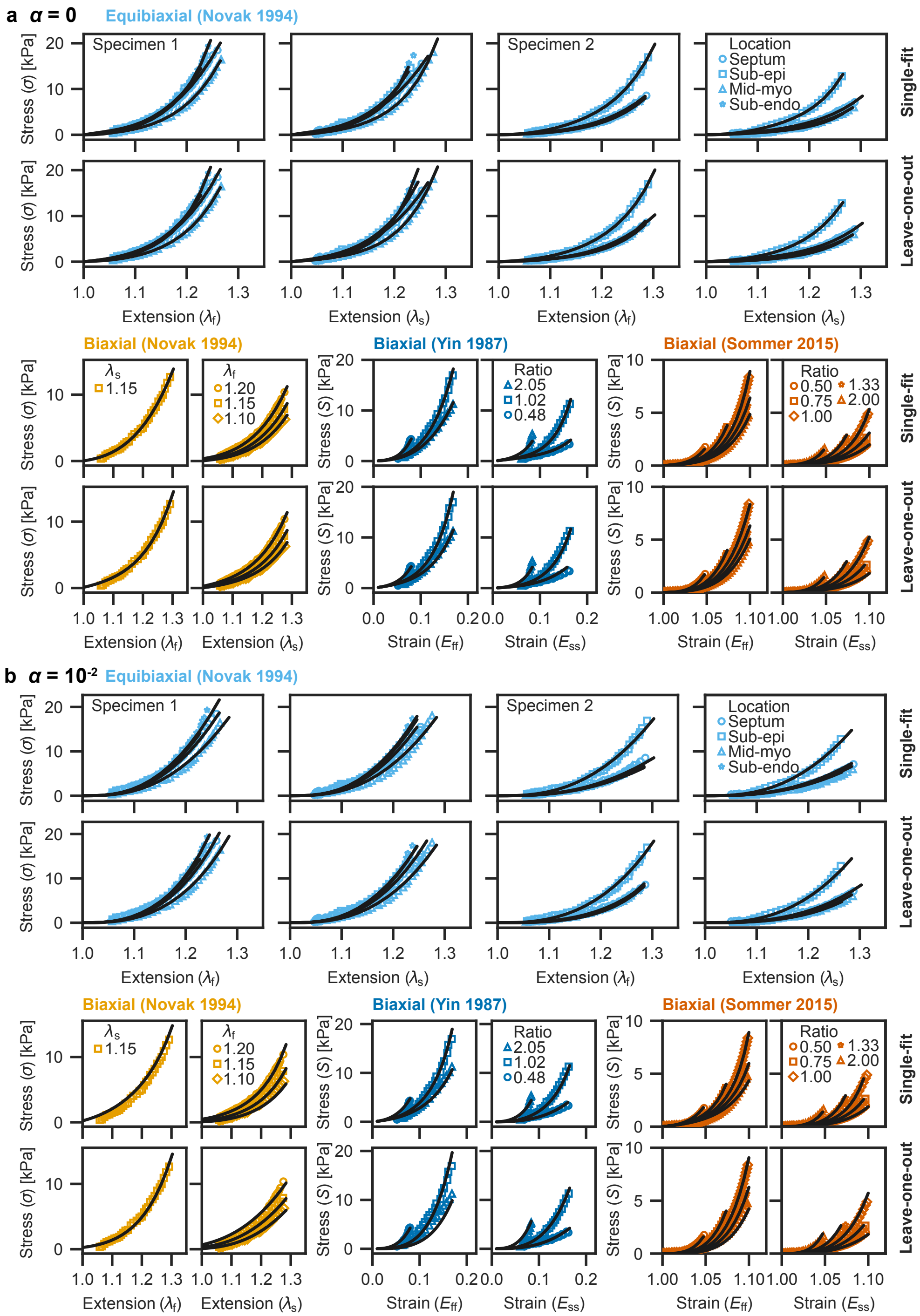

**Supplementary Figure S1: Comparison of strain energy function fits to the biaxial and equibiaxial datasets in the cross validated leave-one-out and single-fit scenarios. a** Without any function length penalty ($\alpha = 0$) **b** With the function length penalty set to $\alpha = 10^{-2}$. Solid lines represent model results whereas the markers indicate experimental data. $\sigma$ = Cauchy stress, $S$ = Second Piola-Kirchhoff stress, $E$ = Green-Lagrange strain, $\lambda$ = extension ratio. Directions (f) = fibre, (s) = sheet, (n) = sheet-normal.

| $\alpha$ | Dataset | Single-Fit | $l_{\text{eq}}$ | Leave-one-out | $l_{\text{eq}}$ |
|---|---|---|---|---|---|
| 0 | Biaxial (Yin 1987) | $p_1 p_2 \tilde{I}_{5\text{s}}(p_3 + \tilde{I}_{4\text{f}}\tilde{I}_{4\text{s}}) + p_5(p_4 + \tilde{I}_{5\text{f}})(p_8 + p_9\tilde{I}_1 + \tilde{I}_{4\text{f}} + \tilde{I}_{5\text{n}}(p_6 + p_7\tilde{I}_{5\text{s}})) + \tilde{I}_1\tilde{I}_{5\text{n}} + \tilde{I}_{4\text{s}}$ | 39 | $\tilde{I}_{8\text{fs}} p_{11}(p_{12} p_1 5(p_{14}\tilde{I}_{5\text{f}}(\tilde{I}_{8\text{fs}} + p_{13}) + \tilde{I}_{4\text{f}} + 2\tilde{I}_{5\text{s}}) + \exp(\tilde{I}_{5\text{s}})) + p_{10}\tilde{I}_{5\text{f}} + p_5 p_9(p_8 + \tilde{I}_{5\text{s}}\exp(p_6(\tilde{I}_{8\text{ns}} + \tilde{I}_1)) + \exp(p_7\tilde{I}_{5\text{f}})) + (p_4 + \tilde{I}_1)(p_1\exp(\tilde{I}_3) + p_2\tilde{I}_{5\text{f}} + p_3)$ | 63 |
| | Biaxial (Novak 1994) | $p_1\tilde{I}_{4\text{n}}(p_2 + p_3\tilde{I}_2 + \tilde{I}_{4\text{f}})\exp(\tilde{I}_{8\text{fn}}) + p_1 7(p_4 p_5\tilde{I}_{5\text{f}}(p_6 + \tilde{I}_{5\text{n}}) + p_7 + p_8 + \tilde{I}_{5\text{f}} + \tilde{I}_{5\text{s}}(p_{14} p_9(p_{11} p_{12}\tilde{I}_{4\text{n}}\tilde{I}_{4\text{s}}(p_{10} + \tilde{I}_1) + p_{13}) + p_1 5 + p_1 6\tilde{I}_{5\text{s}}))$ | 58 | $(p_1 + p_2\tilde{I}_{5\text{f}} + p_3 + p_4\tilde{I}_1)(\tilde{I}_{8\text{ns}}\tilde{I}_1 + p_5 + p_9(\tilde{I}_{8\text{fn}} p_6\tilde{I}_{4\text{f}} + \tilde{I}_{8\text{fs}} + p_7\tilde{I}_{5\text{f}} + p_8 + \tilde{I}_3 + \tilde{I}_{4\text{f}}) + \tilde{I}_1 + \tilde{I}_{4\text{s}})$ | 41 |
| | Equibiaxial (Novak 1994) | $p_{13}(p_{10} p_9\tilde{I}_1 + p_{11} + p_{12}) + (\tilde{I}_{8\text{fs}} + \tilde{I}_{4\text{n}})(p_1 p_2\tilde{I}_{4\text{f}} + p_6 + p_8(p_7 + \tilde{I}_1\tilde{I}_{5\text{s}}) + (p_3 + \tilde{I}_{4\text{n}})(p_4 + p_5\tilde{I}_{5\text{f}}))$ | 41 | $(\tilde{I}_{4\text{f}}(p_5 + 1)(p_6 + p_7(\tilde{I}_{8\text{fs}} + \exp(\tilde{I}_{8\text{fs}} + p_9\tilde{I}_1))(p_8\tilde{I}_{4\text{n}} + \tilde{I}_{5\text{f}})) + (p_1\tilde{I}_{4\text{s}} + p_2)(p_3(\tilde{I}_{8\text{fs}} + p_4) + \tilde{I}_1)$ | 38 |
| | Biaxial (Sommer 2015) | $(p_4\tilde{I}_{4\text{f}}(p_5\tilde{I}_{4\text{n}} + \tilde{I}_{4\text{f}})(\tilde{I}_1 + \tilde{I}_{4\text{s}}) + p_6(p_7 + p_8(\tilde{I}_{8\text{fn}} + \tilde{I}_1) + \tilde{I}_{4\text{f}} + 2\tilde{I}_{5\text{f}} + \tilde{I}_{5\text{s}}) + p_9)(p_1\tilde{I}_{5\text{f}} + p_2\tilde{I}_{4\text{n}} + p_3(\tilde{I}_{8\text{fn}} + \tilde{I}_1) + \tilde{I}_{4\text{n}} + \tilde{I}_{5\text{f}} + \tilde{I}_{5\text{s}})$ | 53 | $\tilde{I}_{5\text{f}}(p_5\exp(p_6\tilde{I}_3) + p_7(\tilde{I}_{8\text{fs}} p_8 + \tilde{I}_1 + \tilde{I}_{5\text{f}}) + p_9(\tilde{I}_{8\text{ns}} + \tilde{I}_{5\text{s}})) + (p_3 + \tilde{I}_1)(p_1\tilde{I}_{5\text{n}} + p_2)(p_4 + \tilde{I}_3 + \tilde{I}_{5\text{s}})$ | 40 |
| | Shear (Sommer 2015) | $(p_6 + p_8(p_9 + \tilde{I}_2)(\tilde{I}_{8\text{ns}} p_7 + \tilde{I}_2) + (I4\text{s} - 1)^4)(\tilde{I}_{8\text{ns}} + p_4 + p_5\tilde{I}_{4\text{f}} + (\tilde{I}_{8\text{fn}} p_3 + \tilde{I}_{8\text{fs}})(p_1\tilde{I}_1\tilde{I}_{5\text{f}} + p_2))$ | 39 | $(p_{14} + p_1 5 + \tilde{I}_1 + \tilde{I}_{5\text{f}})(\tilde{I}_{8\text{fs}}(\tilde{I}_{8\text{fs}} + p_2 + p_4\tilde{I}_{4\text{s}}(p_3 + \tilde{I}_{4\text{s}}) + (p_5\tilde{I}_1 + p_6 + p_7(\tilde{I}_{8\text{fs}} + \tilde{I}_{5\text{f}}))\exp(\tilde{I}_{8\text{ns}})) + p_1 + p_{13}(p_{10}\tilde{I}_1 + p_{11} + p_{12}(\tilde{I}_{8\text{fs}} + \tilde{I}_{5\text{f}})) + p_9\tilde{I}_{4\text{s}}(p_8 + \tilde{I}_{4\text{s}}))$ | 60 |
| | Shear (Dokos 2002) | $(p_1 p_2\tilde{I}_1 + (p_3 + \tilde{I}_{5\text{s}})(p_4\tilde{I}_{5\text{f}} + p_5))(\tilde{I}_{8\text{fs}} + \tilde{I}_{8\text{ns}} p_{10}(p_{11} + \tilde{I}_{4\text{s}}) + p_{12} p_{13}\tilde{I}_{5\text{f}} + p_6 + 2p_7 p_8\tilde{I}_1\tilde{I}_{4\text{n}} + p_9 + \tilde{I}_{5\text{f}})$ | 47 | $p_5(p_1 + (\tilde{I}_{8\text{ns}}\exp(\tilde{I}_{4\text{s}}\tilde{I}_{5\text{n}}) + p_2 p_3\tilde{I}_{5\text{s}} + \tilde{I}_{4\text{f}})\exp(\tilde{I}_{8\text{fn}}\tilde{I}_{8\text{fs}}))(\tilde{I}_{8\text{fs}} + p_6 + 2\tilde{I}_3 + \tilde{I}_{4\text{f}} + \tilde{I}_{5\text{f}})\exp(p_4\tilde{I}_1)$ | 40 |
| $10^{-4}$ | Biaxial (Yin 1987) | $(p_1 + \tilde{I}_{5\text{s}})(p_2 + p_3\tilde{I}_{4\text{f}}(p_7(p_6 + \tilde{I}_{4\text{f}} + \tilde{I}_{4\text{n}}) + p_8 + (p_4 + \tilde{I}_{5\text{s}})(p_5 + \tilde{I}_{5\text{s}})))$ | 27 | $p_6(\tilde{I}_{8\text{fs}} + p_4)(p_5 + \tilde{I}_1)(p_1\tilde{I}_{5\text{f}} + p_3(p_2 + \tilde{I}_1) + \tilde{I}_{5\text{s}})$ | 21 |
| | Biaxial (Novak 1994) | $p_3(p_1 + \tilde{I}_2)(p_4 + \tilde{I}_{5\text{s}})(p_2 + \tilde{I}_2 + \tilde{I}_{5\text{f}})$ | 15 | $p_5(\tilde{I}_{8\text{fs}} + p_4)(p_3 + \tilde{I}_1)(p_2(p_1 + \tilde{I}_{5\text{f}}) + \tilde{I}_{5s})$ | 17 |
| | Equibiaxial (Novak 1994) | $(p_3 + p_4\tilde{I}_{4\text{s}})(p_1 + p_2\tilde{I}_1 + \tilde{I}_{5\text{f}})$ | 13 | $p_1(\tilde{I}_{8\text{fs}} + (p_2 + \tilde{I}_{5\text{f}})(p_5\tilde{I}_1 + \exp(\tilde{I}_{8\text{fs}}))(p_3 + p_4\tilde{I}_{5\text{s}} + \tilde{I}_{5\text{f}}))$ | 22 |
| | Biaxial (Sommer 2015) | $(p_1 + \tilde{I}_{4\text{f}}\tilde{I}_{5\text{f}})(p_2 + \tilde{I}_{5\text{n}} + \exp(p_3 + p_4\tilde{I}_2))$ | 16 | $p_1 + p_5(p_6\tilde{I}_{4\text{f}} + \tilde{I}_{4\text{s}})(p_2 p_3\tilde{I}_{4\text{f}} + p_4 + \tilde{I}_1) + p_7\tilde{I}_1(\tilde{I}_{8\text{fs}} + p_8)$ | 27 |
| | Shear (Sommer 2015) | $p_8((\tilde{I}_{8\text{fs}} + p_6 + \tilde{I}_{4\text{s}})(\tilde{I}_{8\text{fn}} + p_7 + \tilde{I}_2 + \tilde{I}_3) + \exp(p_1 + \tilde{I}_3 + (\tilde{I}_{8\text{fs}} p_4 + p_5)(\tilde{I}_{8\text{ns}} p_2 + p_3\tilde{I}_1\tilde{I}_2 + \tilde{I}_{4\text{f}})))$ | 38 | $(p_1\tilde{I}_{4\text{s}} + p_2\tilde{I}_{4\text{f}} + p_3\tilde{I}_1)(\tilde{I}_{8\text{fs}} + p_4 + p_5\tilde{I}_1(\tilde{I}_{8\text{fs}} + p_6) + p_7\tilde{I}_{4\text{f}} + \tilde{I}_1)$ | 29 |
| | Shear (Dokos 2002) | $(p_1 + \tilde{I}_{5\text{f}})(\tilde{I}_{8\text{fs}} + p_2\tilde{I}_{5\text{s}}(p_3 + \tilde{I}_{5\text{n}}) + p_4 + p_5\tilde{I}_2)$ | 19 | $(p_5\tilde{I}_{4\text{f}} + p_6 + p_7\tilde{I}_1\tilde{I}_{5\text{s}})(\tilde{I}_{8\text{fs}} + p_1 p_2\tilde{I}_1 + p_3\tilde{I}_{4\text{f}} + p_4 + \tilde{I}_{4\text{s}})$ | 27 |
| $10^{-3}$ | Biaxial (Yin 1987) | $(p_3 + \tilde{I}_{4\text{f}})(p_1\tilde{I}_{5\text{s}} + p_2\tilde{I}_{4\text{f}})$ | 11 | $p_1(p_4 + \tilde{I}_{5\text{f}})(p_2 + p_3\tilde{I}_1 + \tilde{I}_{5\text{s}})$ | 13 |
| | Biaxial (Novak 1994) | $\exp[p_1\tilde{I}_{4\text{s}} + \tilde{I}_1 + \tilde{I}_{4\text{f}}]$ | 8 | $p_4(p_2 + \tilde{I}_{5\text{f}})(p_3 + \tilde{I}_1)(\tilde{I}_{8\text{fs}} + p_1 + \tilde{I}_{5\text{s}})$ | 15 |
| | Equibiaxial (Novak 1994) | $(p_1 + \tilde{I}_{5\text{f}})(p_2\tilde{I}_{5\text{s}} + p_3)$ | 9 | $(p_1 + p_2\tilde{I}_{5\text{s}})(p_3 + \tilde{I}_1)(p_4 + \tilde{I}_{5\text{f}})$ | 13 |
| | Biaxial (Sommer 2015) | $(p_1 + \tilde{I}_2)\,(p_3 + \exp[p_2\tilde{I}_{5\text{f}}])$ | 10 | $p_1(p_2 + \tilde{I}_{5\text{s}})(p_3 + \tilde{I}_1)(p_4 + \tilde{I}_{5\text{f}})$ | 13 |
| | Shear (Sommer 2015) | $\exp[p_1\tilde{I}_{4\text{f}} + \tilde{I}_2\,(p_2 + \tilde{I}_{5\text{s}})]$ | 10 | $(p_4 + \tilde{I}_{5\text{f}})(p_1\tilde{I}_1 + p_2\tilde{I}_{5\text{s}} + p_3)$ | 13 |
| | Shear (Dokos 2002) | $\tilde{I}_1 + \exp[p_2(p_1 + \tilde{I}_{5\text{s}})(\tilde{I}_1 + \tilde{I}_{5\text{f}})]$ | 12 | $(p_1 + p_2\tilde{I}_1)(p_3 + p_4\tilde{I}_{4\text{s}} + \tilde{I}_{4\text{f}})$ | 13 |
| $10^{-2}$ | Biaxial (Yin 1987) | $\exp((p_1 + \tilde{I}_1)(p_2 + \tilde{I}_{5\text{f}}))$ | 8 | $\tilde{I}_1(p_1 + p_2\tilde{I}_{4\text{f}})$ | 7 |
| | Biaxial (Novak 1994) | $\tilde{I}_{4\text{f}} + \exp(\exp(\tilde{I}_1))$ | 5 | $\tilde{I}_1(p_1 + p_2\tilde{I}_{4\text{f}})$ | 7 |
| | Equibiaxial (Novak 1994) | $p_1\tilde{I}_1$ | 3 | $\tilde{I}_1(p_1 + p_2\tilde{I}_{4\text{f}})$ | 7 |
| | Biaxial (Sommer 2015) | $p_2\exp(p_1\tilde{I}_1 + \tilde{I}_{5\text{f}})$ | 8 | $p_1\tilde{I}_1(p_2 + \tilde{I}_{4\text{f}})$ | 7 |
| | Shear (Sommer 2015) | $p_2\tilde{I}_2(p_1 + \tilde{I}_{4\text{f}})$ | 7 | $\tilde{I}_1(p_1 + p_2\tilde{I}_{4\text{f}})$ | 7 |
| | Shear (Dokos 2002) | $\exp((p_1 + \tilde{I}_{5\text{f}})\exp(\tilde{I}_{5\text{s}}))$ | 7 | $p_2\tilde{I}_1(p_1 + \tilde{I}_{4\text{f}})$ | 7 |

**Supplementary Table S3: Strain energy functions generated from cross-validation tests**. *Single-fit* refers to the function obtained from a single dataset, whereas *leave-one-out* functions were obtained from all the other datasets.

| Penalty ($\alpha$) | Length ($l_{\text{eq}}$) | Goodness of Fit ($f_{\text{GoF}}$) | Strain Energy Function ($\psi$) | Symbol |
|---|---|---|---|---|
| 0 | 24 | $7.3 \times 10^{-3}$ | $(p_1 + p_2(p_3 + \tilde{I}_{5\text{s}})(p_4(\tilde{I}_{8\text{fs}} + p_5 + \tilde{I}_{5\text{f}})\exp(p_6\tilde{I}_{5\text{f}}) + \tilde{I}_1)$ | - |
| $10^{-4}$ | 25 | $6.1 \times 10^{-3}$ | $(p_1(\tilde{I}_{8\text{ns}} + \tilde{I}_{4\text{n}} + \tilde{I}_{5\text{f}}) + p_2(p_3 + \tilde{I}_{5\text{s}}))(p_4 + p_5(\tilde{I}_{8\text{fs}} + \tilde{I}_{5\text{f}}) + \tilde{I}_1)$ | - |
| $2 \times 10^{-4}$ | 17 | $8.2 \times 10^{-3}$ | $(p_1\tilde{I}_{5\text{s}} + p_2)(p_3 + \tilde{I}_{5\text{f}})(p_4 + \tilde{I}_{5\text{f}} + \tilde{I}_{5\text{n}})$ | - |
| $3 \times 10^{-4}$ | 17 | $8.9 \times 10^{-3}$ | $p_1(p_2 + \tilde{I}_1)(p_3(\tilde{I}_{8\text{fs}} + p_4) + p_5\tilde{I}_{5\text{f}} + \tilde{I}_{5\text{s}})$ | - |
| $5 \times 10^{-4}$ | 13 | $9.8 \times 10^{-3}$ | $(p_1 + \tilde{I}_{5\text{f}})(p_2 + \tilde{I}_{5\text{s}})(p_3 + p_4\tilde{I}_1)$ | - |
| **$10^{-3}$** | **13** | **$9.7 \times 10^{-3}$** | **$p_1(p_2 + \tilde{I}_{5f})(p_3 + \tilde{I}_1)(p_4 + \tilde{I}_{5\text{s}})$** | $\psi_{\text{CH}_2}$ |
| **$5 \times 10^{-3}$** | **11** | **$1.5 \times 10^{-2}$** | **$(p_1 + \tilde{I}_1)(p_2 + p_3(\tilde{I}_{8\text{fs}} + \tilde{I}_{5\text{f}}))$** | $\psi_{\text{CH}_1}$ |
| $10^{-2}$ | 7 | $3.5 \times 10^{-2}$ | $\tilde{I}_1(p_1 + p_2\tilde{I}_{4\text{f}})$ | - |

**Supplementary Table S4: Strain energy functions learned from all experimental datasets collectively.** The two simplest functions with fully orthotropic properties ($\psi_{\text{CH}_1}$, $\psi_{\text{CH}_2}$) are marked in teal and shown with their original numbered parameters $p_1, \ldots p_4$ before we renamed them.

| Strain Energy Function | Parameters ($n_p$) | Invariants ($n_I$) | Exponentials ($n_{\exp}$) | Goodness of Fit ($f_{\text{GoF}}$) |
|---|---|---|---|---|
| $\psi_{\text{CH}_1}$ | **3** | **3** | **0** | $1.5 \times 10^{-2}$ |
| $\psi_{\text{CH}_2}$ | 4 | **3** | **0** | $9.7 \times 10^{-3}$ |
| $\psi_{\text{MA}}$ | 5 | **3** | 3 | $2.3 \times 10^{-2}$ |
| $\psi_{\text{HO}}$ | 8 | 4 | 4 | $6.5 \times 10^{-3}$ |
| $\psi_{\text{CL}}$ | 7 | - | 1 | $8.3 \times 10^{-3}$ |
| $\psi_{\text{SFL}}$ | 12 | - | 6 | $\mathbf{5.3 \times 10^{-3}}$ |
| $\psi_{\text{PZL}}$ | 12 | - | **0** | $6.0 \times 10^{-3}$ |

**Supplementary Table S5: Comparison of the CHESRA energy functions $\psi_{\text{CH}_1}$, $\psi_{\text{CH}_2}$ to energy functions from previous literature.** Quality metrics shown are the number of material parameters $n_p$, number of unique invariants $n_I$, number of exponential functions $n_{\exp}$, and the goodness of fit $f_{\text{GoF}}$ across all of the experimental datasets considered in this study. Bold numbers highlight the function(s) achieving the best result for each metric.

| $\psi_{\text{CH}_1}$ | | | $\psi_{\text{CH}_2}$ | | |
|---|---|---|---|---|---|
| **Removed Term** | $f_{\text{GoF}}$ | $f_{\text{fit}}$ | **Removed Term** | $f_{\text{GoF}}$ | $f_{\text{fit}}$ |
| - | $1.5 \times 10^{-2}$ | $7.0 \times 10^{-2}$ | - | $9.7 \times 10^{-3}$ | $2.3 \times 10^{-2}$ |
| $p_{\text{f-fs}}$ | $2.5 \times 10^{-2}$ | $7.0 \times 10^{-2}$ | $p_{\text{glob}}$ | $1.3 \times 10^{0}$ | $1.3 \times 10^{0}$ |
| $p_{\text{iso}}$ | $3.3 \times 10^{-1}$ | $3.8 \times 10^{-1}$ | $p_{\text{s-iso}}$ | $1.7 \times 10^{-1}$ | $1.8 \times 10^{-1}$ |
| $p_{\text{coup}}$ | $1.2 \times 10^{-1}$ | $1.7 \times 10^{-1}$ | $p_{\text{f-s}}$ | $2.6 \times 10^{-2}$ | $3.7 \times 10^{-2}$ |
| $\tilde{I}_1$ | $6.1 \times 10^{-1}$ | $6.6 \times 10^{-1}$ | $p_{\text{f-iso}}$ | $5.7 \times 10^{-1}$ | $5.8 \times 10^{-1}$ |
| $\tilde{I}_{8\text{fs}}$ | $2.5 \times 10^{-2}$ | $7.0 \times 10^{-2}$ | $\tilde{I}_{5\text{f}}$ | $2.3 \times 10^{-1}$ | $2.4 \times 10^{-1}$ |
| $\tilde{I}_{5\text{f}}$ | $1.9 \times 10^{-1}$ | $2.4 \times 10^{-1}$ | $\tilde{I}_1$ | $7.4 \times 10^{-2}$ | $8.5 \times 10^{-2}$ |
| | | | $\tilde{I}_{5\text{s}}$ | $2.5 \times 10^{-2}$ | $2.6 \times 10^{-2}$ |

**Supplementary Table S6: Local optimality analysis of $\psi_{\text{CH}_1}$ and $\psi_{\text{CH}_2}$ for all of the experimental datasets.** The first line shows the original goodness of fit $f_{\text{GoF}}$ and fitness $f_{\text{fit}}$. In the following lines a term is removed from the function and the remaining parameters are re-optimized. In each case, removing a term worsens the goodness of fit and fails to improve the fitness, demonstrating the local optimality of $\psi_{\text{CH}_1}$ and $\psi_{\text{CH}_2}$.

| | $\psi_{\mathrm{CH}_1}$ | | | $\psi_{\mathrm{CH}_2}$ | | | |
|---|---|---|---|---|---|---|---|
| **Dataset** | $\boldsymbol{p}_{\text{f-fs}}$ | $\boldsymbol{p}_{\text{iso}}$ | $\boldsymbol{p}_{\text{coup}}$ | $\boldsymbol{p}_{\text{glob}}$ | $\boldsymbol{p}_{\text{s-iso}}$ | $\boldsymbol{p}_{\text{f-s}}$ | $\boldsymbol{p}_{\text{f-iso}}$ |
| Biaxial (Yin 1987) | 0.033 | 10.045 | 7.978 | 2.073 | 0.804 | 0.06 | 3.323 |
| Biaxial (Novak 1994) | 0.222 | 3.68 | 0.205 | 0.043 | 1.578 | 1.971 | 2.518 |
| Biaxial (Sommer 2015) | 0.003 | 12.937 | 48.013 | 6.613 | 0.256 | 0.004 | 7.082 |
| Shear (Dokos 2002) | 0.028 | 3.132 | 15.118 | 27.385 | 0.078 | 0.085 | 0.428 |
| Shear (Sommer 2015) | 0.038 | 5.138 | 2.687 | 0.527 | 1.518 | 0.047 | 5.855 |
| Equibiaxial (Novak 1994) | | | | | | | |
| **sub-endo.** sp. 1 | 0.001 | 6.556 | 0.361 | 0.001 | 1.045 | 445.234 | 1.204 |
| **sub-endo.** sp. 2 | 0.328 | 1.312 | 0.088 | 0.005 | 6.835 | 1.29 | 13.145 |
| **mid-myo.** sp. 1 | 0.001 | 4.736 | 0.133 | 0.001 | 1.9 | 112.846 | 2.085 |
| **mid-myo.** sp. 2 | 0.603 | 1.482 | 0.065 | 0.002 | 14.866 | 1.356 | 36.397 |
| **sub-epi.** sp. 1 | 30.823 | 7.018 | 0.001 | 0.017 | 10.216 | 1.676 | 11.19 |
| **sub-epi.** sp. 2 | 0.001 | 4.049 | 0.06 | 0.006 | 15.656 | 1.726 | 15.376 |
| **mid-sept.** sp. 1 | 73.933 | 6.06 | 0.001 | 0.003 | 22.027 | 3.873 | 27.21 |
| **mid-sept.** sp. 2 | 0.587 | 1.885 | 0.038 | 0.001 | 19.731 | 2.979 | 27.219 |

**Supplementary Table S7: Material parameters of $\psi_{\mathrm{CH}_1}$ and $\psi_{\mathrm{CH}_2}$ fitted to experimental datasets.** The last eight rows specify differing specimens and sample locations (sub-endocardium, mid-myocardium, sub-epicardium, or mid-septum) for the equibiaxial dataset of Novak et al. [2].

| | | Line Search | | Nelder-Mead | | |
|---|---|---|---|---|---|---|
| **Scenario** | **Energy Function** | Optimal $\boldsymbol{p}$ | $\mathcal{L}_{VOL}$ (mL$^2$) | Lower Bound $\boldsymbol{p}$ | Optimal $\boldsymbol{p}$ | $\mathcal{L}_{VOL}$ (mL$^2$) |
| **ex vivo** | $\psi_{\mathrm{CH}_1}$ | 1.5 | 74.58 | 0.15 | $p_{\text{f-fs}} = 0.233$<br>$p_{\text{iso}} = 1.399$<br>$p_{\text{coup}} = 2.720$ | 15.2 |
| | $\psi_{\mathrm{CH}_2}$ | 0.6 | 13.29 | 0.06 | $p_{\text{glob}} = 0.596$<br>$p_{\text{s-iso}} = 0.710$<br>$p_{\text{f-s}} = 0.397$<br>$p_{\text{f-iso}} = 0.613$ | 10.0 |
| | $\psi_{\mathrm{MA}}$ | 2.5 | 43.24 | 0.875 | $\mu = 0.885$<br>$a_{\mathrm{f}} = 1.073$ kPa<br>$b_{\mathrm{f}} = 5.712$<br>$a_{\mathrm{n}} = 0.875$ kPa<br>$b_{\mathrm{n}} = 7.528$ | 10.07 |
| | $\psi_{\mathrm{HO}}$ | 2.5 | 140.56 | 0.625 | $a = 0.625$ kPa<br>$b = 2.482$<br>$a_{\mathrm{f}} = 0.625$ kPa<br>$b_{\mathrm{f}} = 1.349$<br>$a_{\mathrm{s}} = 1.53$ kPa<br>$b_{\mathrm{s}} = 47.165$<br>$a_{\mathrm{fs}} = 2.130$ kPa<br>$b_{\mathrm{fs}} = 44.021$ | 13.28 |
| **in vivo** | $\psi_{\mathrm{CH}_1}$ | 0.8 | 16.42 | 0.24 | $p_{\text{f-fs}} = 1.137$<br>$p_{\text{iso}} = 0.731$<br>$p_{\text{coup}} = 0.414$ | 13.94 |
| | $\psi_{\mathrm{CH}_2}$ | 2.0 | 128.4 | 0.045 | $p_{\text{glob}} = 0.045$<br>$p_{\text{s-iso}} = 0.045$<br>$p_{\text{f-s}} = 5.701$<br>$p_{\text{f-iso}} = 2.586$ | 63.47 |
| | $\psi_{\mathrm{MA}}$ | 2.0 | 34.45 | 0.1 | $\mu = 1.940$<br>$a_{\mathrm{f}} = 2.465$ kPa<br>$b_{\mathrm{f}} = 0.476$<br>$a_{\mathrm{n}} = 2.664$ kPa<br>$b_{\mathrm{n}} = 2.634$ | 11.86 |
| | $\psi_{\mathrm{HO}}$ | 1.0 | 96.46 | 0.5 | $a = 5.007$ kPa<br>$b = 1.0$<br>$a_{\mathrm{f}} = 2.732$ kPa<br>$b_{\mathrm{f}} = 3.069$<br>$a_{\mathrm{s}} = 2.003$ kPa<br>$b_{\mathrm{s}} = 1.018$<br>$a_{\mathrm{fs}} = 1.529$ kPa<br>$b_{\mathrm{fs}} = 2.442$ | 9.63 |

**Supplementary Table S8: Optimal volume square loss $\mathcal{L}_{VOL}$ and parameter values $\boldsymbol{p}$ obtained during the two step pressure-free geometry estimation.** The lower bounds were empirically determined during the Nelder-Mead step to ensure numerical stability.

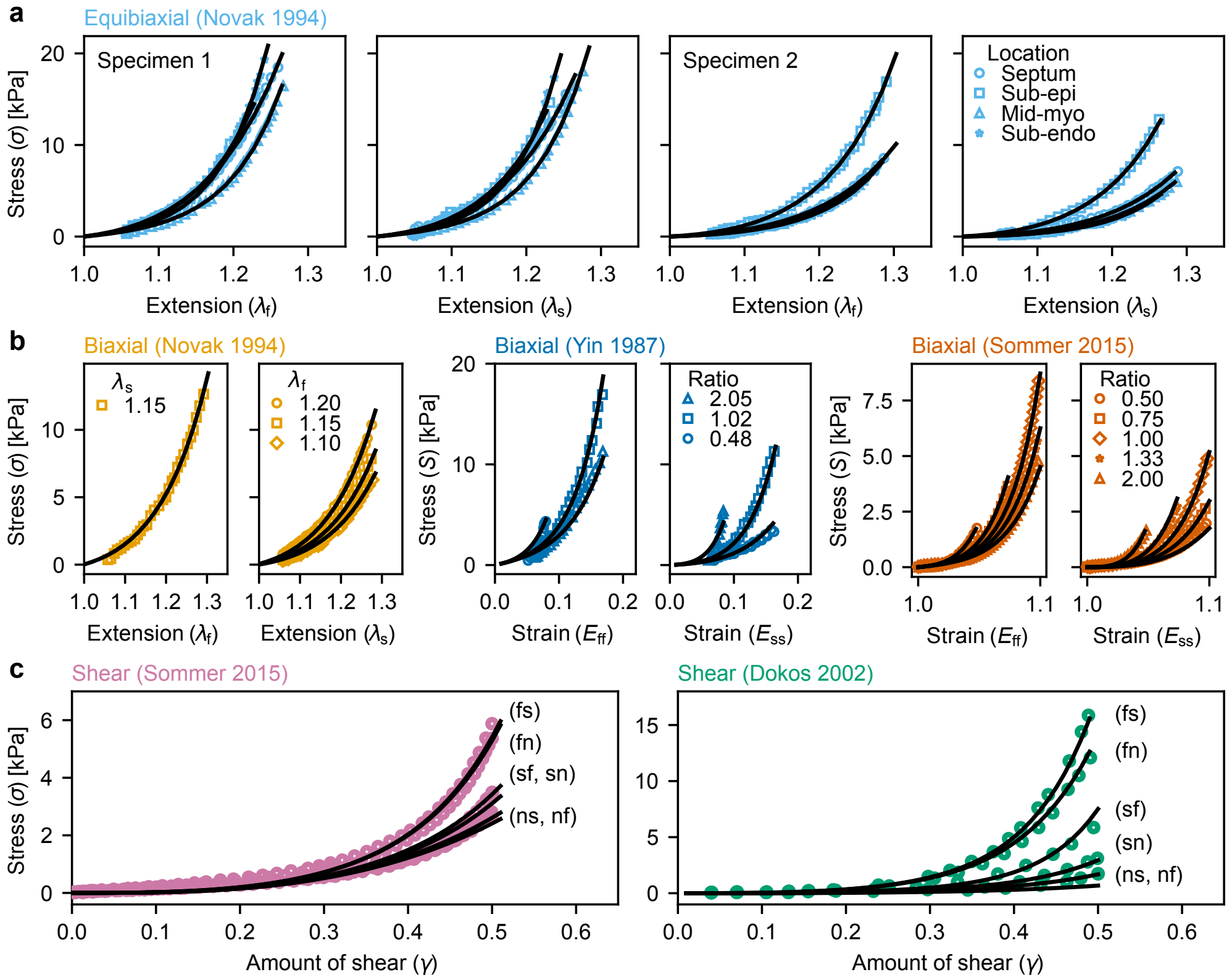

**Supplementary Figure S2: The CHESRA function $\psi_{\mathrm{CH}_2}$ closely fits all of the various biaxial and shear datasets considered in our study. a** Equibiaxial stretch [2], **b** biaxial stretch [1–3], and **c** triaxial shear [3, 4] data. Solid lines represent model results whereas the markers indicate experimental data. $\sigma$ = Cauchy stress, $S$ = Second Piola-Kirchhoff stress, $E$ = Green-Lagrange strain, $\lambda$ = extension ratio, $\gamma$ = simple shear angle. The labels (fs, fn, sf, sn, ns, nf) indicate the direction of shear, with the first coordinate indicating the normal of the tissue cube face that is translated by the shear, and the second the shearing direction in which that face is shifted. Directions (f) = fibre, (s) = sheet, (n) = sheet-normal.

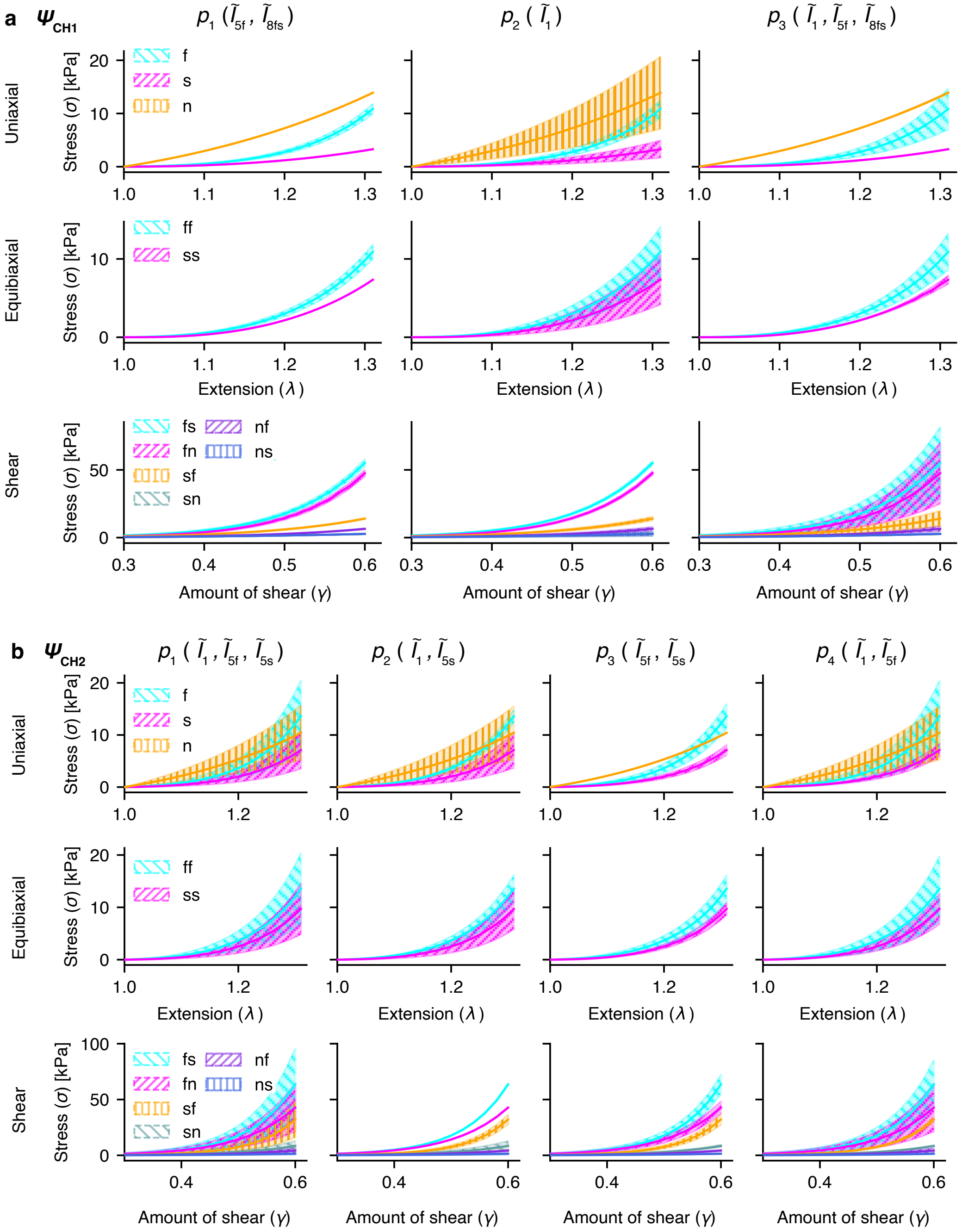

**Supplementary Figure S3: Sensitivity study of individual CHESRA strain energy function (SEF) parameters in uniaxial, biaxial and simple shear loading scenarios. a** Sensitivity study of the parameters $(p_1, p_2, p_3)$ of the SEF $\psi_{\mathrm{CH}_1}$. **b** Sensitivity study of the parameters $(p_1, p_2, p_3, p_4)$ of the SEF $\psi_{\mathrm{CH}_2}$. To assess parameter sensitivity, we varied each material parameter independently from its baseline value (solid lines) by ±50% (hashed area) and evaluated the resulting changes in the Cauchy stress $(\sigma)$ under uniaxial, equibiaxial, and shear loading. As baseline parameters for the uni- and equibiaxial tests, we used the fitted material parameters from the equibiaxial dataset of Novak et al. (specimen 2, mid-myocardium) [2], whereas for the shear tests we employed the parameters fitted to the data of Dokos et al. [4] (see Supplementary Table S7). The invariants $(\tilde{I}_1, \tilde{I}_{5\mathrm{f}}, \tilde{I}_{5\mathrm{s}}, \tilde{I}_{8\mathrm{fs}})$ listed in brackets for each parameter indicate which invariants are affected by that parameter in the expanded form (S2) of the CHESRA SEFs $\psi_{\mathrm{CH}_1}$ and $\psi_{\mathrm{CH}_2}$. $\lambda$ = extension ratio, $\gamma$ = simple shear angle. The labels (fs, fn, sf, sn, ns, nf) indicate the direction of shear, with the first coordinate indicating the normal of the tissue cube face that is translated by the shear, and the second the shearing direction in which that face is shifted. Directions (f) = fibre, (s) = sheet, (n) = sheet-normal.

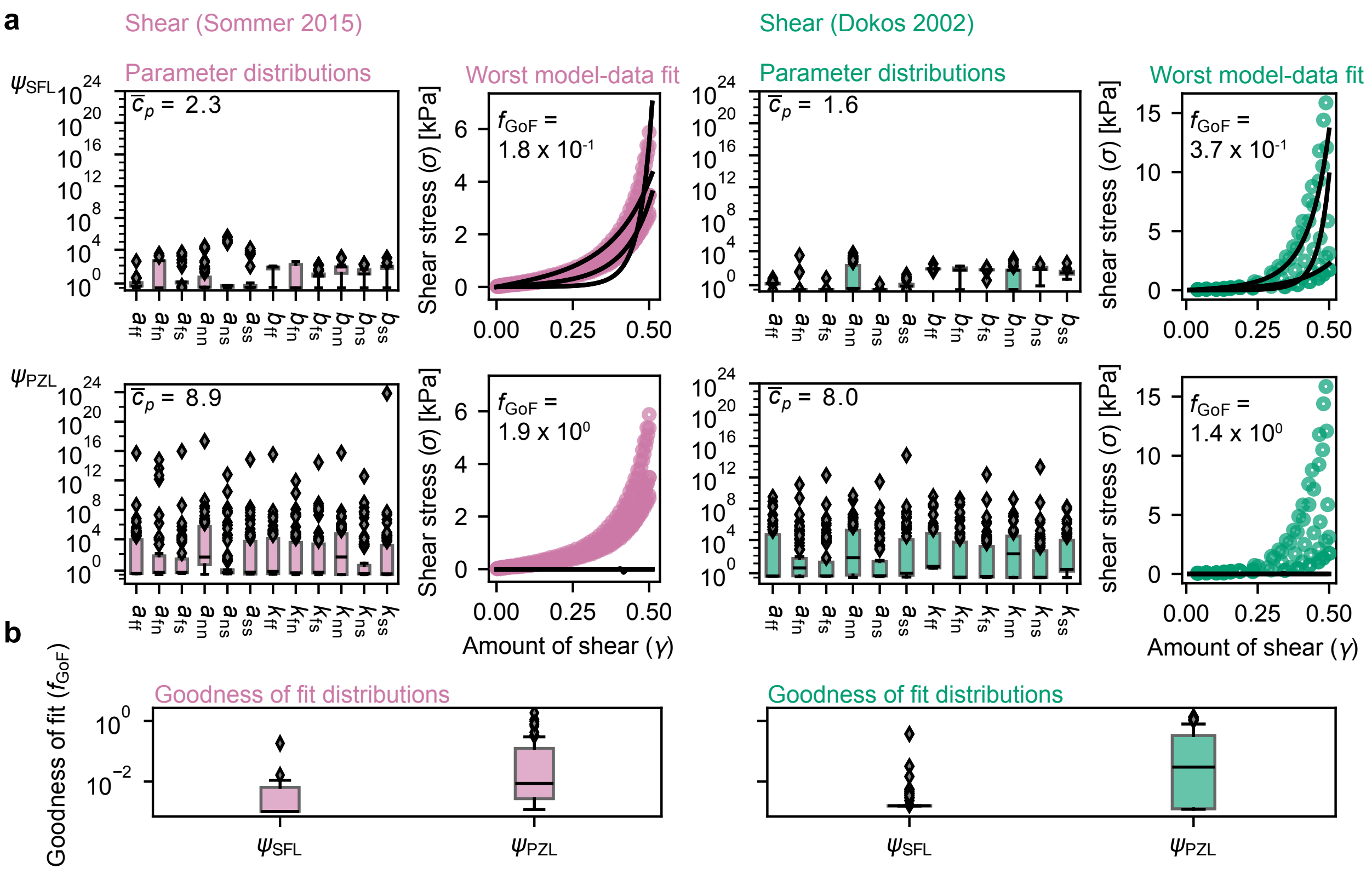

**Supplementary Figure S4: Results of the parameter variability benchmark with experimental tissue data for the strain energy functions $\psi_{PZL}$ and $\psi_{SFL}$. a** Distributions of estimated material parameters resulting from least-squares fitting to shear data with 100 random initializations, along with corresponding worst model-data fits. In the model-data fit panels solid lines represent model results whereas the markers indicate experimental data. **b** Corresponding goodness of fit ($f_{GoF}$) distributions. $\bar{c}_p$ = average model parameter coefficient of variation, $\sigma$ = Cauchy stress, $\gamma$ = simple shear angle.

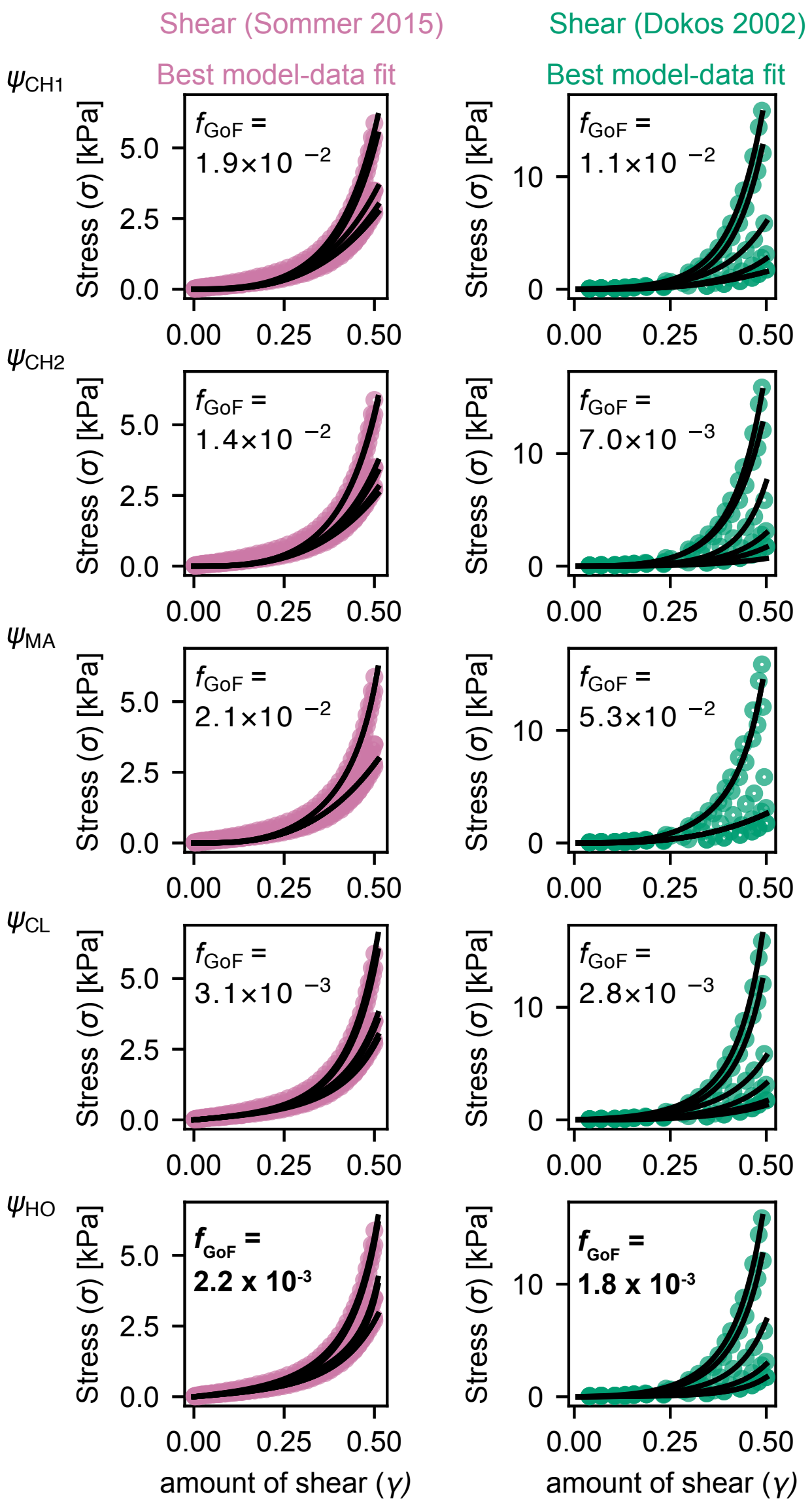

**Supplementary Figure S5: Best fitting parameterizations of the strain energy functions ($\psi_{\mathrm{CH_1}}, \psi_{\mathrm{CH_2}}, \psi_{\mathrm{MA}}, \psi_{\mathrm{CL}}, \psi_{\mathrm{HO}}$) in the parameter variability benchmark with experimental tissue data.** Solid lines represent model results whereas the markers indicate experimental data. Bold face indicates the best fitting result among the strain energy functions. $\sigma$ = Cauchy stress, $\gamma$ = simple shear angle, $f_{\mathrm{GoF}}$ = goodness of fit.

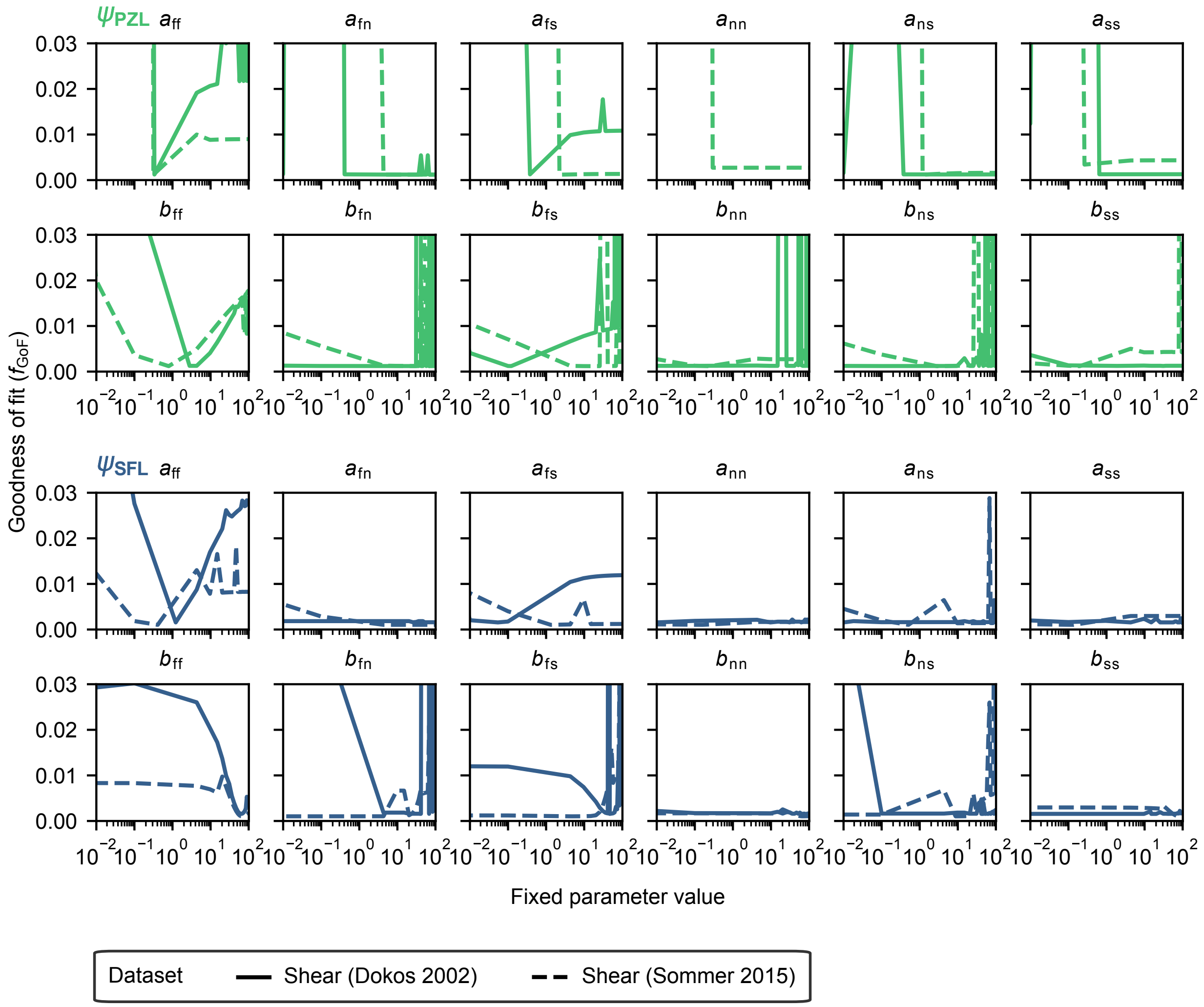

**Supplementary Figure S6: Goodness of fit ($f_{GoF}$) landscapes for the strain energy functions $\psi_{PZL}$ and $\psi_{SFL}$.** In each panel the indicated parameter is fixed for a range of values while the other parameters are re-optimized.

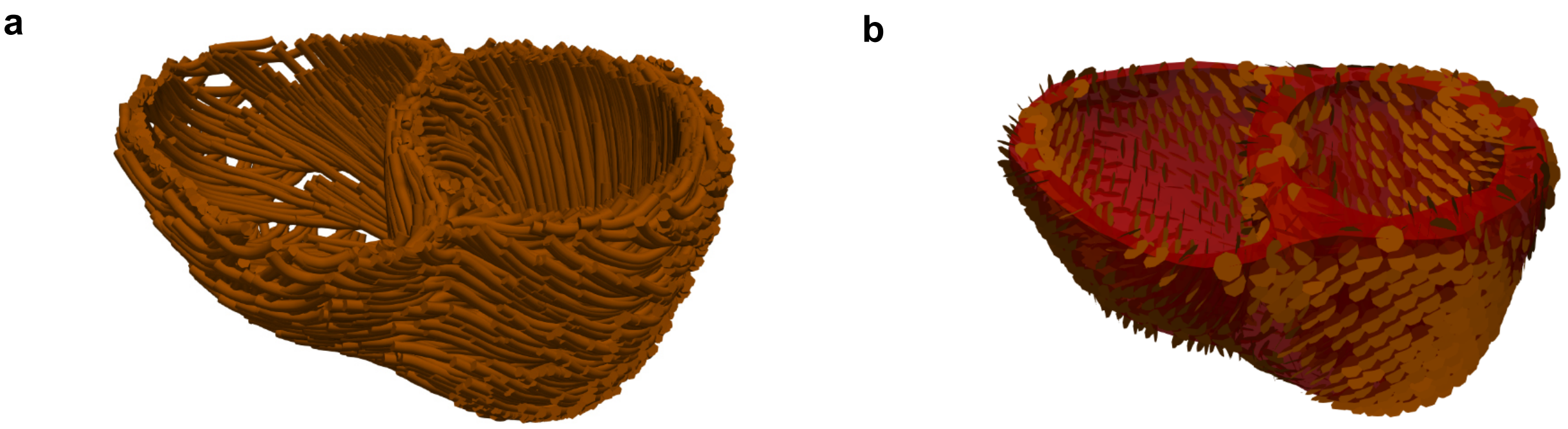

**Supplementary Figure S7: Synthetic cardiac muscle microstructures. a** Tube representation of the local fiber direction in the patient left ventricle geometry. **b** Circle representation of the local muscle sheet plane orientations.

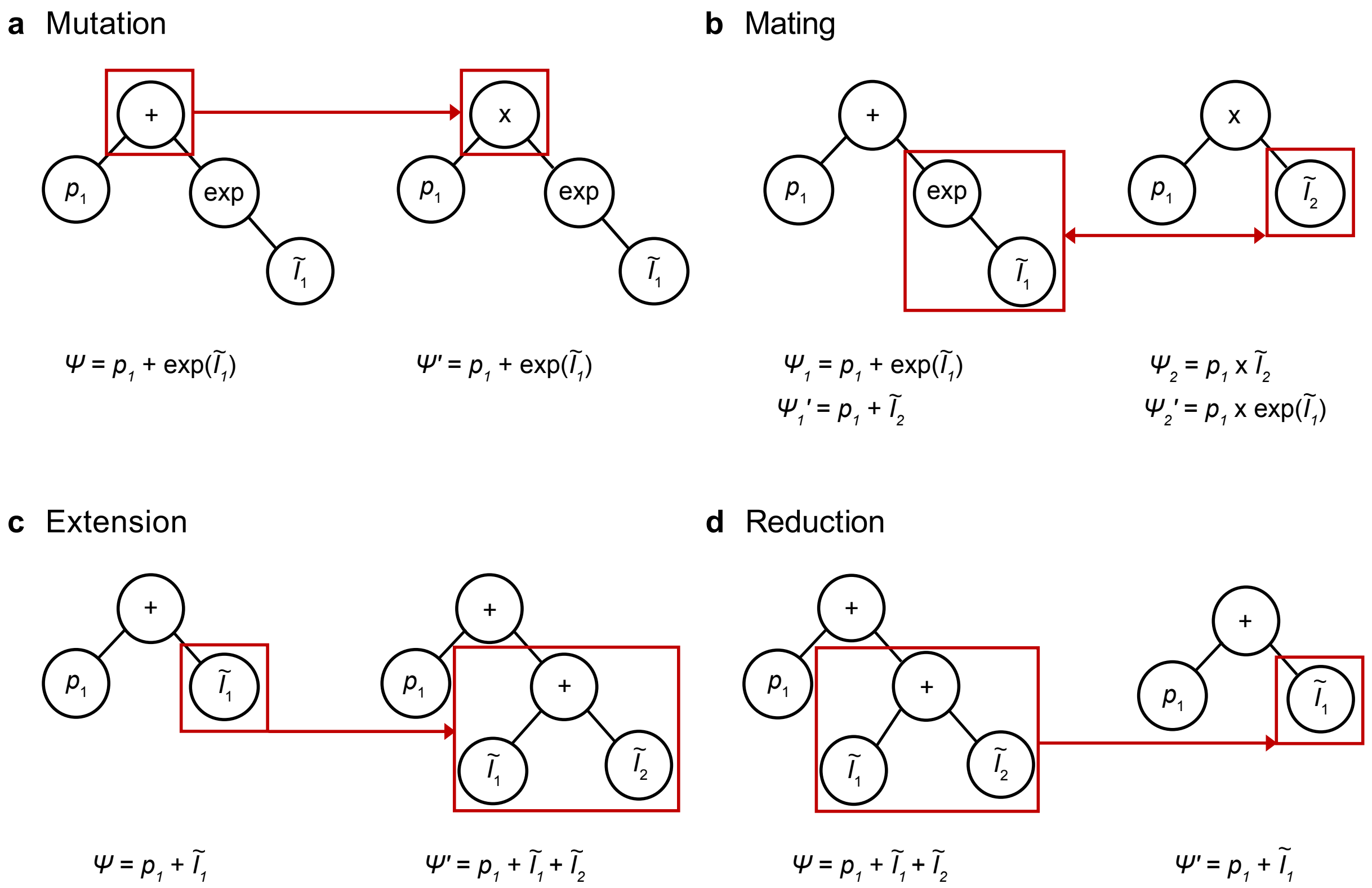

**Supplementary Figure S8: Example evolutionary changes on a strain energy function (SEF) with symbol $\psi$ and strain invariant arguments $(\tilde{I}_1, \tilde{I}_2)$ and material parameter $p_1$. a** The SEF $\psi$, represented by a function tree, is *mutated* to $\psi'$ by choosing a random node of the tree and replacing it with a random node of the same type. **b** Two parent SEFs, $\psi_1$ and $\psi_2$, *mate* to generate two child SEFs, $\psi_1'$ and $\psi_2'$, by swapping two random sub-trees. **c** $\psi$ is *extended* by adding a random function tree to a randomly selected node. **d** $\psi$ is *reduced* by deleting a random sub-tree of the corresponding function tree.

## References

1. Yin, F. C., Strumpf, R. K., Chew, P. H. & Zeger, S. L. Quantification of the mechanical properties of noncontracting canine myocardium under simultaneous biaxial loading. *J. Biomech.* **20,** 577–589 (1987).
2. Novak, V. P., Yin, F. & Humphrey, J. Regional mechanical properties of passive myocardium. *J. Biomech.* **27,** 403–412 (1994).
3. Sommer, G. *et al.* Biomechanical properties and microstructure of human ventricular myocardium. *Acta Biomater.* **24,** 172–192 (2015).
4. Dokos, S., Smaill, B. H., Young, A. A. & LeGrice, I. J. Shear properties of passive ventricular myocardium. *Am. J. Physiol. Heart Circ. Physiol.* **283,** H2650–H2659 (2002).